\documentclass[a4paper,11pt]{article}
\pdfoutput=1
\usepackage{jheppub}
\usepackage{amssymb,amsmath}
\usepackage{mathtools}
\usepackage[normalem]{ulem}
\usepackage[utf8x]{inputenc}
\usepackage{slashed}
\usepackage{graphicx}
\usepackage{multirow}
\usepackage{csquotes} 
\usepackage{comment}
\usepackage{mathrsfs}
\usepackage{float}
\usepackage{enumerate}
\usepackage{cleveref}
\usepackage{hhline}
\numberwithin{equation}{section} 

\definecolor{ForestGreen}{RGB}{0, 155, 85}

\newcommand{\blue}[1]{\textcolor{blue}{#1}}
\newcommand{\green}[1]{\textcolor{ForestGreen}{#1}}
\newcommand{\black}[1]{\textcolor{black}{#1}}
\usepackage{colortbl}

\preprint{
\begin{minipage}{5cm}
\small
\flushright
KEK-TH-2583\\ KYUSHU-HET-276
\end{minipage}}

\title{Autoencoder-Driven Clustering of Intersecting D-brane Models via Tadpole Charge}

\author{Keiya Ishiguro$^{1}$,} 
\author{Satsuki Nishimura$^{2}$ and}
\author{Hajime Otsuka$^{2}$} 
\affiliation{
$^1$ Graduate University for Advanced Studies (Sokendai), 1-1 Oho, Tsukuba, Ibaraki 305-0801, Japan\\
$^2$ Department of Physics, Kyushu University, 744 Motooka, Nishi-ku, Fukuoka 819-0395, Japan\\
}
\emailAdd{ishigu@post.kek.jp}
\emailAdd{nishimura.satsuki@phys.kyushu-u.ac.jp}
\emailAdd{otsuka.hajime@phys.kyushu-u.ac.jp}

\abstract{
We study the well-known type IIA intersecting D-brane models on the $T^6/(\mathbb{Z}_2 \times \mathbb{Z}'_2)$ orientifold via a machine-learning approach. We apply several autoencoder models with and without positional encoding to the D6-brane configurations satisfying certain concrete models described in Ref. \cite{Gmeiner:2005vz} and attempt to extract some features which the configurations possess. We observe that the configurations cluster in two-dimensional latent layers of the autoencoder models and analyze which physical quantities are relevant to the clustering. As a result, it is found that tadpole charges of hidden D6-branes characterize the clustering. We expect that there is another important factor because a checkerboard pattern in two-dimensional latent layers is observed in the clustering.
}
\makeatletter
\gdef\@fpheader{}
\makeatother

\begin{document}

\maketitle

\section{Introduction}
\label{sec:Introduction}
String Landscape on the toroidal backgrounds has been studied for decades.
Due to its geometrically simple and controllable structure, the background functions as a desirable environment for studying string phenomenology.
As for the type IIA intersecting D6-brane models \cite{Berkooz:1996km} (see for reviews, e.g., \cite{Blumenhagen:2006ci, Ibanez:2012zz, Blumenhagen:2005mu, Lust:2007kw}),
enormous effort has been made (see for early works on phenomenological models, \cite{Ibanez:2001nd, Cvetic:2001nr, Cvetic:2001tj, Cvetic:2002pj, Aldazabal:2000cn, Aldazabal:2000dg}) while there is another viewpoint from statistics of the landscape.
Indeed, the degrees of freedom of the D-brane configurations are numerous.
Let us focus on the $T^6/(\mathbb{Z}_2 \times \mathbb{Z}'_2)$ orientifold, which is one of the most intensively studied. 
Although the number of physically-independent supersymmetric (SUSY) D-brane configurations is shown to be finite, the exact amount which is counted in Ref. \cite{Loges:2022mao} reaches as many as 134 billion.
Following the discussion in Ref. \cite{Gmeiner:2005vz}, there will be $\mathcal{O}(10)$ phenomenological models (at least) but the probability for three families of quarks and leptons is one in a billion.
It is difficult to find phenomenologically viable D-brane configuration explicitly because of the problem being the Diophantine type, and thus the statistical study is performed in Ref. \cite{Douglas:2006xy} while deterministic brute-force approaches e.g., Ref. \cite{He:2021gug} has also been made.

One may consider the possibility to study the problem via the machine-learning approaches considering their recent powerful applications in various fields.
Indeed, there exist fruitful studies using \textit{reinforcement learning} method in Ref. \cite{Halverson:2019tkf} and \textit{genetic algorithm} in Ref. \cite{Loges:2021hvn} on the intersecting D-brane models to discover new phenomenological configurations.
Thanks to the many efforts, it should be true that the understanding of the phenomenological intersecting D-brane models on the $T^6/(\mathbb{Z}_2 \times \mathbb{Z}'_2)$ orientifold became profound.

On the other hand, we wonder what the characteristic features of the phenomenological configurations are.
We wonder if the difference between configurations that satisfy some phenomenological models and ones that do not is reflected in some fundamental physical quantities in string compactifications aside from the resulting generations of chiral fermions etc. 
The statistical analyses adopted in the literature e.g., Refs. \cite{Douglas:2006xy, Gmeiner:2005vz} exactly aimed to characterize the phenomenological configurations.
The results allow us to focus on the region with efficient phenomenological configurations, whereas the statistical methods pose a difficulty in choosing which quantity to focus on.
This is an issue we address in this paper.
Thanks to the developments of machine learning methods, we may use them to extract the features of phenomenological configurations.
\textit{Autoencoder} is a candidate to perform the feature extraction.
It is an unsupervised neural network whose middle layer, which is called the latent layer, has small dimensions.
Therefore, a well-trained autoencoder will contain the features of input data through clustering in its latent layer, 
and we would like to analyze the output of the latent layer to clarify which physical quantities are weighted by the neural network.
Note that, in the context of type IIA intersecting D-branes, the autoencoder method was used to visualize the distributions of configuration in Refs. \cite{Li:2019nvi,Mansha:2023kwq}.
Our aim in this paper is to discover the differences between clusters.
This method has also been adopted on different string vacua. (See for the applications in heterotic string theory, Refs. \cite{Mutter:2018sra,Otsuka:2020nsk,Deen:2020dlf,Escalante-Notario:2022fik} where clustering is also observed.) 

Unfortunately, there are several limitations in this approach and we have to consider workarounds.
\begin{enumerate}
    \item   The difficulties in realizing phenomenological configurations\\
            We cannot generate an exact phenomenological configuration due to its small probability of existence.
            We instead weaken the meaning of ``phenomenological'', and we will consider only the configurations where the generations of quarks and leptons are the same in the concrete models on which statistics is discussed in Ref. \cite{Gmeiner:2005vz}.
    \item   The curse of dimensionality\\
            The dimensions of the latent layer should be very small (e.g., two) to analyze the hidden features.
            Naively, the phenomenological configurations and not phenomenological ones are expected to cluster on different islands in the latent layer.
            The possibility that we can extract features depends on the possibility that we can analyze the clustering in practice. 
            The curse of dimensionality will appear for higher dimensional latent layers, and we have to fix the dimensions to be small.
            Thus, it is impossible to extract all relevant features.
            Instead, we will obtain the most important features characterizing the configurations.
    \item   Robustness \\
            Since there are huge degrees of freedom to architect neural networks, the robustness might be problematic in the machine learning approaches.
            We will address this kind of problem via introducing \textit{positional encoding} technique, which was invented in \textit{Transformer} \cite{vaswaniAttentionAllYou2017} as a central architecture of famous \textit{ChatGPT}, in a modified way.
    \item   Interpretability \\
            It is very difficult to make the autoencoder models explain which physical quantities (or in general, their non-linear combinations) are important in a direct way.
Thus we again have to study statistics on the latent layer, but the region to search becomes narrowed.
\end{enumerate}

Keeping in mind these limitations, we perform the machine learning with the autoencoder models in the following. 
It turns out that the expected clustering which distinguishes the phenomenological configurations from the non-phenomenological ones does not occur, but their parts are gathered in a nontrivial way and those sets form the clusters in the latent layer. 

This paper is organized as follows: In Section \ref{sec:2}, we will briefly review ingredients of the type IIA intersecting D-brane models on $T^6/(\mathbb{Z}_2 \times \mathbb{Z}'_2)$ which is needed for our purpose.
We cite the concrete models in Ref. \cite{Gmeiner:2005vz} there.
In Section \ref{sec:computer search}, we will briefly review the method to generate D-brane configurations in Ref. \cite{Gmeiner:2005vz} and summarize our datasets.
Then we define the architectures of the autoencoder models in Section \ref{sec:autoencoder and preprocessing}.
We prepare multiple neural networks so that we can ensure the robustness of our discussion.
We firstly focus on Model 1 (which will be explained later) and compare the results of the various methods in Section \ref{sec:Comparison of Methods} to determine the latent layers that we try to extract the learned features.
This procedure is explained in Section \ref{sec:feature extraction of Model 1} in a transparent way.
In Section \ref{sec:feature extraction of Models 2 and 3}, we examine whether the feature found in the previous section is also extracted in the case of Models 2 and 3. 
Finally, we conclude our paper in Section \ref{sec:con}.

\section{Configurations of Type IIA intersecting D6-branes}
\label{sec:2}

In this section, we briefly review type IIA superstring theory with intersecting D6-branes on toroidal orientifolds (for detailed reviews, see e.g., Refs. \cite{Blumenhagen:2006ci, Ibanez:2012zz}). 
After summarizing generic ingredients of type IIA intersecting D6-brane models in Section \ref{sec:2_1}, we review intersecting D6-branes systems on the $T^6/(\mathbb{Z}_2 \times \mathbb{Z}'_2)$ orientifold in Section \ref{sec:2_2}.
In Section \ref{sec:concrete models}, we briefly cite the concrete phenomenological models considered in Ref. \cite{Gmeiner:2005vz}.
Throughout this paper, our notation is based on Ref. \cite{Gmeiner:2005vz}.

\subsection{Model building constraints}
\label{sec:2_1}
There are some consistency conditions which intersecting D6-branes models have to satisfy.
These consistency conditions are nicely summarized in Refs. \cite{Blumenhagen:2006ab, Gmeiner:2005vz} for example.
\paragraph{RR Tadpole Cancellation Condition} \mbox{}\\
The consistent D-brane models require the existence of orientifold planes (O-planes) whose Ramond-Ramond (RR) charges have the opposite sign of that of D-branes. Geometrically, these O-planes are defined as fixed regions under the orientifold action and possible orbifold actions. The orientifold action $\Omega{\cal R}$ is determined by an anti-holomorphic involution on the extra 6-dimensional (6d) space ${\cal R}$ and the worldsheet parity $\Omega$. Then RR charge cancellation condition, which is called the tadpole cancellation condition, is expressed in terms of the homology of D6-branes and O6-planes as
\begin{align}
   \sum_{a} N_a \left( \left[ \Pi_a \right] +  \left[ \Pi_a' \right] \right) = 4 \left[ \Pi_{O6} \right],
\end{align}
where $N_{a}$ denotes the number of D6-branes in the $a$-th stack of D-branes, and $\left[ \Pi_a' \right]$ denotes the mirror cycle of D-branes wrapping $\left[ \Pi_a \right]$ with respect to the O6-planes.

In general, an open string having one of its two endpoints on $a$-th stacks of D6-branes leads a set of four-dimensional (4d) chiral fermions which transform as $\mathbf{N}_{a}$ fundamental representation (with $+1$ $U(1)$ charge) of $U(N_a)$ group.
Hence we can construct massless states in bifundamental representations by assuming open strings which localize at an intersecting region of the $a$- and $b$-th stacks.
The multiplicity of the chiral fermions in the representation is given by the intersecting number $I_{ab}$ of two D6-brane cycles, where $I_{ab}$ becomes a homological quantity $I_{ab} = \left[ \Pi_a \right] \circ \left[ \Pi_b \right]$.
In orientifolds, there are in general intersections of D6-branes and O6-planes, and different representations emerge where $I_{a O6}$ also affects their multiplicities.
The chiral massless spectrum is determined by these representations and the corresponding intersection numbers.
We summarize the general structure of the spectrum in Table \ref{tab:chiralspectrum}.

\begin{table}[h]
\centering
\begin{tabular}{c|c}
\hline
Representation                       & Multiplicity                     \\\hline
$(\mathbf{N}_a, \overline{\mathbf{N}}_b)$ & $I_{ab}$                          \\
$(\mathbf{N}_a, \mathbf{N}_b)$       & $I_{ab'}$                         \\
${\rm Sym}^2_{a}$ (symmetric)   & $\frac{1}{2}(I_{aa'} - I_{aO6})$ \\
$\Lambda^2_{a}$ (anti-symmetric)     & $\frac{1}{2}(I_{aa'} + I_{aO6})$ \\\hline
\end{tabular}
\caption{Multiplicities of massless states in each chiral representation.}
\label{tab:chiralspectrum}
\end{table}

Of course, we need some geometric details of these cycles to calculate the intersection numbers. On the manifold ${\cal M}$ with the special geometry, it has a homology group $H_3({\cal M})$ and it is decomposed into orientifold even($+$) and odd($-$) groups $H_3^{\pm}({\cal M})$ and there is a symplectic basis ${\alpha_I, \beta_J}$ with $\alpha_I \in H_3^{+}({\cal M})$ and $\beta_J \in H_3^{-}$ ($I, J \in \{0, 1, \dots, h^{2,1}\}$) which satisfies $\alpha_I \circ \beta_J = \delta_{IJ}, \alpha_{I} \circ \alpha_{J} = 0$ and $\beta_{I} \circ \beta_{J} = 0$. Then generic three-cycles on ${\cal M}$ can be expanded as
\begin{align}
    \left[ \Pi_a \right] = \vec{X}_a \vec{\alpha} + \vec{Y}_{a} \vec{\beta},~
    \left[ \Pi_a' \right] = \vec{X}_a \vec{\alpha} - \vec{Y}_{a} \vec{\beta},~
    \left[ \Pi_{O6} \right] = \frac{1}{2} \vec{L} \vec{\alpha}, \label{eq:three-cycle expansion}
\end{align}
with $\vec{X_a} \vec{\alpha} = \sum_{I=0}^{h^{2, 1}} X^I_a \alpha_I$. Here, the coefficients are considered as integers. Then, for instance, the intersection number between $I_{ab}$ is given by
\begin{align}
    I_{ab} = \left[ \Pi_a \right] \circ \left[ \Pi_b \right] = \vec{X}_{a} \vec{Y}_b - \vec{Y}_{a} \vec{X}_b. 
\end{align}

\paragraph{SUSY conditions}\mbox{}\\
If one consider stable configurations of D6-branes preserving 4d ${\rm N} = 1$ SUSY, D6-branes should wrap special Lagrangian cycles which satisfy \cite{Becker:1995kb}:
\begin{align}
    J|_{\Pi_a} = 0,~  {\rm Im}(e^{i\phi_a} \Omega_3)|_{\Pi_a} = 0.
\end{align}
Here, $J$ and $\Omega$ are respectively the K\"ahler form and the unique holomorphic 3-form of Calabi-Yau orientifolds ${\cal M}$, and $|_{\Pi_a}$ represents a restriction on them to the cycle $\Pi_a$. $\phi_a$ is a phase which determines preserved supersymmetry by corresponding D-branes, and when SUSY is conserved it appears in the volume of the cycle $\int_{\Pi_a} {\rm Re}\left(e^{i \phi_a} \Omega_3 \right)$, but this should be zero for all $a$ because of the presence of O6-planes. 
In the orientifold that we will study in this paper, the special Lagrangian (sLag) condition ensures that D6-branes on the special Lagrangian cycles correspond to straight lines on each $T^2$ of factorizable toroidal orbifolds.

Using the special geometry, the complex-structure moduli and derivatives of the prepotential are defined as
\begin{align}
    U_I = \int_{\alpha_I} \Omega_3,
    \qquad
    F_I = \int_{\beta_I} \Omega_3.
\end{align}
Then the sLag condition falls into
\begin{align}
    \vec{Y}_a \vec{F}(U) = 0, 
    \qquad
    \vec{X}_a \vec{U} > 0.
\end{align}
Note that the elements of $\vec{F}$ are $F_I$ ($\vec{F}$ is not a prepotential).

\paragraph{K-theory charge cancellation}\mbox{}\\
It is known that there are additional hidden charges of D-branes which can be described by K-theory \cite{Witten:1998cd}.
They are $\mathbb{Z}_2$-valued quantities and their cancellations are equivalent to the constraint that global $Sp(2N)$ anomalies are canceled out \cite{Witten:1982fp}. Since a $Sp(2) \simeq SU(2)$ brane\footnote{In this paper, $Sp(2N)$ denotes the unitary symplectic group $USp(2N)$.} behaves as a probe of the anomalies \cite{Uranga:2000xp}, for all possibilities to introduce such D-branes, we have to ensure that the K-theory charge is canceled out.
Schematically, the form of the condition is
\begin{align}
    \sum_{a} N_a \left[ \Pi_a \right] \circ \left[ \Pi_{{\rm probe~}Sp(2)} \right] \equiv 0 ~ ({\rm mod }\,2).
\end{align}
Such D6-branes with $Sp(2N)$ gauge factor are ones on top of O6-planes.
Thus we have to find out some cycles which are invariant under each of the orientifold times the orbifold actions.
Of course, the cycle structures which support such D-branes vary depending on background orbifold actions. We will come back to this point later in the specific orbifold.  

\paragraph{Gauge Anomalies and Massless U(1)}\mbox{}\\
The tadpole cancellation condition ensures that there is no cubic $SU(N)$ gauge anomaly, and the Green-Schwartz mechanism also does the absence of mixed- and cubic-abelian anomalies. Hence the theory becomes consistent, but in these cancellations, one can find that anomaly-free $U(1)$s can be massive via the Stueckelberg mechanism. From a phenomenological point of view, one needs to know a massless linear combination of these factors. The condition for the existence of such a linear combination (see \cite{Ibanez:2001nd} for more detailed discussion)
\begin{align}
    U(1)_{\rm massless}= \sum_a f_a U(1)_a
\end{align}
is given by
\begin{align}
    \sum_a f_a N_a \vec{Y}_a = 0,
\end{align}
where $\vec{Y}_a \propto \left[\Pi_a\right] \circ \vec{\alpha}$.

\subsection{The $T^6/(\mathbb{Z}_2 \times \mathbb{Z}'_2)$ orientifold}
\label{sec:2_2}
\paragraph{Background Orbifold Geometry}\mbox{}\\
Throughout this paper, we deal with a type IIA factorizable toroidal orientifold theory, which is based on the $T^6/(\mathbb{Z}_2 \times \mathbb{Z}'_2)$ orbifold.
The theory is characterized by the following actions on its complex coordinates;
\begin{align}
    \theta: (z_1,z_2, z_3) \rightarrow (-z_1, -z_2, z_3), 
    \qquad
    \omega: (z_1, z_2, z_3) \rightarrow (z_1, -z_2, -z_3). \label{eq:orbifold action}
\end{align}
There is an additional possibility to include the discrete torsion but we consider the case without it.
In type IIA theory, the orientifold action $\Omega {\cal R}$ acts as
\begin{align}
    \Omega {\cal R}: (z_1, z_2, z_3) \rightarrow (\bar{z}_1, \bar{z}_2, \bar{z}_3). \label{eq:orientifold action}
\end{align}
With a complex-coordinate basis ${e^x_i, e^y_i}$, the coordinate of $i$-th Torus is expanded as $z_i = x_i e^x_i + y_i e^y_i$ and imposed on the periodic conditions of the underlying torus as $z_i \sim z_i + k e^x_i + l e^y_i ~ (k, l \in \mathbb{Z})$. From the viewpoint of complex plane $\mathbb{C}$, the basis is defined as
\begin{align}
    e^x_i = 2\pi (R^x_i + i b_i R^y_i),
    \qquad
    e^y_i = i 2\pi R^y_i.
\end{align}
Indeed, one can show that there are only two possibilities for values of $b_i$ to be compatible with the orientifold action (\ref{eq:orientifold action}) and they must be $0$ or $\frac{1}{2}$ for each torus. The torus with nonzero $b_i$ is called tilted. We will consider only the non-tilted case in the following discussion, while it is pointed out that configurations on non-tilted tori dominate those on tilted tori \cite{Gmeiner:2005vz} in the models which are learned by autoencoder models in the following.\footnote{In addition to the tilt, there are additional degrees of freedom of including shifts in the orbifold actions (\ref{eq:orbifold action}), which we ignore in this paper.}. 

To compute the chiral spectrum, we only need to know the homology classes corresponding to one-cycles wrapped by D6-branes on each torus. Since we consider the orbifolding, we need to treat these cycles carefully not to include ill-defined ones. 
We will come back to this point later, but let us start by introducing cycles on the underlying tori. More details and complete discussions can be found in Refs. \cite{Blumenhagen:2005tn, Blumenhagen:2006ab, Forste:2010gw}.
The one-cycles are parameterized by a pair of coprime numbers called the wrapping number of the one-cycles $(n, m)$. We denote the basic one-cycles of the $i$-th torus by $\{[\pi_{2i-1}], [\pi_{2i}]\}$ which wrap along $e^x_i$ and $e^y_i$, respectively. In other words, the one-cycle on the $i$-th torus with the wrapping number $(n_i, m_i)$ is expressed as $n_i [\pi_{2i-1}] + m_i [\pi_{2i}]$. 
Note that these one-cycles behave as Grassmann numbers: $[\pi_{2i-1}] \circ [\pi_{2j}] = - [\pi_{2j}] \circ [\pi_{2i-1}] = -\delta_{ij}, ~{\rm others} = 0$.
The overall three-cycle is a multiplication of these three one-cycles. For example, let us consider a three-cycle wrapped by D6-branes of the $a$-th stack. Then, the three-cycle is
\begin{align}
    \left[\Pi_a^T \right] = \bigotimes_{i=1}^3 (n^a_i [\pi_{2i-1}] + m^a_i [\pi_{2i}]),
\end{align}
and quantities that were introduced in the last section are written in terms of the wrapping numbers. Here, the subscript $T$ means that the corresponding cycle is defined on underlying tori (not orbifold).  
Before moving to details, let us introduce a useful basis of one-cycles. The orientifold action (\ref{eq:orientifold action}) acts on the basic one-cycles as
\begin{align}
    \Omega {\cal R}: [\pi_{2i-1}] \rightarrow [\pi_{2i-1}] - 2b_i [\pi_{2i}],
    \qquad
    \Omega {\cal R}: [\pi_{2i}] \rightarrow - [\pi_{2i}]. 
\end{align}
One can define a new $\Omega {\cal R}$ even one-cycle as
\begin{align}
    \left[\tilde{\pi}_{2i-1}\right] = [\pi_{2i-1}] - b_i [\pi_{2i}]
\end{align}
and change the basis to $\left\{ \left[\tilde{\pi}_{2i-1}\right], [\pi_{2i}] \right\}$. The new basis satisfies the same algebra: $\left[\tilde{\pi}_{2i-1}\right] \circ [\pi_{2j}] = - [\pi_{2j}] \circ \left[\tilde{\pi}_{2i-1}\right] = -\delta_{ij}, ~{\rm others} = 0$.
Note that $[\pi_{2i-1}]$s are along $e^x_i$ for both cases $b_i = {0, \frac{1}{2}}$ and $[\pi_{2i-1}]$s along the real axis of the underlying $i$-th torus. Of course, changing to a new basis does not affect the overall three-cycles;
\begin{align}
    \left[\Pi_a^T \right] = \bigotimes_{i=1}^3 (n^a_i [\pi_{2i-1}] + m^a_i [\pi_{2i}]) = \bigotimes_{i=1}^3 \left(n^a_i [\tilde{\pi}_{2i-1}] + \tilde{m}^a_i [\pi_{2i}] \right)
\end{align}
with $\tilde{m}_i = m_i + b_i n_i$. 
The quantity $2^{{\rm Total~number~of~tilted~tori}}$ which is given by $\hat{b} = \prod_{i=1}^{3} (1-b_i)^{-1}$ will appear frequently.
The problem is that $\tilde{m}_i$s are not integers in general. Since we would like to express the three-cycle in the manner of Eq. (\ref{eq:three-cycle expansion}) with integer coefficients, we need to modify the definition of basic cycles again. 
Originally, the symplectic basis $\{\alpha_I, \beta_J\}$ is defined as
\begin{align}
\begin{aligned}
    \alpha^0 &= [\pi_1] [\pi_3] [\pi_5],            & \beta^i &= [\pi_{2i}] [\pi_{2j-1}] [\pi_{2k-1}],  \\
    \alpha^i &= [\pi_{2i-1}] [\pi_{2j}] [\pi_{2k}], & \beta^0 &= [\pi_{2}] [\pi_{4}] [\pi_{6}], 
\end{aligned}
\end{align}
and corresponding coefficients $\vec{X}$ and $\vec{Y}$ are
\begin{align}
\begin{aligned}
    X^0 &= n_1 n_2 n_3,  & Y^i &= m_i n_j n_k, \\
    X^i &= n_i m_j m_k,  & Y^0 &= m_1 m_2 m_3. 
\end{aligned}
\label{eq:defXY}
\end{align}
Here $\{i, j, k\}$ is a cyclic permutation of $\{1, 2, 3\}$ (e.g. $X^2= n_2 m_3 m_1$) and $\left[ \Pi_a \right] = \vec{X}_a \vec{\alpha} + \vec{Y}_{a} \vec{\beta}$ holds. 
On the other hand, we define new three-cycle basis $\{\tilde{\alpha}_I, \tilde{\beta}_I\}$ as
\begin{align}
\begin{aligned}
    \tilde{\alpha}^0 &= [\tilde{\pi}_1'] [\tilde{\pi}_3'] [\tilde{\pi}_5'], & \tilde{\beta}^i &= -[\pi_{2i}'] [\tilde{\pi}_{2j-1}'] [\tilde{\pi}_{2k-1}'], \\
    \tilde{\alpha}^i &= -[\tilde{\pi}_{2i-1}'] [\pi_{2j}'] [\pi_{2k}'], & \tilde{\beta}^0  &= [\pi_{2}'] [\pi_{4}'] [\pi_{6}'],
\end{aligned}
\end{align}
with $[\tilde{\pi}_{2i-1}'] = \hat{b}^{-\frac{1}{3}} [\tilde{\pi}_{2i-1}]$ and $[\pi_{2i}'] = \hat{b}^{-\frac{1}{3}} [\pi_{2i}]$. We denote $\hat{\vec{\alpha}} = (\tilde{\alpha}^0,\tilde{\alpha}^1,\tilde{\alpha}^2,\tilde{\alpha}^3)$.
Correspondingly, we can define integer coefficients as
\begin{align}
\begin{aligned}
    \hat{X}^0 &= \hat{b} n_1 n_2 n_3,  & \hat{Y}^i &= -\hat{b} \tilde{m}_i n_j n_k, \nonumber \\
    \hat{X}^i &= -\hat{b} n_i \tilde{m}_j \tilde{m}_k,  & \hat{Y}^0 &= \hat{b} \tilde{m}_1 \tilde{m}_2 \tilde{m}_3.
\end{aligned}
\end{align}
Note that $\hat{X}^0$ is always a multiple of $\hat{b}$.
The modifications change the algebra slightly;
\begin{align}
    \left[\tilde{\pi}_{2i-1}'\right] \circ [\pi_{2j}'] &= - [\pi_{2j}'] \circ \left[\tilde{\pi}_{2i-1}'\right] = - \hat{b}^{-\frac{2}{3}} \delta_{ij}, \nonumber \\
    \tilde{\alpha}_I \circ \tilde{\beta}_J &= - \tilde{\beta}_I \circ \tilde{\alpha}_J = \hat{b}^{-2} \delta_{IJ}.
\end{align}
Then the overall three-cycle is expressed as
\begin{align}
    \left[\Pi_a^T \right] = \hat{\vec{X}}_a \hat{\vec{\alpha}} + \hat{\vec{Y}}_a \hat{\vec{\beta}}, 
\end{align}
and the intersection number $I^T_{ab}$ on the tori becomes
\begin{align}
\begin{aligned}
    I^T_{ab} &= \left[\Pi^T_a\right] \circ \left[\Pi^T_b \right] = {\vec{X}}_a {\vec{Y}}_b - {\vec{Y}}_a {\vec{X}}_b = \hat{b}^{-2} (\hat{\vec{X}}_a \hat{\vec{Y}}_b - \hat{\vec{Y}}_a \hat{\vec{X}}_b), \\
    I^T_{aa'} &=  - {\vec{X}}_a {\vec{Y}}_a - {\vec{Y}}_a {\vec{X}}_{a} = - \hat{b}^{-2} (\hat{\vec{X}}_a \hat{\vec{Y}}_a + \hat{\vec{Y}}_a \hat{\vec{X}}_a). 
\end{aligned}
\end{align}
The following algebraic relations are turned out to be useful later:
\begin{align}
\begin{aligned}
    \hat{X}^I \hat{Y}^I &= \hat{X}^J \hat{Y}^J ~ (\forall I, J), & & & & \\
    \hat{X}^I \hat{X}^J &= - \hat{Y}^K \hat{Y}^L, ~& \hat{X}^L (\hat{Y}^L)^2 &= - \hat{X}^I \hat{X}^J \hat{X}^K, ~& \hat{Y}^L (\hat{X}^L)^2 = -\hat{Y}^I \hat{Y}^J \hat{Y}^K,
\end{aligned}
\label{eq:hatXYalgebra}
\end{align}
with $I, J, K, L$ on the second line is a permutation of $\{0, 1, 2, 3\}$. These relations also hold for $X, Y$s without hats.

\paragraph{O6-planes}\mbox{}\\
Furthermore, we need the information on cycles wrapped by the O6-planes. Since the O-plane is defined as fixed loci under an orientifold action times for each generator of orbifold actions, the O6-planes in this background are classified into four types. The corresponding three-cycles are
\begin{align}
\begin{aligned}
       \Omega {\cal R} &: 2^3 [\tilde{\pi}_1][\tilde{\pi}_3][\tilde{\pi}_5], & \Omega {\cal R} \omega &= -2^{3- 2(b_2 + b_3)} [\tilde{\pi}_1] [\pi_4] [\pi_6], \\
    \Omega {\cal R} \theta &: -2^{3- 2(b_1+b_2)} [\pi_2][\pi_4][\tilde{\pi}_5], & \Omega {\cal R} \theta \omega &= -2^{3- 2(b_1 + b_2)} [\pi_2] [\tilde{\pi}_3] [\pi_6], \label{eq:orientifoldinvariantcycles}
\end{aligned}
\end{align}
and we define the total homology class $[\Pi_{O6}^T]$ as
\begin{align}
    \left[\Pi^T_{O6}\right] =  2^3( [\tilde{\pi}_1][\tilde{\pi}_3][\tilde{\pi}_5]-2^{- 2(b_2 + b_3)} [\tilde{\pi}_1] [\pi_4] [\pi_6]-2^{- 2(b_1+b_2)} [\pi_2][\pi_4][\tilde{\pi}_5]-2^{- 2(b_3 + b_1)} [\pi_2] [\tilde{\pi}_3] [\pi_6]).
\end{align}
Note that we included the number of orientifold planes in the above definition and the relative signs of each three-cycle are set to make the O6-plane charge negative. After rescaling, we obtain $\left[\Pi_{O6}^T\right] = \frac{1}{2} \hat{\vec{L}} \hat{\vec{\alpha}}$ with
\begin{align}
    \hat{\vec{L}} = 16 \left(\hat{b}, \frac{1}{1-b_i}\right) \label{eq:defL}.
\end{align}
In addition, the intersection number $I^T_{a O6}$ is given by
\begin{align}
    I^T_{a O_6} = -2^3 \hat{b}^{-2} \left(\hat{Y}^0_a \hat{b} + \sum_{i} \hat{Y}^i_a \frac{1}{1-b_i}\right).
\end{align}

\paragraph{D-branes and O-planes on the orbifold}\mbox{}\\
As we mentioned, we have to distinguish cycles on the bulk of the orbifold from ones on the underlying tori.
Since there are three orbifold images for each factorizable three-cycle, we sum up these contributions to construct the corresponding three-cycle on the orbifold. For the $T^6/{(\mathbb{Z}_2 \times \mathbb{Z}'_2)}$ case, the images have a common set of wrapping numbers.
Thus the bulk three-cycle $\left[\Pi^{\rm B}_{a}\right]$ which corresponds to the $\left[\Pi^{\rm T}_{a}\right]$ is given by
\begin{align}
    \left[\Pi^{\rm B}_{a}\right] = 4 \left[\Pi^{\rm T}_{a}\right] = 4 \bigotimes_{i=1}^3 \left(n^a_i [\tilde{\pi}_{2i-1}] + \tilde{m}^a_i [\pi_{2i}] \right). \label{eq:bulkthreecycle}
\end{align}
The definition of intersection numbers between bulk three-cycles changes slightly as
\begin{align}
    I^{\rm B}_{ab} = I_{a} \circ_{\rm B} I_{b} = \frac{1}{4} \left(4 \left[\Pi^{\rm T}_{a}\right] \right) \circ \left(4 \left[\Pi^{\rm T}_{b}\right] \right) = 4 I^{\rm T}_{ab}, \label{eq:bulkintersectingnumber}
\end{align}
where $\circ_{\rm B}$ denotes the topological intersection on the orbifold. The additional $\frac{1}{4}$ factor is included so as not to overcount intersecting loci on the orbifold.  

Since the intersection number $I^B_{ab}$ (\ref{eq:bulkintersectingnumber}) indicates that the form of bulk three-cycles (\ref{eq:bulkthreecycle}) does not span the whole of $H_3(T^6/(\mathbb{Z}_2 \times \mathbb{Z}'_2), \mathbb{Z})$, this implies that there exist shorter three-cycles on the orbifold. Indeed, the fractional (bulk) three-cycle is defined as
\begin{align}
    \left[\Pi^{\rm F}_{a}\right] = \frac{1}{2} \left[\Pi^{\rm B}_{a}\right] = 2 \left[ \Pi_a^{\rm T} \right]
\end{align}
and satisfies $\left[\Pi^{\rm F}_{a}\right] \circ_B \left[\Pi^{\rm F}_{b}\right] = \left[\Pi^{\rm T}_{a}\right] \circ \left[\Pi^{\rm T}_{b}\right]$. 
Hence, the most general cycles
\footnote{We assumed the case without the discrete torsion. Hence, exceptional cycles which correspond to the twisted sector do not appear.} 
on the orbifold are the fractional bulk cycles. We call D6-branes wrapping these cycle fractional D6-branes, and these are fundamental ingredients in model buildings with intersecting D6-branes. 

For the O6-planes, the definition that they are fixed loci of the orientifold and orbifold images freezes their dynamical degree of freedom. Hence, we do not need to distinguish the O6-planes from their images. Thus the O6-plane three-cycle on the tori $\left[ \Pi_{O6}^{\rm T} \right]$ and that on the orbifold $\left[ \Pi_{O6} \right]$ are identical. That is, 
\begin{align}
    \left[ \Pi_{O6} \right] = \frac{1}{4} \left[ \Pi_{O6}^{\rm B} \right] =\left[ \Pi_{O6}^{\rm T} \right].
\end{align}

\paragraph{Explicit form of the intersection numbers}
As mentioned, the chiral massless spectrum is determined by the intersection numbers.
These numbers should be computed on the orbifold (which the subscript ${\rm O}$ below means), and it is convenient to summarize the result of intersection numbers:  
\begin{align}
\begin{aligned}
    I^{{\rm O}}_{ab} &= \left[ \Pi^{\rm F}_a \right] \circ_B \left[ \Pi^{\rm F}_b \right] = \frac{1}{4} I^{\rm B}_{ab} = I^{\rm T}_{ab} = {\vec{X}}_a {\vec{Y}}_b - {\vec{Y}}_a {\vec{X}}_b = \hat{b}^{-2} (\hat{\vec{X}}_a \hat{\vec{Y}}_b - \hat{\vec{Y}}_a \hat{\vec{X}}_b),\\
    I^{{\rm O}}_{aa'} &=  - \hat{b}^{-2} (\hat{\vec{X}}_a \hat{\vec{Y}}_a + \hat{\vec{Y}}_a \hat{\vec{X}}_a) = - 2 \hat{b}^{-2} \hat{\vec{X}}_a \hat{\vec{Y}}_a, \\
    I^{{\rm O}}_{aO6} &= \frac{1}{2} I^{\rm T}_{aO6} = -2^2 \hat{b}^{-2} \left(\hat{Y}^0_a \hat{b} + \sum_{i} \hat{Y}^i_a \frac{1}{1-b_i}\right).
\end{aligned}
\end{align}

\paragraph{Tadpole cancellation condition}\mbox{}\\
Using the results of the previous paragraph, we can compute the explicit form of the tadpole cancellation condition on this background. All three-cycles in the condition should be defined on the orbifold. Particularly the D6-branes are fractional D6-branes. Thus
\begin{align}
    \sum_a N_a \hat{X}^I_a = \frac{1}{2}\hat{L}^I, \label{eq:tccorbifold}
\end{align}
or equivalently, 
\begin{align}
\begin{aligned}
    \sum_{a} N_a \hat{X}^0_a &= 8 \hat{b},\\
    \sum_{a} N_a \hat{X}^i_a &= \frac{8}{1-b_i}.   
\end{aligned}
\end{align}

\paragraph{SUSY conditions}\mbox{}\\
On the $T^6/{(\mathbb{Z}_2 \times \mathbb{Z}_2^\prime)}$ orientifold, the integrals $\int_{\alpha, \beta} \Omega_3$ leads to
\begin{align}
    \sum_{I} \hat{Y}^I \frac{1}{U^I} = 0,
    \qquad
    \sum_{I} \hat{X}^I {U_I} > 0. \label{eq:susycond}
\end{align}
\paragraph{K-theory constraints}\mbox{}\\
As mentioned, the probe $Sp(2)$ D6-branes wrap cycles which are invariant under the orientifold times the orbifold actions. These $\Omega {\cal R} \theta^k \omega^l ~ k, l \in \{0, 1\}$-invariant cycles\footnote{Here, the normalization is determined so that a stack of branes wrapping $N$ times these invariant cycles gives $USp(2N)$ gauge group. Similarly, $N$ D6-branes in general leads a $U(N)$ group.} can be read off from Eq. (\ref{eq:orientifoldinvariantcycles}):
\begin{align}
\begin{aligned}
    \left[\Pi_{\Omega {\cal R}}^{\rm T} \right] &= \hat{b} \left[ \tilde{\pi_1} \right] \left[ \tilde{\pi_3} \right] \left[ \tilde{\pi_5} \right], \\
    \left[\Pi_{\Omega {\cal R} \omega }^{\rm T} \right] &= -\frac{1}{1-b_1}  \left[ \tilde{\pi}_1 \right] \left[ \pi_4 \right] \left[ \pi_6 \right], \\
    \left[\Pi_{\Omega {\cal R} \theta \omega }^{\rm T} \right] &= -\frac{1}{1-b_2}  \left[ \pi_2 \right] \left[ \tilde{\pi}_3 \right] \left[ \pi_6 \right], \\
    \left[\Pi_{\Omega {\cal R} \theta }^{\rm T} \right] &= -\frac{1}{1-b_3}  \left[ \pi_2 \right] \left[ \pi_4 \right] \left[ \tilde{\pi}_5 \right]. \label{eq:Sp(2)cycles}
\end{aligned}
\end{align}
These cycles correspond to wrapping numbers:
\begin{align}
\begin{aligned}
     \left[\Pi_{\Omega {\cal R}}^{\rm T} \right] &\leftrightarrow \hat{X}^0 = \hat{b}^2, \hat{X}^{1, 2, 3} =0, \\
     \left[\Pi_{\Omega {\cal R} \omega }^{\rm T} \right] &\leftrightarrow \hat{X}^1 = \frac{\hat{b}}{1-b_1}, \hat{X}^{0, 2, 3} =0, \\
     \left[\Pi_{\Omega {\cal R} \theta \omega }^{\rm T} \right] &\leftrightarrow \hat{X}^2 = \frac{\hat{b}}{1-b_2}, \hat{X}^{0, 1, 3} =0, \\
     \left[\Pi_{\Omega {\cal R} \theta }^{\rm T} \right] &\leftrightarrow \hat{X}^3 = \frac{\hat{b}}{1-b_3}, \hat{X}^{0, 1, 2} =0,
\end{aligned}
\end{align}
and $Y^{I}= 0 ~ ({}^\forall I)$ for all these invariant cycles.
Note that the prefactors enter in because the orientifold invariant cycles are closed in different ways for the untilted and tilted cases. 
Since $Sp(2)$ cycles are invariant, the definition of the fractional cycles and the torus cycles are the same. The K-theory constraints $\sum_{a} N_{a} \left[ \Pi_a^{\rm F} \right] \circ  \left[\Pi_{Sp(2)}^{\rm F} \right] \equiv 0~({\rm mod ~}{2}) $ becomes
\begin{align}
    \sum_a N_a \hat{Y}^0_a &\equiv 0, \\
    (1-b_j) (1-b_k) \sum_a N_a \hat{Y}^i_a &\equiv 0
\end{align}
with $\{i, j, k\}$ being a cyclic permutation of $\{1, 2, 3\}$. 
In the formulae, it is sufficient to consider that the index $a$ denotes only $U(N_a)$ branes since $\hat{Y}^I_{Sp(2)} = 0$ does not contribute to the summations.

\paragraph{Eliminating exotic chiral matters}\mbox{}\\
For simplicity, we do not impose the non-existence of exotic matters.
From a phenomenological viewpoint, however, one might have to eliminate the exotic chiral matters which transform as (anti-)symmetric representations.
The multiplicities of massless states on Table \ref{tab:chiralspectrum} imply that $I^{\rm O}_{aa'} = I^{\rm O}_{a O_6}$ and the multiplicity of $\Lambda^2_a$ becomes $I^{\rm O}_{aa'} = -2 \hat{b}^{-2} \hat{\vec{X}}_a \hat{\vec{Y}}_a$.
As we will explain later, D6-brane configurations consistent with SUSY are classified into three types: no vanishing, two vanishing, and three vanishing $\hat{X}^I$s.
For the three vanishing case which corresponds to $Sp(2)$ branes, the multiplicity is already zero.
For other cases, this reduces to $I^{\rm O}_{aa'} = - 8 \hat{b}^{-2} \hat{X}^0_a \hat{Y}^0_a$ for the case of non vanishing $\hat{X}$, while it reduces to $I^{\rm O}_{aa'} = 0$ for the latter case via Eq. (\ref{eq:hatXYalgebra}).

To check our notation, let us consider two cases: (i) absent symmetric and anti-symmetric reps. and (ii) absent symmetric rep. but present anti-symmetric reps. 
Then, these conditions that are consistent with SUSY configurations are summarized as follows.
\begin{enumerate}[(i)]
    \item Absent symmetric and anti-symmetric reps. \\
    From the algebraic relations and the SUSY condition, it must be satisfied that
    \begin{align}
        \begin{aligned}
            &\hat{Y}^I_a = \hat{Y}^J_a = \hat{X}^K_a = \hat{X}^L_a = 0, \\
            &\hat{X}^I_a \hat{X}^J_a = - \hat{Y}^K \hat{Y}^L \neq 0, \\
            &\hat{Y}^K_a \hat{L}^K + \hat{Y}^L_a \hat{L}^L = 0, \\
            &\hat{L}^K U_K = \hat{L}^L U_L ~ {\rm (no~sum)},
        \end{aligned} \label{eq:nosymnoasymcond}
    \end{align}
    for some permutations $I, J, K, L$ of $\{0, 1, 2, 3\}$.
    Thus the two vanishing $\hat{X}^I$ case is a candidate for the case.
    \item Absent symmetric rep. but present anti-symmetric reps.\\
    In this case, all $\hat{X}^I$s should be nonzero and 
    \begin{align}
        \sum_I \frac{1}{\hat{X}^I_a} \hat{L}^I = 32 \label{eq:nosymcond}
    \end{align}
    with $\hat{\vec{L}}$ in Eq. (\ref{eq:defL}). This ensures that the anti-symmetric reps. do not vanish. Note that it is independent of the SUSY condition. 
\end{enumerate}
These two conditions are consistent with the conditions described in Ref. \cite{Gmeiner:2005vz}.

\paragraph{Coprime condition}\mbox{}\\
Although we use $X$ and $Y$ as variables, the fundamental ingredients are the wrapping number $(n, m)$ and two integers $n$ and $m$ must be coprime. This condition should be encoded in a new condition that should be satisfied by $(X, Y)$. Indeed one can see from Eq. (\ref{eq:defXY}) that 
\begin{align}
    (Y^0)^2 = \prod_{i=1}^3 {\rm gcd}(X^i, Y^0) \label{eq:coprime condition}
\end{align}
must be satisfied for the non vanishing $\hat{X}^I$ case. 

\subsection{Concrete MSSM-like models on $T^6/(\mathbb{Z}_2 \times \mathbb{Z}'_2)$}
\label{sec:concrete models}
Here, we cite the concrete minimal supersymmetric Standard Model (MSSM)-like models built in Ref. \cite{Gmeiner:2005vz}.
We note the models from which we could find configurations where generations of quark and leptons are the same: ${\rm gen}(Q) = {\rm gen}(L)$.
We will call those configurations ``aligned'' ones.
As pointed out in Ref. \cite{Gmeiner:2005vz}, finding three generations models is an extremely difficult task, at least in this set-up.
We will compare the configurations in which ${\rm gen}(Q) = {\rm gen}(L)$ with ones in which ${\rm gen}(Q) \neq {\rm gen}(L)$.
The models which are relevant to our paper are as follows:
\begin{itemize}
    \item Model 1: $U(3)_a \times USp(2)_b \times U(1)_c \times U(1)_d$ in the visible sector with a hypercharge $Q_Y^S$
\begin{align*}
\begin{array}{|c||c||c||}
\hline \text { Matter } & \multicolumn{1}{c||}{\text{Representation}} & \text { Multiplicity } \\
\hline Q_{L} & (\mathbf{3}, \mathbf{2})_{0,0} & I_{a b} \\
\hline u_{R} & (\overline{\mathbf{3}}, 1)_{-1,0}+(\overline{\mathbf{3}}, 1)_{0,-1} & I_{a^{\prime} c}+I_{a^{\prime} d} \\
\hline d_{R} & (\overline{\mathbf{3}}, 1)_{1,0}+(\overline{\mathbf{3}}, 1)_{0,1} & I_{a^{\prime} c^{\prime}}+I_{a^{\prime} d^{\prime}} \\
d_{R} & \left(\overline{\mathbf{3}}_{A}, 1\right)_{0,0} & \frac{1}{2}\left(I_{a a^{\prime}}+I_{a O6}\right) \\
\hline L & (1, \mathbf{2})_{-1,0}+(1, \mathbf{2})_{0,-1} & I_{b c}+I_{b d} \\
\hline e_{R} & (\mathbf{1}, \mathbf{1})_{2,0} & \frac{1}{2}\left(I_{c c^{\prime}}-I_{c O 6}\right) \\
e_{R} & (\mathbf{1}, \mathbf{1})_{0,2} & \frac{1}{2}\left(I_{d d^{\prime}}-I_{d O6}\right) \\
e_{R} & (\mathbf{1}, \mathbf{1})_{1,1} & I_{c d^{\prime}} \\
\hline
\end{array}
\end{align*}
The generations of quarks and leptons are given by
\begin{align}
    \begin{split}
         {\rm gen} (Q_L) &= I_{ab}, \\
         {\rm gen} (Q_R) &= I_{a'c} + I_{a'd} = I_{a'c'} + I_{a'd'} + \frac{1}{2}(I_{aa'} + I_{a O6}), \\
         {\rm gen} (L_L) &= I_{bc} + I_{bd},  \\
         {\rm gen} (L_R) &= \frac{1}{2}(I_{cc'} - I_{c O6}) +  \frac{1}{2}\left(I_{d d^{\prime}}-I_{d O6}\right) + I_{c d^{\prime}}.
    \end{split}
\end{align}
\item Model 2: $U(3)_a \times U(2)_b \times U(1)_c \times U(1)_d$ in the visible sector with a hypercharge $Q_Y^S$

\begin{align*}
\begin{array}{|c||c||c||}
\hline Q_{L} & (\mathbf{3}, \overline{\mathbf{2}})_{0,0} & I_{a b} \\
Q_{L} & (\mathbf{3}, \mathbf{2})_{0,0} & I_{a b^{\prime}} \\
\hline u_{R} & (\overline{\mathbf{3}}, 1)_{-1,0}+(\overline{\mathbf{3}}, 1)_{0,-1} & I_{a^{\prime} c}+I_{a^{\prime} d} \\
\hline d_{R} & (\overline{\mathbf{3}}, 1)_{1,0}+(\overline{\mathbf{3}}, 1)_{0,1} & I_{a^{\prime} c^{\prime}}+I_{a^{\prime} d^{\prime}} \\
d_{R} & \left(\overline{\mathbf{3}}_{A}, 1\right)_{0,0} & \frac{1}{2}\left(I_{a a^{\prime}}+I_{a {O}6}\right) \\
\hline L & (1, \mathbf{2})_{-1,0}+(1, \mathbf{2})_{0,-1} & I_{b c}+I_{b d} \\
L & (1, \overline{\mathbf{2}})_{-1,0}+(1, \overline{\mathbf{2}})_{0,-1} & I_{b^{\prime} c}+I_{b^{\prime} d} \\
\hline e_{R} & (\mathbf{1}, \mathbf{1})_{2,0} & \frac{1}{2}\left(I_{c c^{\prime}}-I_{c {O} 6}\right) \\
e_{R} & (\mathbf{1}, \mathbf{1})_{0,2} & \frac{1}{2}\left(I_{d d^{\prime}}-I_{d {O} 6}\right) \\
e_{R} & (\mathbf{1}, \mathbf{1})_{1,1} & I_{c d^{\prime}} \\
\hline
\end{array}
\end{align*}

The generations of quarks and leptons are given by
\begin{align}
    \begin{split}
        {\rm gen} (Q_L) &= I_{ab} + I_{ab'}, \\
         {\rm gen} (Q_R) &= I_{a'c} + I_{a'd} = I_{a'c'} + I_{a'd'} = \frac{1}{2}(I_{aa'} + I_{aO6}), \\
         {\rm gen} (L_L) &= (I_{bc} + I_{bd}) + (I_{b'c} + I_{b'd}), \\
         {\rm gen} (L_R) &= \frac{1}{2}(I_{cc'} - I_{c O6}) +  \frac{1}{2}\left(I_{d d^{\prime}}-I_{d O6}\right) + I_{c d^{\prime}}.
    \end{split}
\end{align}

\item Model 3: $U(3)_a \times U(2)_b \times U(1)_c \times U(1)_d$ in the visible sector with a different hypercharge $Q_Y^{(1)}$

\begin{align*}
\begin{array}{|c||c||c||}
\hline
Q_{L} & (\mathbf{3}, \overline{\mathbf{2}})_{0,0} & I_{a b} \\
\hline
u_{R} & (\overline{\mathbf{3}}_A, 1)_{0,0} & \frac{1}{2}\left(I_{a a^{\prime}}+I_{a {O}6}\right) \\
\hline
d_{R} & (\overline{\mathbf{3}}, 1)_{-1,0}+(\overline{\mathbf{3}}, 1)_{0,-1} & I_{a^{\prime} c}+I_{a^{\prime} d} \\
d_{R} & (\overline{\mathbf{3}}, 1)_{1,0}+(\overline{\mathbf{3}}, 1)_{0,1} & I_{a^{\prime} c^{\prime}}+I_{a^{\prime} d^{\prime}} \\
\hline
L & (1, \mathbf{2})_{-1,0}+(1, \mathbf{2})_{0,-1} & I_{b c}+I_{b d} \\
L & (1, \overline{\mathbf{2}})_{-1,0}+(1, \overline{\mathbf{2}})_{0,-1} & I_{b c^{\prime}}+I_{b d^{\prime}} \\
\hline
e_{R} & (\mathbf{1}, \overline{\mathbf{1}}_A)_{0,0} & -\frac{1}{2}\left(I_{b b^{\prime}} + I_{b O6}\right) \\
\hline
\end{array}
\end{align*}
The generations of quarks and leptons are given by
\begin{align}
    \begin{split}
       {\rm gen} (Q_L) &= I_{ab}, \\
       {\rm gen} (Q_R) &= \frac{1}{2}(I_{aa'} + I_{aO6}) = (I_{a'c} + I_{a'd}) + (I_{a'c'} + I_{a'd'}), \\
       {\rm gen} (L_L) &= (I_{bc} + I_{bd}) + (I_{bc'} + I_{bd'}),  \\
       {\rm gen} (L_R) &= -\frac{1}{2}(I_{bb'} - I_{b O6}).
    \end{split}
\end{align}
\end{itemize}
The hypercharges are given as $Q_Y^{S} = \frac{1}{6} Q_a + \frac{1}{2}Q_c + \frac{1}{2}Q_d$ in models 1 and 2 and $Q_Y^{(1)} = -\frac{1}{3} Q_a - \frac{1}{2}Q_b$ in model 3, respectively. The massless $U(1)_Y$ conditions are $\hat{Y}_a + \hat{Y}_ b + \hat{Y}_d = 0$ and  $\hat{Y}_a + \hat{Y}_b = 0$.
\section{Computer Search}
\label{sec:computer search}

In Section \ref{sec:method}, we introduce the method to perform a computer search over possible D-brane configurations for fixed complex-structure moduli, following Refs. \cite{Gmeiner:2005vz,Blumenhagen:2004xx}. Based on this method, we summarize the generated configurations used in our analysis in Section \ref{sec:dataset}.

\subsection{Method}
\label{sec:method}
\paragraph{SUSY and Tadpole conditions}\mbox{}\\
We can reduce the SUSY and tadpole cancellation conditions \cite{Gmeiner:2005vz}  to 
\begin{align}
    \sum_a S_a &= C, \\
    \sum_{I} \frac{1}{2} \hat{L}^I U_I &\geq \sum_I \hat{X}^I_a U_I > 0 ~ (\forall a), \label{eq:SUSYineq}\\
    \sum_{I} \hat{Y_a}^I \frac{1}{U_I} &= 0 ~ (\forall a), \label{eq:SUSYYeq}
\end{align}
with
\begin{align}
    C &= \sum_I \frac{1}{2} \hat{L}^I U_I, \\
    S_a &= \sum_I N_a \hat{X}^I_a U_I.
\end{align}

\paragraph{Exact expressions of $\hat{X}$ and $\hat{Y}$}\mbox{}\\
In the computer search, we generate $\hat{X}_a^I$s which satisfy Eq. (\ref{eq:SUSYineq}) for each $a$. 
For $\hat{Y}^I_a$s, we obtain them by the algebra (\ref{eq:hatXYalgebra}) instead of generating them independently. Taking into account the SUSY condition (\ref{eq:SUSYYeq}) for $\hat{Y}^I_a$s, one can see that the expressions for each $a$ are classified into the following four cases \cite{Blumenhagen:2004xx};
\begin{itemize}
    \item No $\hat{X}^I$ vanishing\\
    We can compute all $\hat{Y}$s using the algebraic relations: 
    \begin{align}
        (\hat{Y}^L)^2 = - \frac{\hat{X}^I \hat{X}^J \hat{X}^K}{\hat{X}^L}. \label{eq:YfornovanishingX}
    \end{align}
    Thus all $Y$s are nonzero and only odd numbers of negative $X$s are allowed.
    To fix relative signs of $Y$s, we determine $Y^0$ by taking the square root of Eq. (\ref{eq:YfornovanishingX}) and $Y^i$ via the algebraic relation
    \begin{align}
        {\hat{Y}^i}= \frac{\hat{X}^0}{\hat{X}^i} \hat{Y}^0.
    \end{align}
    One can obtain massless spectra with asymmetric but no symmetric representations by imposing additional constraints in Eq. (\ref{eq:nosymcond}) for this case. If one takes the SUSY condition into account, one can show 
    that only one negative $\hat{X}$ is allowed and $\hat{X}$s should satisfy both of
    \begin{align}
        \frac{1}{\hat{X}^0 U_0} + \frac{1}{\hat{X}^1 U_1} + \frac{1}{\hat{X}^2 U_2} + \frac{1}{\hat{X}^3 U_3} &= 0,\\
        \frac{1}{\hat{X}^0} \hat{L}^0 + \frac{1}{\hat{X}^1} \hat{L}^1 + \frac{1}{\hat{X}^2} \hat{L}^2 + \frac{1}{\hat{X}^3} \hat{L}^3 &=32.
    \end{align}
    Hence it is required that $\vec{U}^{-1}\equiv (U^{-1}_0,U^{-1}_1,U^{-1}_2,U^{-1}_3) $ and $\hat{\vec{L}}$ are linearly independent for such a massless chiral spectrum.
    
    \item One $\hat{X}^I$ vanishing\\
    This case is inconsistent with the algebra (\ref{eq:hatXYalgebra}), hence we ignore it.
    \item Two $\hat{X}^I$s vanishing\\
    Let us assume $\hat{X}^I, \hat{X}^J = 0$. Then the algebraic relation
    \begin{align}
        \hat{X}^K \hat{X}^L = - \hat{Y}^I \hat{Y}^J
    \end{align}
    implies $\hat{Y}^I, \hat{Y}^J \neq 0$. On the other hand, the pair of each nonzero $\hat{X}$ becomes zero since
    \begin{align}
        \hat{X}^K \hat{Y}^K = \hat{X}^I \hat{Y}^I = 0.
    \end{align}
    Thus the SUSY relation (\ref{eq:SUSYYeq}) is reduced to
    \begin{align}
        \hat{Y}^I U_J + \hat{Y}^J U_I = 0 ~ ({\rm no~sum.}),
    \end{align}
    and the ratio $\frac{U_I}{U_J}$ is restricted to a rational number. The $\hat{Y}s$ are computed as
    \begin{align}
        \begin{aligned}
            \hat{Y}^I &= \pm \sqrt{\frac{U_I}{U_J}} \sqrt{\hat{X}^K \hat{X}^L}, \\
            \hat{Y}^J &= \mp \sqrt{\frac{U_J}{U_I}} \sqrt{\hat{X}^K \hat{X}^L}.
        \end{aligned}
        \label{eq:YfortwovanishingX}
    \end{align}
    Note that, if $\frac{U_J}{U_I}$ is a squared integer, the product $\hat{X}^K \hat{X}^L$ has also to be a squared integer. To eliminate exotic chiral matter, $\frac{U_J}{U_I} = \frac{\hat{L}^I}{\hat{L}^J}$ and this is exactly equal to one for the non-tilted tori. 
    
    In any case, the problem here is the coprime condition. The constraint that two $\hat{X}$s are zero implies that at least two $X$s vanish.
    For our purpose, it is sufficient to assume all $b_i = 0$.
    Then exactly two $\hat{X}$s become zero, and the coprime condition (\ref{eq:coprime condition}) is modified to take the following forms:
    \begin{itemize}
        \item $X^0 = 0$ and $X^i = 0$ case\\
        This implies $n_i = 0$, and thus $m_i$ must be one if we fix the number of the corresponding D-branes. The coprime condition becomes
        \begin{align}
            {\rm gcd}(X^j, Y^0) {\rm gcd}(X^k, Y^0) = |Y^0|.
        \end{align}
        \item $X^i = 0$ and $X^j = 0$ case\\
        This implies $m_k=0~(k \neq i, j)$, and thus $n_k$ must be one for the same reason. The coprime condition becomes
        \begin{align}
            {\rm gcd}(X^0, Y^i) {\rm gcd}(X^0, Y^j) = |X^0|.
        \end{align}
    \end{itemize}
    The coprime condition in these forms turns out to be very restrictive in the actual search.
    \item Three $\hat{X}^I$s vanishing\\
    The D6-branes with $\hat{X}^I_a$s in the case are exactly $Sp(2)$ branes. Thus the only allowed wrapping numbers are those corresponding to Eq. (\ref{eq:Sp(2)cycles}). 
\end{itemize}

\paragraph{Finiteness of brane configurations}\mbox{}\\
Let us review the finiteness of brane configurations \cite{Blumenhagen:2004xx} for fixed complex-structure moduli.
As we mentioned, there are only two possible cases for $\hat{X}^I$s except for $Sp(2)$ branes: two vanishing or no vanishing $\hat{X}^I$s.
One can see that the possible configurations are finite for each cases with given complex-structure moduli\footnote{More generally, it was shown in Ref. \cite{Douglas:2006xy} that the whole possible SUSY configurations are finite. We follow Ref. \cite{Blumenhagen:2004xx} here so that we would like to generate the concrete configurations for later discussion.}, as follows;
\begin{itemize}
    \item No $\hat{X}^I$ vanishing \mbox{}\\
    Since the number of negative $\hat{X}^I$s must be odd, let us suppose $(\hat{X}^A)(\hat{X}^{I \neq A}) < 0$.
    Then, the SUSY condition on $\hat{Y}^I$ \ref{eq:susycond} becomes
    \begin{align}
        0 = \frac{\hat{Y}^A}{U_A} \left( 1 + \sum_{I \neq A} \frac{U_A \hat{Y^I}}{U_I \hat{Y}^A} \right) = \frac{\hat{Y}^A}{U_A} \left( 1 + \sum_{I \neq A} \frac{U_A \hat{Y}^I}{U_I \hat{Y}^A} \right) = \frac{\hat{Y}^A}{U_A} \left( 1 + \sum_{I \neq A} \frac{U_A \hat{X}^A}{U_I \hat{X}^I} \right), 
    \end{align}
    and it implies $|U_A \hat{X}^A|/|U_I \hat{X}^I| < 1$ (no sum and $\forall I \neq A$).
    Taking into account the SUSY condition on $\hat{X}^I$ \ref{eq:susycond}, one can see that $\hat{X}^A < 0$ is the only possible candidate.
    Combining with the tadpole cancellation condition, one can see that the inequality
    \begin{align}
        0 < \hat{X}^{I \neq A} U_I \leq \frac{1}{2} \sum_{J} \hat{L}^{J} U_{J}
    \end{align}
    holds where there is no summation in the middle of this inequality.
    Hence $\hat{X}^I$ is in the range of
    \begin{align}
        1 \leq \hat{X}^{I \neq A} \leq \frac{1}{2} \sum_{J} \hat{L}^{J} \frac{U_{J}}{U_{I}}. 
    \end{align}
    Thus, the candidates for $\hat{X}^{I \neq A}$ are finite for fixed $U_I$.
    The remaining $\hat{X}^{A}$ and $\hat{Y}^I$s are determined by the algebra (\ref{eq:hatXYalgebra}), and the D6-brane configurations wrapping this type of cycle are finite.
    Note that, since $\hat{X}^0$ is a multiple of $\hat{b}$ and $\hat{X}^i$ is that of $\frac{1}{1-b_i}$, in the actual search we modify the inequality.
    \item Two $\hat{X}^I$s vanishing \mbox{}\\
    The algebra $\hat{X}^I \hat{X}^J = - \hat{Y}^K \hat{Y}^L$ (\ref{eq:hatXYalgebra}) and the SUSY condition (\ref{eq:susycond}) imply that $\hat{Y}^K \hat{Y}^L < 0$, where $\hat{X}^K = 0 = \hat{X}^L$ is assumed.
    Together with the SUSY condition on $\hat{X}^I$, it leads $\hat{X}^I > 0, \hat{X}^J > 0$, and it boils down to the same discussion as the previous case.
\end{itemize}

\subsection{The Dataset}
\label{sec:dataset}
Following the previous computing method, and restricting us within the range $U^0 = 1, |U| \coloneqq \sqrt{(U^1)^2 + (U^2)^2 + (U^3)^2} \leq 12$ with $U^1 \leq U^2 \leq U^3$\footnote{Hence, the dataset does not possess the permutation symmetry among $I = 0, 1, 2, 3$.} and $b_{1,2,3} = 0$, we obtain the whole configurations which satisfies the algebra.
For each model in \Cref{sec:concrete models}, we fix the number of D-branes in the visible sector as assigned by the model.
We also impose the corresponding massless $U(1)_Y$ condition.
Ignoring possible permutation, the number of configurations that we obtained is summarized in \Cref{tab:number of configurations for each model}.
Note that we could not obtain a three-generation model in this set-up as predicted in \cite{Gmeiner:2005vz}.

\begin{table}[H]
    \centering
    \begin{tabular}{|c||c c c|}\hline
       Model  & Model 1 & Model 2 & Model 3 \\ \hline
       Aligned/Total & 1350/39551 & 1030/24336 & 816/171908 \\\hline
    \end{tabular}
    \caption{The number of configurations for each model. Those configurations satisfy the algebra of $X, Y$ and the corresponding massless $U(1)_Y$ condition. Here, each of the right-side numbers denotes the total number of the configurations, and each of the left-side numbers denotes that of the aligned configurations where ${\rm gen} (Q) = {\rm gen} (L)$ i.e., ${\rm gen} (Q_L, Q_R, L_L, L_R)$ have the same value.}
    \label{tab:number of configurations for each model}
\end{table}

As a pre-processing, we perform zero-padding so that the length of data becomes the same for each model and the architectures of neural networks can be fixed.
The maximum numbers of D-brane stacks in the dataset for each model are 13, 11, and 12, respectively.
Note that we will not normalize the data. One may normalize $X$ and $Y$ in a naive way, but there are no criteria to normalize the number of D6-branes $N$. 
\section{Feature Extraction with Autoencoder Models}
\label{sec:feature extraction with Autoencoder Models}
So far, we have prepared a set of D6-brane configurations satisfying the models via the computer method.
In the previous studies, it was discussed that the configurations that constitute the string Landscape have some statistical features.
With a similar but a little different perspective, it is also interesting to examine whether some important features distinguish (a set of) configurations from other (sets of) configurations.
However, it is certainly difficult to achieve that beyond statistical methods that have already been utilized in previous studies.
Possible solutions to find such the features of the Landscape which may surpass the classical methods are the machine learning methods, as mentioned.
In this section, we focus on the autoencoder models that attempt to reproduce given input data with a Deep Neural Network (DNN).
By virtue of the non-linearity of DNN, one often expects that the autoencoder models manage to reproduce the data even if the dimensions of intermediate layers are smaller than that of input data.
We will feed both of the configurations which satisfy the models and do not satisfy any of the models to the autoencoder models and hope that they distinguish those two classes of configurations somehow. 
However, we would like to emphasize again that the autoencoder models usually lack interpretability.
Thus, if one tries to express some features obtained in a form that can be easily understood, one usually needs to rely on statistical methods again to find features which used to separate subsets of the data.
Rather, the naive use of the autoencoder models provides us some implications that state the existence of the features, and it is itself interesting.
Therefore, we start \Cref{sec:autoencoder and preprocessing} by constructing the autoencoder models explicitly and checking whether outputs of the latent layer are clustered or not in \Cref{sec:Comparison of Methods} for Model 1.
As a result, we will see in \Cref{sec:feature extraction} that our autoencoder models do not precisely distinguish the configurations (in the sense described in \Cref{sec:Introduction}, weakly) satisfying the models from those that do not. However, we will also see that the former ones are clustered in several specific islands, and those clustering may be characterized by the D6-brane tadpole charges in the hidden sector.

\subsection{The Autoencoder Models and Additional Data Pre-processing}
\label{sec:autoencoder and preprocessing}
In order to extract hidden features in an interpretable form, the dimensions latent layer is usually taken to be a small integer.
We choose it to be a dense layer with two dimensions and expect that the clustering is observed on the two-dimensional plane. 
However, it is surely a difficult task to extract features from large data on such small dimensions.
Indeed, each configuration has typically $\mathcal{O}(10^2)$ elements.
As a result, we could not observe the clustering in a simple autoencoder model, and thus we had to make experiments with different autoencoders. 
Let us introduce our three autoencoders which we call AE-0, AE-1, and AE-2. 
The architecture of each model is summarized in \Cref{tab:network_AE0,tab:network_AE1,tab:network_AE2}.
In those tables, the upper row shows the structure of the encoder part and the lower row shows the structure of the decoder part, where the hidden layers are denoted as E and D, respectively.
There, the latent layers in different rows represent the same layer, and the decoder and encoder parts are connected via the latent layer.
As mentioned, the length of each element is different for each D-brane model.
In the following, the tables are for Model 1.
We train the AE-0 and the AE-1 only on Model 1 to test the autoencoder models.
On the other hand, we train AE-2 on each D-brane model.
The dimensions of the first layers and the final layers differ between D-brane models, as 13 is changed to 11 and 12 for Models 2 and 3, respectively. 

First, the simplest autoencoder (AE-0) we defined is the traditional one, which is purely composed of serial dense layers (see \Cref{tab:network_AE0}).
The dimensions of the layer become smaller as the position is deeper.
Those dimensions are one of the families of hyperparameters, but we did not choose them carefully.
Rather, we chose them to become smaller constantly.
The activation functions we adopted are SeLU and $\tanh$. We mainly use $\tanh$ as the activation function for the latent layer to normalize the outputs.
One of the reasons we choose SeLU as the activation function for most of the other layers is that it will be robust against saturation.
The D-brane configurations include negative elements as well as positive elements, and it is another reason to choose SeLU.

\begin{table}[H]
    \centering
    \begin{tabular}{|c||c|c|c|c|c|c|c|}\hline
        Layer & Input & E1 & E2 & E3 & E4 & E5 & Latent\\
       \hline
        Dimension & $\mathbb{Z}^{104} + \mathbb{Z}^{13} +\mathbb{Z}^{4}$ & $\mathbb{R}^{88}$ & $\mathbb{R}^{55}$ & $\mathbb{R}^{22}$ & $\mathbb{R}^{11}$ & $\mathbb{R}^{6}$ & $\mathbb{R}^{2}$
       \\
       \hline \hline
       Layer & Latent & D1 & D2 & D3 & D4 & D5 &  Output \\
       \hline
       Dimension & $\mathbb{R}^{2}$ & $\mathbb{R}^{6}$ & $\mathbb{R}^{11}$ & $\mathbb{R}^{22}$ & $\mathbb{R}^{55}$ & $\mathbb{R}^{88}$ & $\mathbb{Z}^{121}$ \\
       \hline
    \end{tabular}
    \caption{The architecture of AE-0. In the neural network, both the input and the output are the same vector whose length is 121, where 121 comes from 13 sets of $(X_a, Y_a)$, 13 $N_a$s, and 4 moduli ($121 = 13 \times 8 + 13 + 4$). The activation functions are $\tanh$ for the latent layer and SeLU for the other layers. The autoencoder tries to fit 121 values between the input layer and the output layer.}
    \label{tab:network_AE0}
\end{table}

As mentioned, we could not find any clustering in its latent layer.
However, it is not correct that our goal of study is to precisely reproduce given input data, but to find important features that are necessary in the reconstruction. 
It is problematic if the learning does not proceed at all, but it is a valid option to give a DNN a few hints and see if clustering occurs in the output of the latent layer.
Keeping in mind this philosophy, we additionally construct two more autoencoder models named AE-1 (\Cref{tab:network_AE1}) and AE-2 (\Cref{tab:network_AE2}) with three patterns of activation functions (\Cref{tab:three types of the activation functions}).
Both models are fed with some hints in learning. Thus they can be regarded as one of the semi-supervised learning models.
There the labels are chosen as the hint, and we shall adopt this approach as well. 
As the labels, we choose the complex-structure moduli since effectively only three elements are relevant to the configurations while they constrain the configurations in the nontrivial ways via the conditions introduced in $\Cref{sec:2}$.
The second (AE-1) model is a simple semi-supervised autoencoder, while the third (AE-2) model has extra pre-/post-dense layers.
Since there are relations between $(N, X, Y, U)$ for each stack of D6-branes, we consider that it is effective to do pre-training for each stack.
The post-dense layers are introduced for the same reason.
The layers that are fed with the complex-structure moduli values are placed on both the encoder and decoder. In this sense, one can consider the third model as a type of \textit{conditional autoencoder} (not variational).

\begin{table}[H]
    \centering
    \begin{tabular}{|c||c|c|c|c|c|c|c|c|c|}\hline
       Layer & Input & Pre & E1 & E2 & E3 & E4 & E5 & E6 & Latent\\
       \hline
       Dimension & $\mathbb{Z}^{104}+\mathbb{Z}^{4}$ & $\mathbb{R}^{108}$ & $\mathbb{R}^{80}$ & $\mathbb{R}^{48}$ & $\mathbb{R}^{20}$ & $\mathbb{R}^{10}$ & $\mathbb{R}^{6}$ & $\mathbb{R}^{4}$ & $\mathbb{R}^{2}$
       \\
       \hline \hline
       Layer & Latent & D1 & D2 & D3 & D4 & D5 & D6 & Output &\\
       \hline
       Dimension & $\mathbb{R}^{2}$ & $\mathbb{R}^{2}+\mathbb{Z}^{4}$ & $\mathbb{R}^{8}$ & $\mathbb{R}^{16}$ & $\mathbb{R}^{24}$ & $\mathbb{R}^{48}$ & $\mathbb{R}^{80}$ & $\mathbb{Z}^{104}+\mathbb{Z}^{4}$ &
       \\
       \hline
    \end{tabular}
    \caption{The architecture of AE-1 without positional encoding.
    In the neural network, both the inputs and the outputs are the 104 sets of $\left(N_a X_a,N_aY_a\right)$ and 4 moduli.
    The activation functions are SeLU for all of the layers except the latent layer and the ``Pre'' layer.
    We use $\tanh$ for the latent layer, and no activation function for ``Pre''.
    The 4 moduli are concatenated as the labels in D1 (denoted as $\mathbb{R}^{2}+\mathbb{Z}^{4}$).
    The autoencoder tries to fit the 104 sets of $\left(N_a X_a, N_a Y_a\right)$ between the input layer and the output layer.
    When the positional encoding is considered, the input and output become real numbers instead of integers.}
    \label{tab:network_AE1}
\end{table}

\begin{table}[H]
    \centering
    \begin{tabular}{|c||c|c|c|c|c|c|c|c|c|}\hline
       Layer & Input & Pre & E1 & E2 & E3 & E4 & E5 & Latent &\\
       \hline
       Dimension & $13\mathbb{Z}^{12}$ & $13\mathbb{R}^{4}$ & $\mathbb{R}^{52}+\mathbb{Z}^{4}$ & $\mathbb{R}^{40}$ & $\mathbb{R}^{20}$ & $\mathbb{R}^{10}$ & $\mathbb{R}^{4}$ & $\mathbb{R}^{2}$ &
       \\
       \hline \hline
       Layer & Latent & D1 & D2 & D3 & D4 & D5 & D6 & Post & Output\\
       \hline
       Dimension & $\mathbb{R}^{2}$ & $\mathbb{R}^{2}+\mathbb{Z}^{4}$ & $\mathbb{R}^{8}$ & $\mathbb{R}^{16}$ & $\mathbb{R}^{24}$ & $\mathbb{R}^{36}$ & $\mathbb{R}^{52}$ & $13\mathbb{R}^{4}$ & $13\mathbb{Z}^{8}$
       \\
       \hline
    \end{tabular}
    \caption{This is the structure of AE-2. In the neural network, the inputs are the 13 sets of $\left(N_a X_a,N_a Y_a,U\right)$, the outputs are the 13 sets of $\left(N_a X_a,N_a Y_a\right)$.
    We will try three patterns of activation functions which are summarized in \Cref{tab:three types of the activation functions}.
    The 4 moduli are input in E1 after the pre-dense layer (denoted as $\mathbb{R}^{52}+\mathbb{Z}^{4}$), and these moduli are used as the labels in D1 (denoted as $\mathbb{R}^{2}+\mathbb{Z}^{4}$). The autoencoder tries to fit 104 sets of $\left(NX,NY\right)$ between the input layer and the output layer.
    When the positional encoding is considered, the input and output become real numbers instead of integers.}
    \label{tab:network_AE2}
\end{table}

\begin{table}[H]
    \centering
    \begin{tabular}{|c||c|c|c|c|c|c|}
    \hline
    Layer&E4& E5 & Latent & D1 & D2 & Others \\
    \hline
    ``$\tanh$-1''      & SeLU      & SeLU & $\tanh$ & SeLU & SeLU & SeLU \\ 
    ``SeLU-1''      & SeLU      & SeLU & SeLU & SeLU & SeLU & SeLU\\ 
    ``$\tanh$-2'' & SeLU & $\tanh$ & $\tanh$ & $\tanh$  & SeLU & SeLU\\
    \hline
    \end{tabular}
    \caption{The activation functions for AE-2. We try the three patterns and compare the results in the following.}
    \label{tab:three types of the activation functions}
\end{table}

As we will show later, the two improved models exhibit clustering in their latent layers, but before describing the results we would like to emphasize that we have to be careful regarding the structure of the input data.
When we give the configurations to the autoencoders, we have to align each data to be a sequence.
Then, the autoencoders indeed interpret each of the input data as a sequence, not a \textit{set}.
The original D6-brane configurations have no information about the orders of the branes, but the sequences do.
To make the autoencoders avoid learning such artificial information, one can naively take three options:
\begin{enumerate}
    \item First, if we manage to construct a model that can learn a set as a set, the problem does not occur.
Indeed, there are several studies on this point, and one of the architectures is suggested in Ref. \cite{zaheer2017deep}.
However, it may not apply to our data, because our data is not an actual set in that sense; it is data with some complex partial-permutation symmetries.
We do not know models with which one can control such partial symmetries easily, thus we give up the first option in this paper.

\item The second option is just giving a whole set of data whose elements are rearranged in accordance with the symmetries, but it is definitely an inefficient way.
In fact, the typical number of D6-branes in our data is around 10. It is extremely difficult to train the autoencoders on $10!$ times as much as the original data.

\item Therefore, the only remaining option seems just to give up training on the property of sets and to take the viability of learning in exchange for the rigorous ``consistency'' (ensuring that the trained models are robust to data permutation).
In the remainder of this paper, we will always take this option, at least on the permutation symmetry of the stacks.
\end{enumerate}

Even ignoring that symmetry, there remains another point which requires attention.
As we have seen in the \Cref{sec:2}, among the stacks, $X^I$s have to satisfy the tadpole cancellation condition for each $I$. 
Similarly, $Y^I$s have to satisfy the K-theory constraint for each $I$. 
However, if we are given a sequence of numbers, it is almost impossible to find correspondence among the positions of the numbers.
The autoencoders are expected to have a very high power to find such relationships due to their non-linearity, but giving hints could be a shortcut.
Thus, we would like to examine another notion, which is called \textit{positional encoding}.

The positional encoding was originally introduced in a very different context\footnote{See for a review of this kind notion, \cite{dufterPositionInformationTransformers2022a}}: Natural Language Processing (NLP).
For example, \textit{Chat Generative Pre-trained Transformer} (i.e. ChatGPT) is one of the most famous services. 
\textit{Transformer} is used mainly in the region of NLP and overcomes difficulties of \textit{Recurrent Neural Network} (RNN).
Then, the positional encoding is invented for the precision of 
the Transformer in Ref. \cite{vaswaniAttentionAllYou2017}. 
In this technique, numerical periodic values are added to the original elements of the data, depending on the positions they appear in the data.
Usually, the periodic values are chosen to be generated by a certain combination of sine and cosine as in Ref. \cite{vaswaniAttentionAllYou2017} since a translation in the position space can be expressed as a linear combination of the values measured at the original position.
If the Transformer does not use the positional encoding, the order of the components of the input sentence (\textit{token}) is naively neglected.
In other words, it is a method to add information about those ignored positions to the model.
In this paper, given the fact that positional encoding was important for the Transformer to successfully learn the features of the data structure, we try to apply this method to our autoencoders.
The concrete methods that we examine are the following:
\begin{itemize}
    \item PE-i: Distinguish both the D6-branes stacks and four directions $I = 0, 1, 2, 3$ in each stack (AE-1/AE-2)\\
    Let $X^I = (X^0, X^1, X^2, X^3)$ and $Y^I = (Y^0, Y^1, Y^2, Y^3)$ as an input data of $pos$-th stack.
    Then we add $A\sin \left( \frac{pos}{T^{2i/4}} \right)$ to elements at position $I = 2i$ and $A\cos \left( \frac{pos}{T^{2i/4}} \right)$ to elements at $I = 2i + 1$. Here, $A$ is amplitude. About the AE-i case, we also add these values with $pos = 0$ to $U^I$. On the other hand, we associate $U^I$ with each stack in the AE-2 case, as illustrated. $pos$-th $U^I$s are added the same values for $pos$-th $X, Y$s.
    \item PE-ii: Distinguish only four directions $I = 0, 1, 2, 3$ (AE-2)\\
    In the PE-ii case, we have the pre-training layer for each stack. Thus it is more natural to ignore information about $pos$ in this case. We universally add the values with $pos \equiv 1$ to all $(N X^I, N Y^I, U^I)$.
\end{itemize}
Furthermore, we will vary amplitude $A$ to values other than one and compare the results.
We expect that the positional encoding contributes to the consistency under the permutations, but training autoencoders will be more difficult if they indeed find features because the original values of elements become vague.
Thus we should compare results with and without the encoding. 

\clearpage
\begin{figure}[H]
\centering
    \begin{tabular}{cc}
    \begin{minipage}[t]{0.45\hsize}
        \centering
        \includegraphics[keepaspectratio, width=6.5cm]{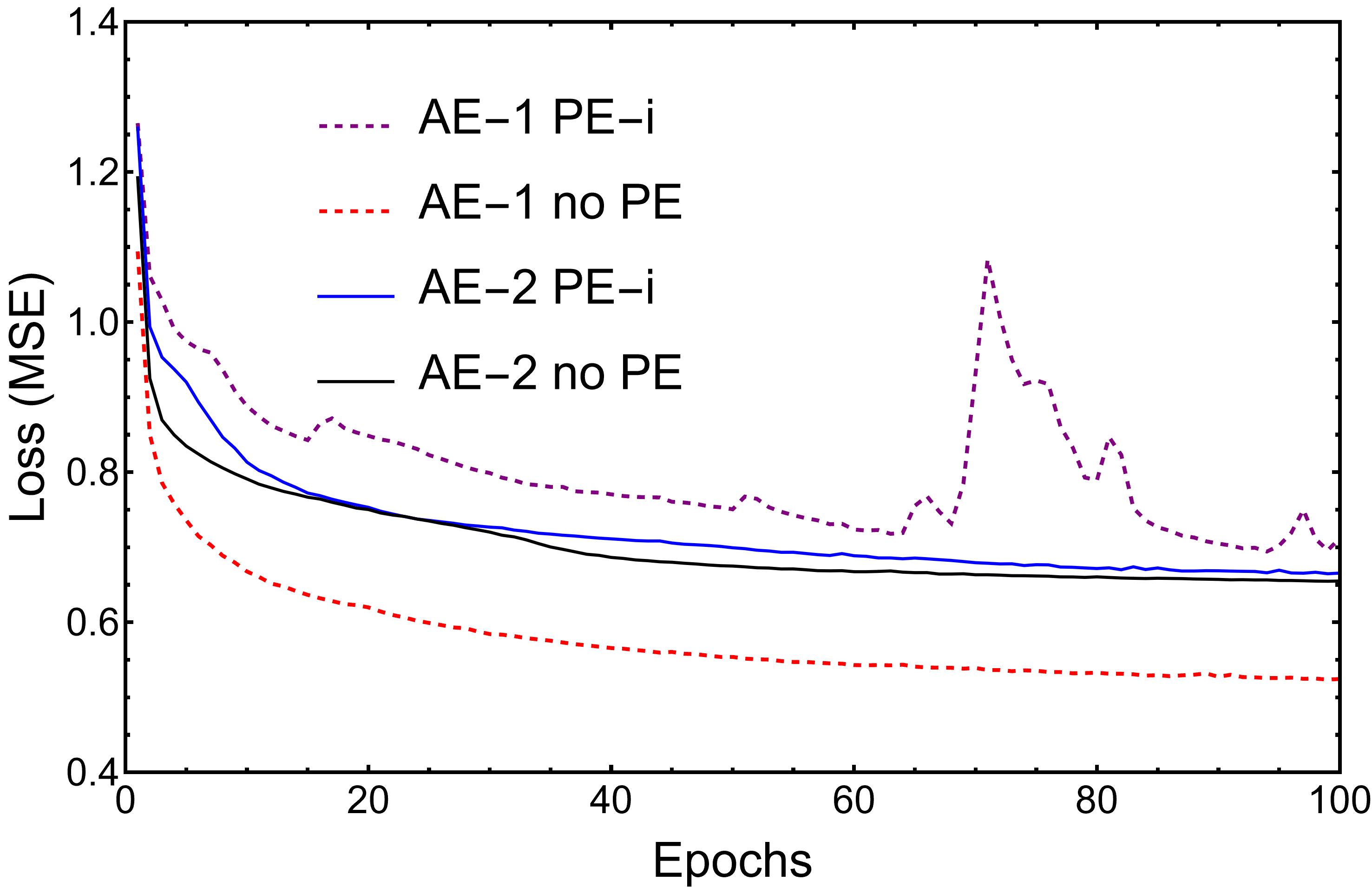}
        \vspace*{-5mm}
	\caption{Loss in training of AE-1/2 with and without PE-i.}
        \label{fig:losses_PE-i}
    \end{minipage} &
    \begin{minipage}[t]{0.45\hsize}
        \centering
        \includegraphics[keepaspectratio, width=6.5cm]{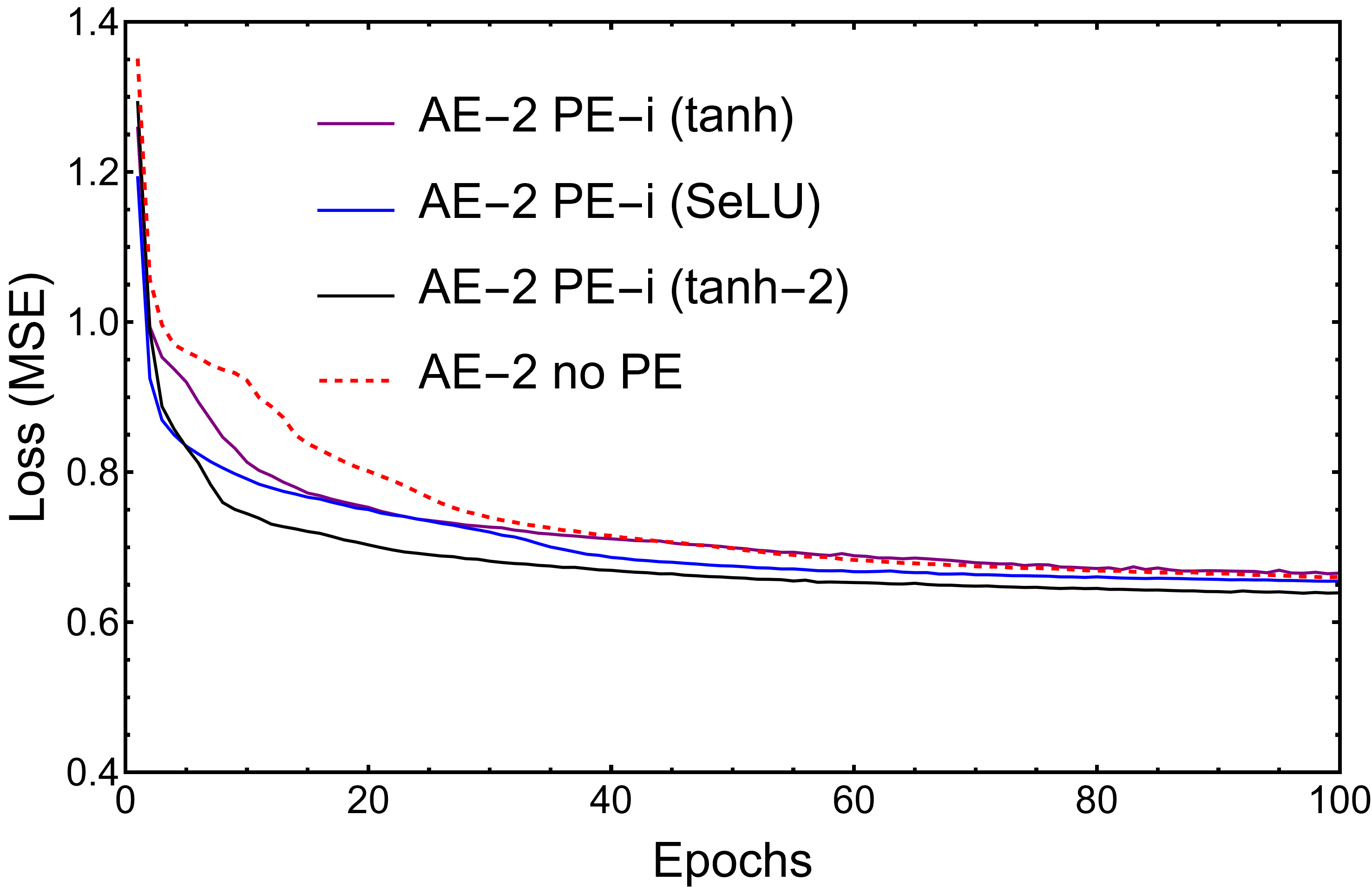}
        \vspace*{-5mm}
	\caption{Loss in training of AE-2 with different activation functions (PE-i).}
        \label{fig:losses_AE-2}
    \end{minipage}
    \end{tabular}
\end{figure}

\begin{figure}[H]
    \centering
    \includegraphics[keepaspectratio, width=7.5cm]{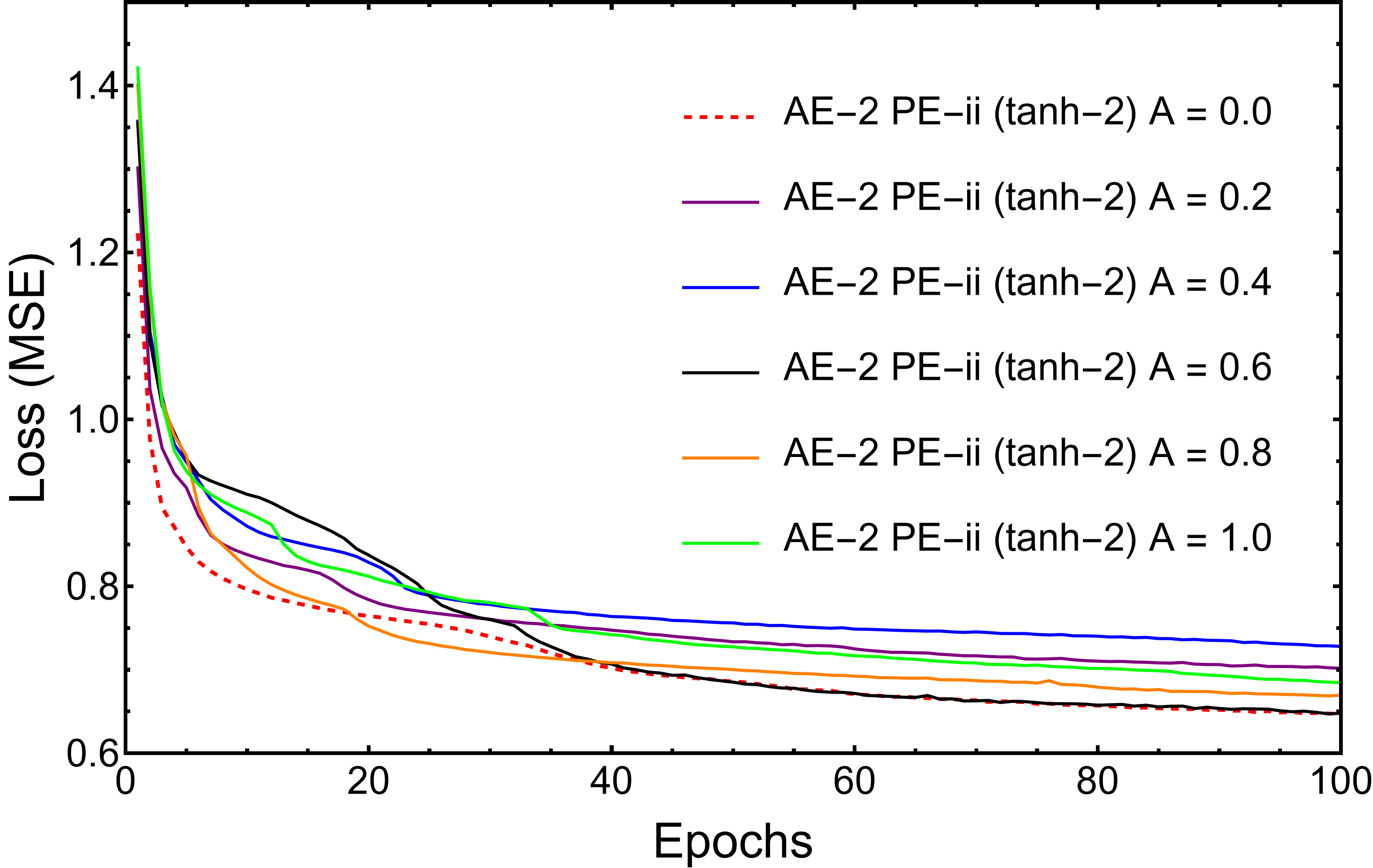}
    \caption{Loss in training of AE-2 with different amplitudes for PE-ii.}
    \label{fig:losses_AE-2-variousA}
\end{figure}

\subsection{Comparison of Methods (Model 1)}
\label{sec:Comparison of Methods}
In this section, we summarize the training results with various methods: two autoencoder models AE-1/2 and various positional encoding methods PE-i/ii with several amplitudes $A$.
Then, we determine which model and method we should adopt to analyze the latent layer.
To examine the various methods, we fix the problem by focusing on Model 1. 

First, we would like to exclude results from training in which losses do not decrease. \Cref{fig:losses_PE-i,fig:losses_AE-2,fig:losses_AE-2-variousA} illustrate the losses.
We simply choose mean squared error as the loss function.
If we focus only on the losses, we might conclude that the most suitable method we can choose is AE-1 without any positional encoding.
However, as one can see in \Cref{fig:losses_PE-i}, after adding the positional encoding PE-i, the loss in AE-1 is not stable while the other model AE-2 seems to be robust to the encoding.
Since we introduced the positional encoding to respect the correspondences between elements of the configurations, we do not consider AE-1 with no positional encoding as the most suitable method to extract hidden features. 
As mentioned, we also changed the activation functions in the model AE-2.
From \Cref{fig:losses_AE-2}, one can see that the losses remain almost the same, but let us focus on the ``tanh-2'' model. 
The robustness under the positional encoding is also observed in the case of PE-ii.
Interestingly, as illustrated in \Cref{fig:losses_AE-2-variousA}, the case PE-ii with $A = 1.0$ scored the slightly lower loss than the case with $A = 0.0$.
In this regard, we consider that the AE-2 model is superior to the AE-1 model.
On the other hand, it is fair to point out the high losses of the AE-2 model.
Since the typical values of elements are around ${\cal O}(1)$, the losses around ${\cal O}(1)$ are quite dangerous.
Indeed, if a configuration consists only of equal numbers of 0s and 1s, the model will achieve quite lower losses just by outputting 0.5 constantly.
However, the situation is somewhat better because the actual data is perhaps a bit more complex.
Taking into account the fact that the losses are less than 1.0, we decided not to exclude any methods at this point.

\begin{figure}[H]
\hspace*{-10mm}
    \begin{tabular}{ccc}
      \begin{minipage}[t]{0.35\hsize}
        \centering
        \includegraphics[keepaspectratio, scale=0.40]{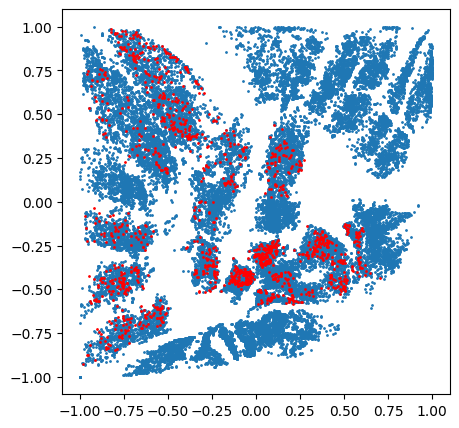}
        \vspace*{-7.0mm}
        \caption{AE-1 no PE.}
        \label{fig:AE-1 no PE}
      \end{minipage} &
      \begin{minipage}[t]{0.35\hsize}
        \centering
        \includegraphics[keepaspectratio, scale=0.40]{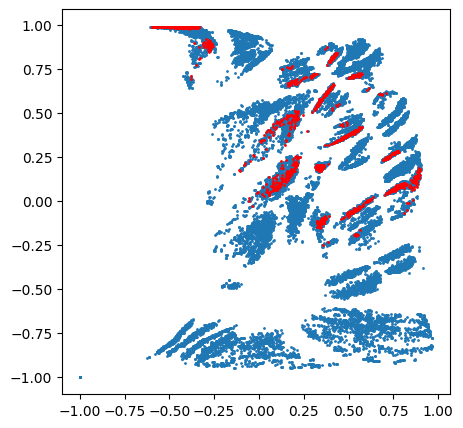}
        \vspace*{-7.0mm}
        \caption{AE-1 PE-i.}
        \label{fig:AE-1 PE-i}
      \end{minipage} &
      \begin{minipage}[t]{0.35\hsize}
        \centering
        \includegraphics[keepaspectratio, scale=0.40]{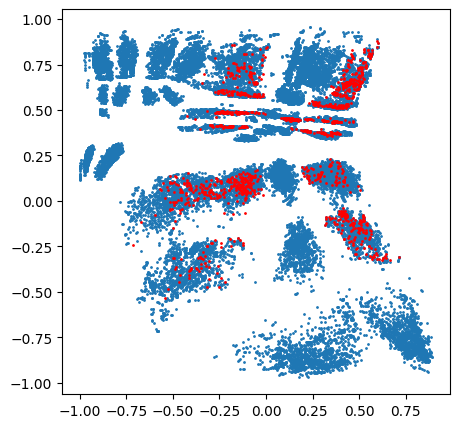}
        \vspace*{-7.0mm}
        \caption{AE-2 no-PE tanh-1.}
        \label{fig:AE-2 no-PE tanh-1}
      \end{minipage}\\
      \begin{minipage}[t]{0.35\hsize}
        \centering
        \includegraphics[keepaspectratio, scale=0.40]{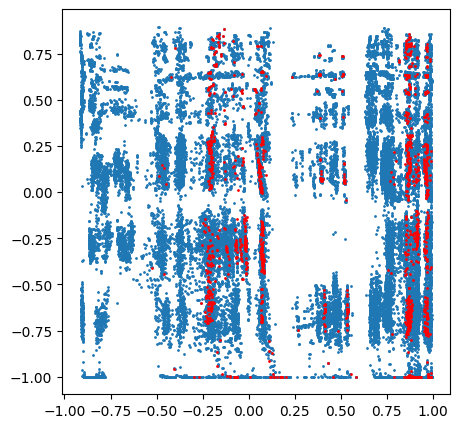}
        \vspace*{-7.0mm}
        \caption{AE-2 PE-i tanh-1.}
        \label{fig:AE-2 PE-i tanh-1}
      \end{minipage} &
      \begin{minipage}[t]{0.35\hsize}
        \centering
        \includegraphics[keepaspectratio, scale=0.40]{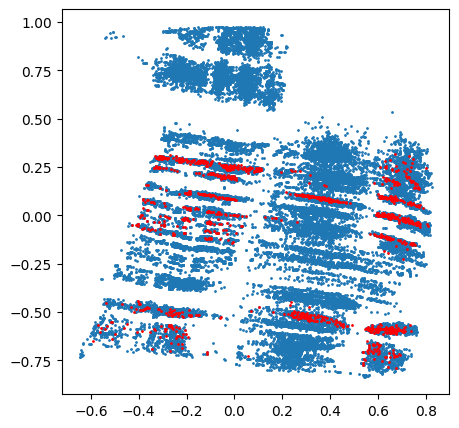}
        \vspace*{-7.0mm}
        \caption{AE-2 PE-i tanh-2.}
        \label{fig:AE-2 PE-i tanh-2}
      \end{minipage} &
      \begin{minipage}[t]{0.35\hsize}
        \centering
        \includegraphics[keepaspectratio, scale=0.40]{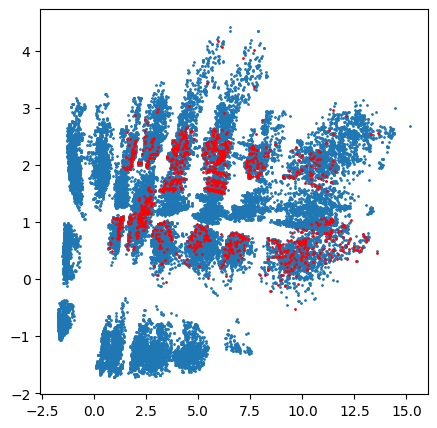}
        \vspace*{-7.0mm}
        \caption{AE-2 PE-i SeLU.}
        \label{fig:AE-2 PE-i SeLU}
      \end{minipage}\\
      \begin{minipage}[t]{0.35\hsize}
        \centering
        \includegraphics[keepaspectratio, scale=0.40]{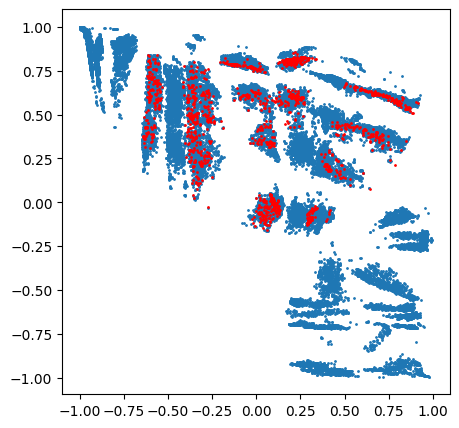}
        \vspace*{-7.0mm}
        \caption{AE-2 PE-ii tanh-1.}
        \label{fig:AE-2 PE-ii tanh-1}
      \end{minipage} &
      \begin{minipage}[t]{0.35\hsize}
        \centering
        \includegraphics[keepaspectratio, scale=0.40]{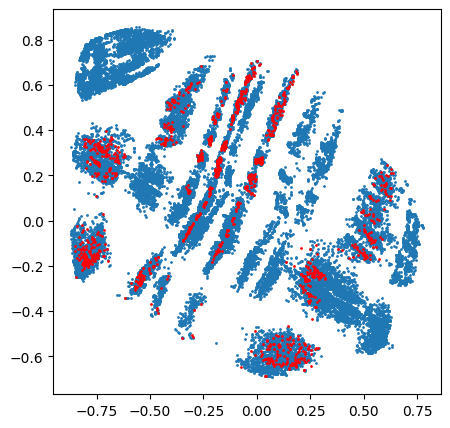}
        \vspace*{-7.0mm}
        \caption{AE-2 PE-ii tanh-2.}
        \label{fig:AE-2 PE-ii tanh-2}
      \end{minipage} &
      \begin{minipage}[t]{0.35\hsize}
        \centering
        \includegraphics[keepaspectratio, scale=0.40]{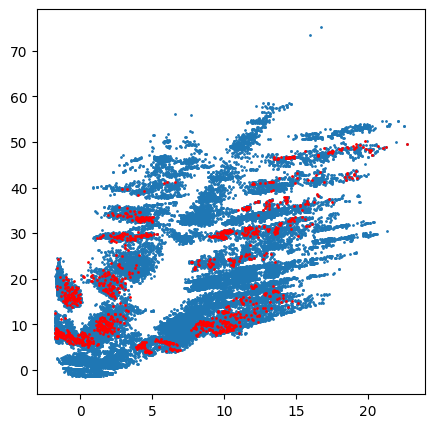}
        \vspace*{-7.0mm}
        \caption{AE-2 PE-ii SeLU.}
        \label{fig:AE-2 PE-ii SeLU}
      \end{minipage}\\
      \begin{minipage}[t]{0.35\hsize}
        \centering
        \includegraphics[keepaspectratio, scale=0.40]{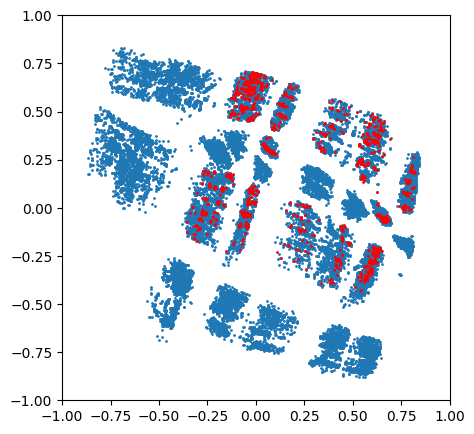}
        \vspace*{-7.0mm}
        \caption{AE-2 PE-ii $A = 0.0$.}
        \label{fig:AE-2 PE-ii A0.0}
      \end{minipage} &
      \begin{minipage}[t]{0.35\hsize}
        \centering
        \includegraphics[keepaspectratio, scale=0.40]{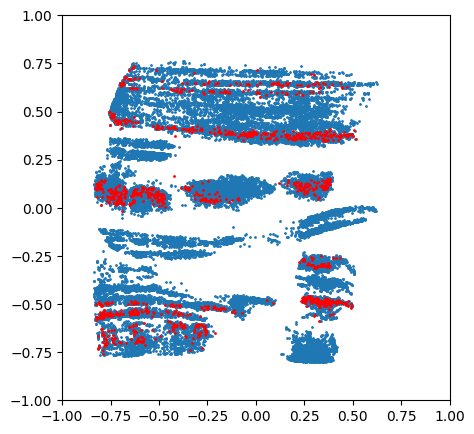}
        \vspace*{-7.0mm}
        \caption{AE-2 PE-ii $A = 0.2$.}
        \label{fig:AE-2 PE-ii A0.2}
      \end{minipage} &
      \begin{minipage}[t]{0.35\hsize}
        \centering
        \includegraphics[keepaspectratio, scale=0.40]{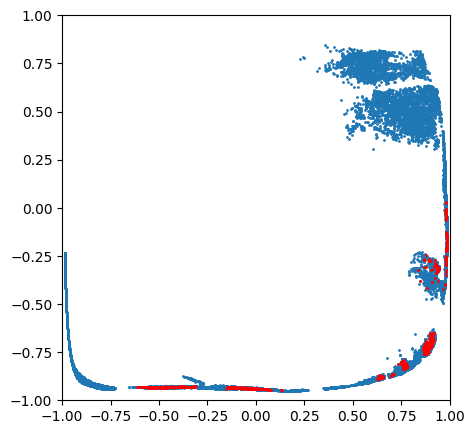}
        \vspace*{-7.0mm}
        \caption{AE-2 PE-ii $A = 0.4$.}
        \label{fig:AE-2 PE-ii A0.4}
      \end{minipage}\\
      \begin{minipage}[t]{0.35\hsize}
        \centering
        \includegraphics[keepaspectratio, scale=0.40]{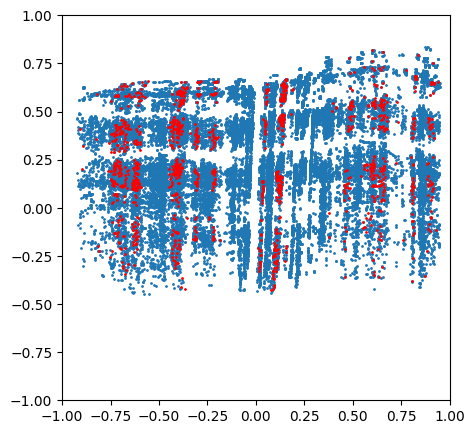}
        \vspace*{-7.0mm}
        \caption{AE-2 PE-ii $A = 0.6$.}
        \label{fig:AE-2 PE-ii A0.6}
      \end{minipage} &
      \begin{minipage}[t]{0.35\hsize}
        \centering
        \includegraphics[keepaspectratio, scale=0.40]{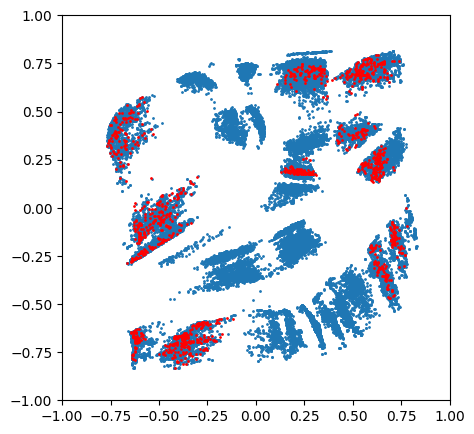}
        \vspace*{-7.0mm}
        \caption{AE-2 PE-ii $A = 0.8$.}
        \label{fig:AE-2 PE-ii A0.8}
      \end{minipage} &
      \begin{minipage}[t]{0.35\hsize}
        \centering
        \includegraphics[keepaspectratio, scale=0.40]{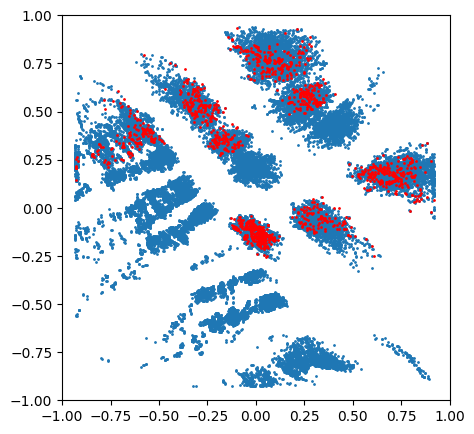}
        \vspace*{-7.0mm}
        \caption{AE-2 PE-ii $A = 1.0$.}
        \label{fig:AE-2 PE-ii A1.0}
      \end{minipage}
    \end{tabular}
\end{figure}

Another criterion for choosing some methods is occurrences of clustering.
We summarize the outputs from the latent layers in \Cref{fig:AE-1 no PE,fig:AE-1 PE-i,fig:AE-2 no-PE tanh-1,fig:AE-2 PE-i tanh-1,fig:AE-2 PE-i tanh-2,fig:AE-2 PE-i SeLU,fig:AE-2 PE-ii tanh-1,fig:AE-2 PE-ii tanh-2,fig:AE-2 PE-ii SeLU,fig:AE-2 PE-ii A0.0,fig:AE-2 PE-ii A0.2,fig:AE-2 PE-ii A0.4,fig:AE-2 PE-ii A0.6,fig:AE-2 PE-ii A0.8,fig:AE-2 PE-ii A1.0}.
The red dots denote the configuration where the generations of leptons and quarks are equal to each other.
One may observe the clustering in several figures, but not in all figures.
For example, \Cref{fig:AE-2 PE-ii A0.4} exhibits the severe saturation while that method scored the same loss compared to other methods.
Since our aim is to find hidden features from the latent layer, we simply ignore such methods.
On the other hand, we can observe clusters clearly\footnote{Although there is no positional encoding with $A = 0$, choosing the AE-2 model is consistent with the observed robustness. Thus, making an experiment with various positional encodings still makes sense. Note that, while changing the amplitude we adopt the $\tanh$-2 model. Therefore, \Cref{fig:AE-2 no-PE tanh-1} and \Cref{fig:AE-2 PE-ii A0.0} are different.} in \Cref{fig:AE-2 PE-ii A0.0}.
We will mainly analyze this latent layer in the following.
Ideally, a feature which might be discovered in a latent layer of a certain model is expected to be found also in latent layers of other models.
We will also examine this point.

\newpage

\newpage

\subsection{Feature Extraction}
\label{sec:feature extraction}
\subsubsection{Model 1}
\label{sec:feature extraction of Model 1}
\begin{figure}[H]
\hspace*{-7mm}
    \begin{tabular}{ccc}
    \begin{minipage}[t]{0.33\hsize}
        \centering
        \includegraphics[keepaspectratio, scale=0.24]{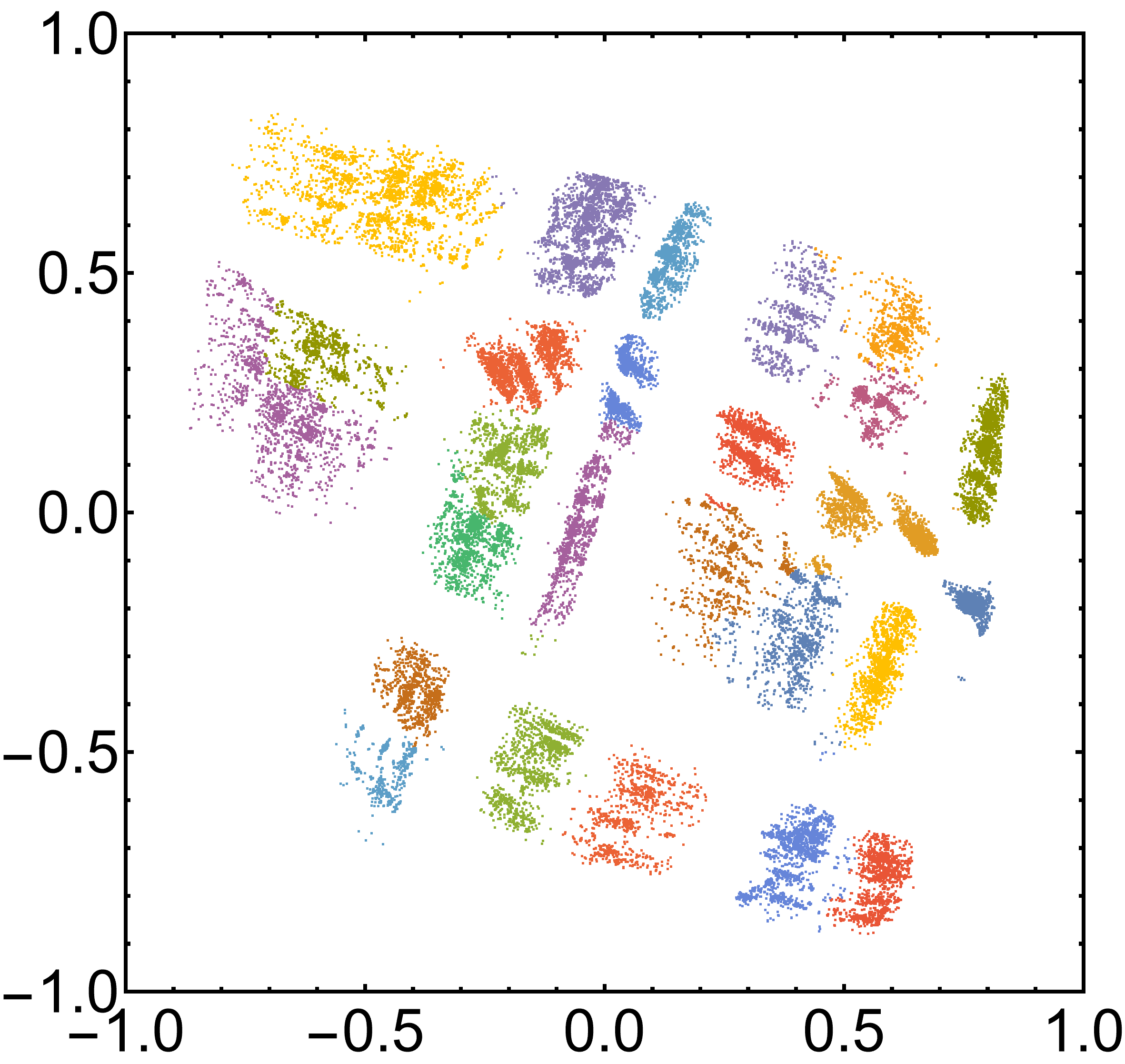}
        \vspace*{-3mm}\caption{\texttt{MeanShift}.}
        \label{fig:MeanShift}
    \end{minipage} &
    \begin{minipage}[t]{0.33\hsize}
        \centering
        \includegraphics[keepaspectratio, scale=0.24]{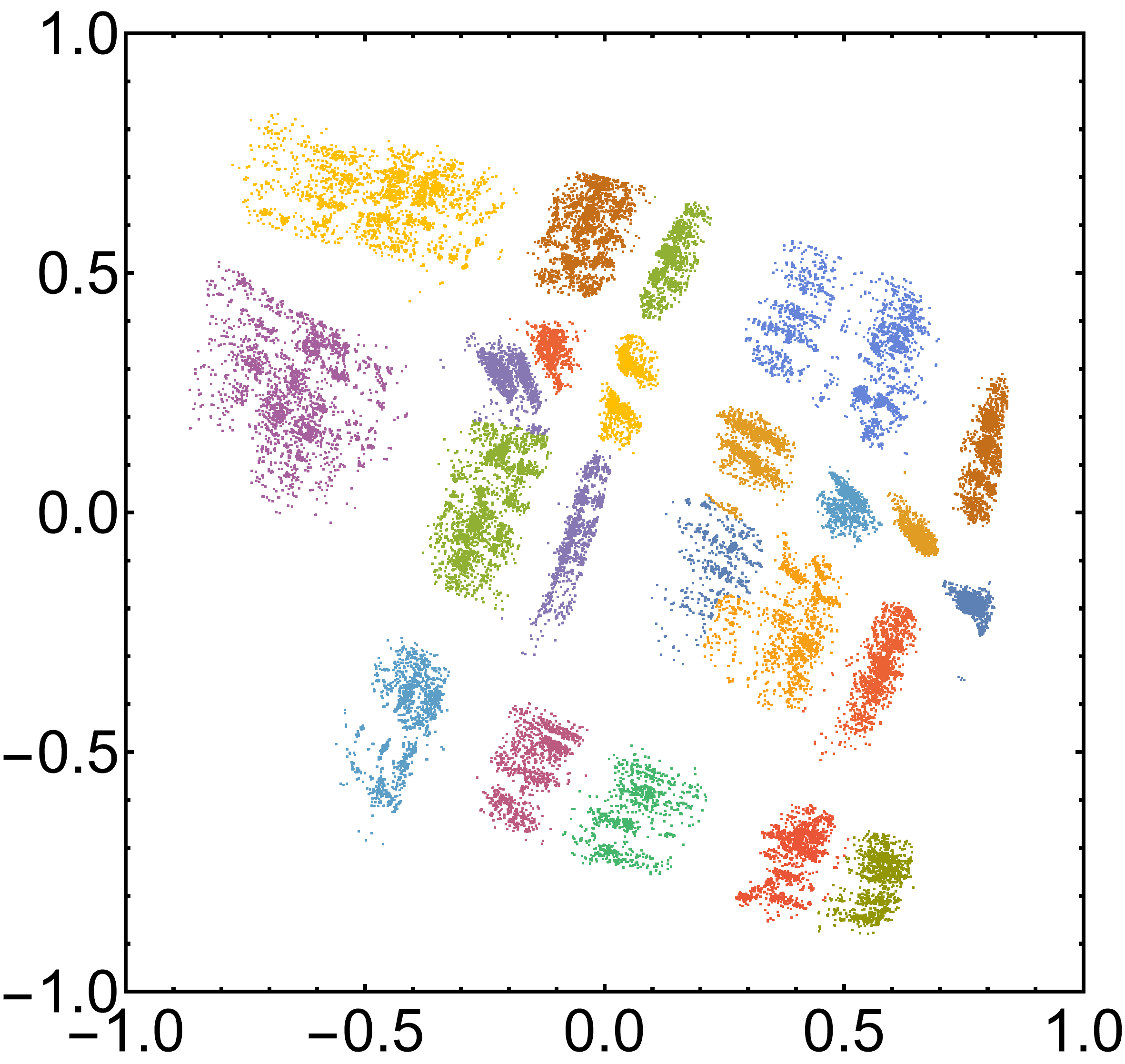}
        \vspace*{-3mm}\caption{\texttt{JarvisPatrick} with radius 0.137.}
        \label{fig:JarvisPatrick}
    \end{minipage} &
    \begin{minipage}[t]{0.33\hsize}
        \centering
        \includegraphics[keepaspectratio, scale=0.24]{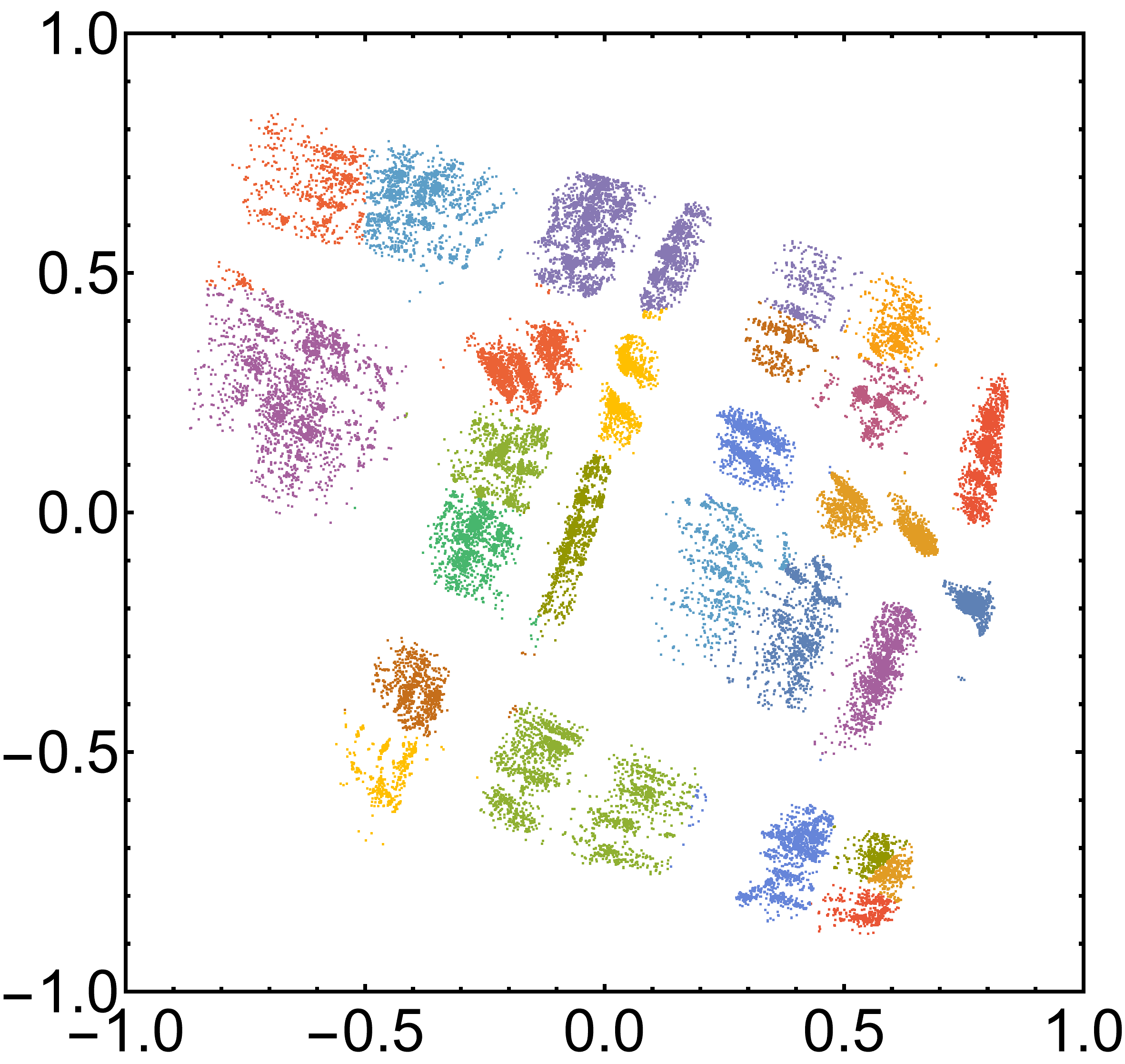}
        \vspace*{-3mm}\caption{\texttt{KMeans} with 27 clusters.}
        \label{fig:KMeans}
    \end{minipage}
    \end{tabular}
\end{figure}

\paragraph{Clustering}\mbox{}\\
For the Model 1, let us start the discussion by focusing on the case AE-2 ($\tanh$-2) with no positional encoding (\Cref{fig:AE-2 PE-ii A0.0}). 
The first difficult task is grouping the configurations into several clusters\footnote{In the following, this grouping procedure is again referred to as clustering.}.
Although we might determine the clustering by hand, if there are some clustering algorithms that can find clusters appropriately, it will be fairer than the rule of thumb.
However, the non-constant densities and sizes of the observed clusters make the problem difficult.
Indeed, tuning parameters of the algorithms changes the results dramatically as illustrated in \Cref{fig:MeanShift,fig:JarvisPatrick,fig:KMeans}\footnote{Here, we used the function \texttt{FindClusters} of Mathematica 11.3. The radius is specified by the \texttt{NeighborhoodRadius} option of the \texttt{JarvisPatrick} algorithm. Other options are left as default.}.
This is a principal difficulty of the problem since the number of configurations for each possible characteristic does not need to be constant, and the clusters should depend on the training progress.
In this problem, choosing a specific algorithm and a set of parameters
would not be fairer than artificial clustering.
Therefore, robustness to changes in clustering methods should always be considered.
In the following analysis, we simply adopt the \texttt{JarvisPatrick} method with a radius of 0.137.
Then, the number of clusters is 20, and the distribution of them is illustrated in \Cref{fig:JarvisPatrick}.

Then, we classify the clusters into two categories: containing or not containing the aligned configurations.
The number of the configurations in each cluster is summarized in \Cref{tab:number of each cluster}.

\begin{table}[H]
\centering
\scalebox{0.7}{
    \begin{tabular}{|c||cccccccccc|}\hline
        Cluster No.& 1& 2& 3*& 4*& 5& 6& 7& 8& 9& 10\\ \hline
        Aligned/Total& 0/1426& 137/1449& 137/3058& 10/2544& 297/3211& 0/1649& 0/2593& 0/2745& 0/1318& 0/1529\\ \hline
        Cluster No.& 11& 12*& 13& 14*& 15& 16& 17& 18& 19& 20 \\ \hline
        Aligned/Total& 209/2384& 86/2186& 0/2202& 11/1868& 56/1344& 134/1881& 95/1351& 75/1877& 0/921& 93/2015\\ \hline
    \end{tabular}
    }
    \caption{The numbers of the configurations in each cluster are summarized. Here, each of the right-side numbers denotes the total number of the configurations in the corresponding cluster, and each of the left-side numbers denotes that of the aligned configurations in the cluster. * denotes the clusters which will be modified in the following.}
    \label{tab:number of each cluster}
\end{table}

\begin{figure}[H]
\centering
    \begin{tabular}{cc}
    \begin{minipage}[t]{0.45\hsize}
        \centering
        \includegraphics[keepaspectratio, width=6.5cm]{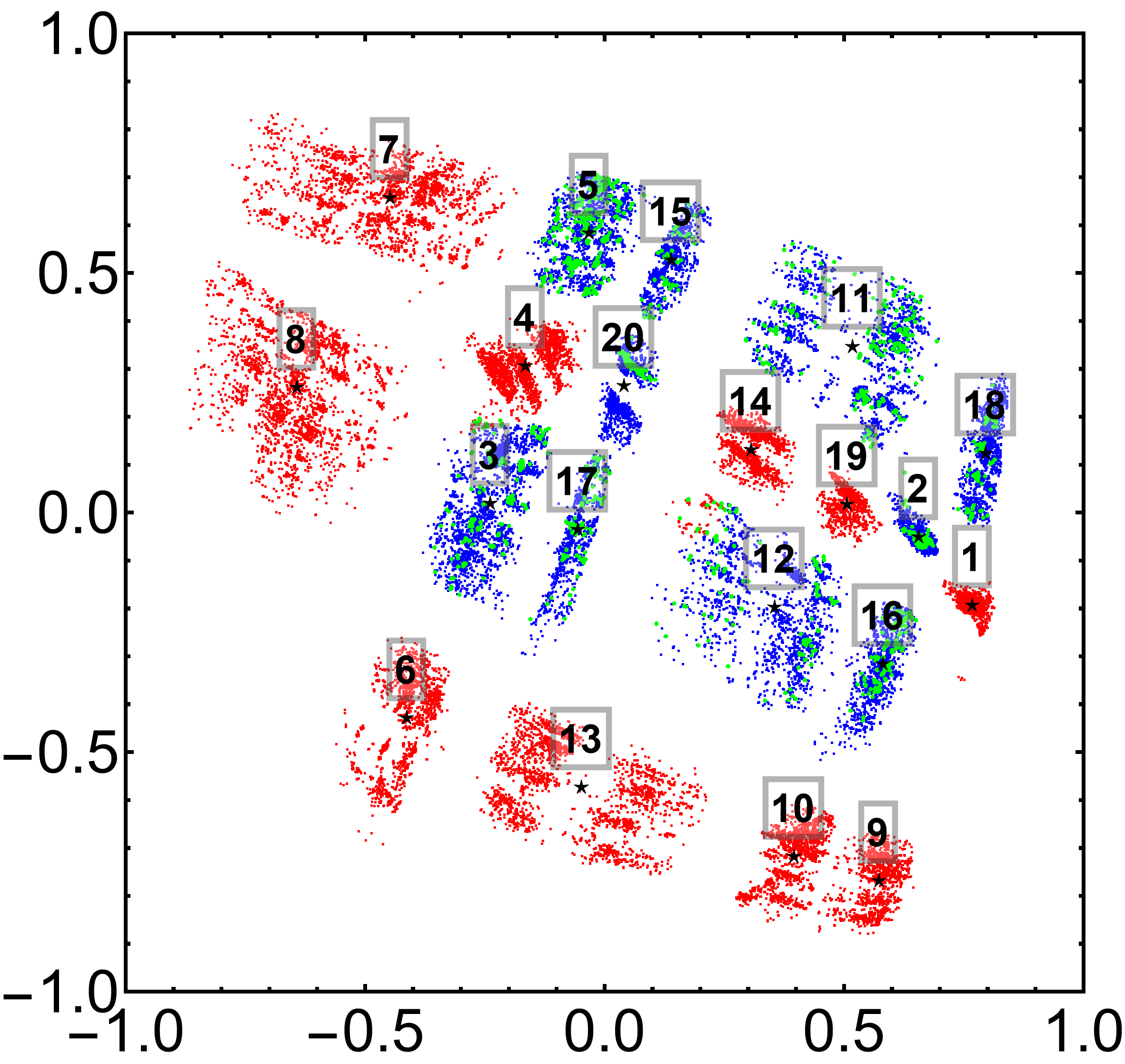}
        \vspace*{-3mm}\caption{The two categories. The blue dots denote the category containing the aligned configurations, while the red dots correspond to the other configurations. The aligned configurations are represented by the green dots. The labels of the clusters correspond to the cluster number (No.) in \Cref{tab:number of each cluster}. Clusters 4 and 14 are colorized as red in this figure.}
        \label{fig:categories}
    \end{minipage} &
    \begin{minipage}[t]{0.45\hsize}
        \centering
        \includegraphics[keepaspectratio, width=6.5cm]{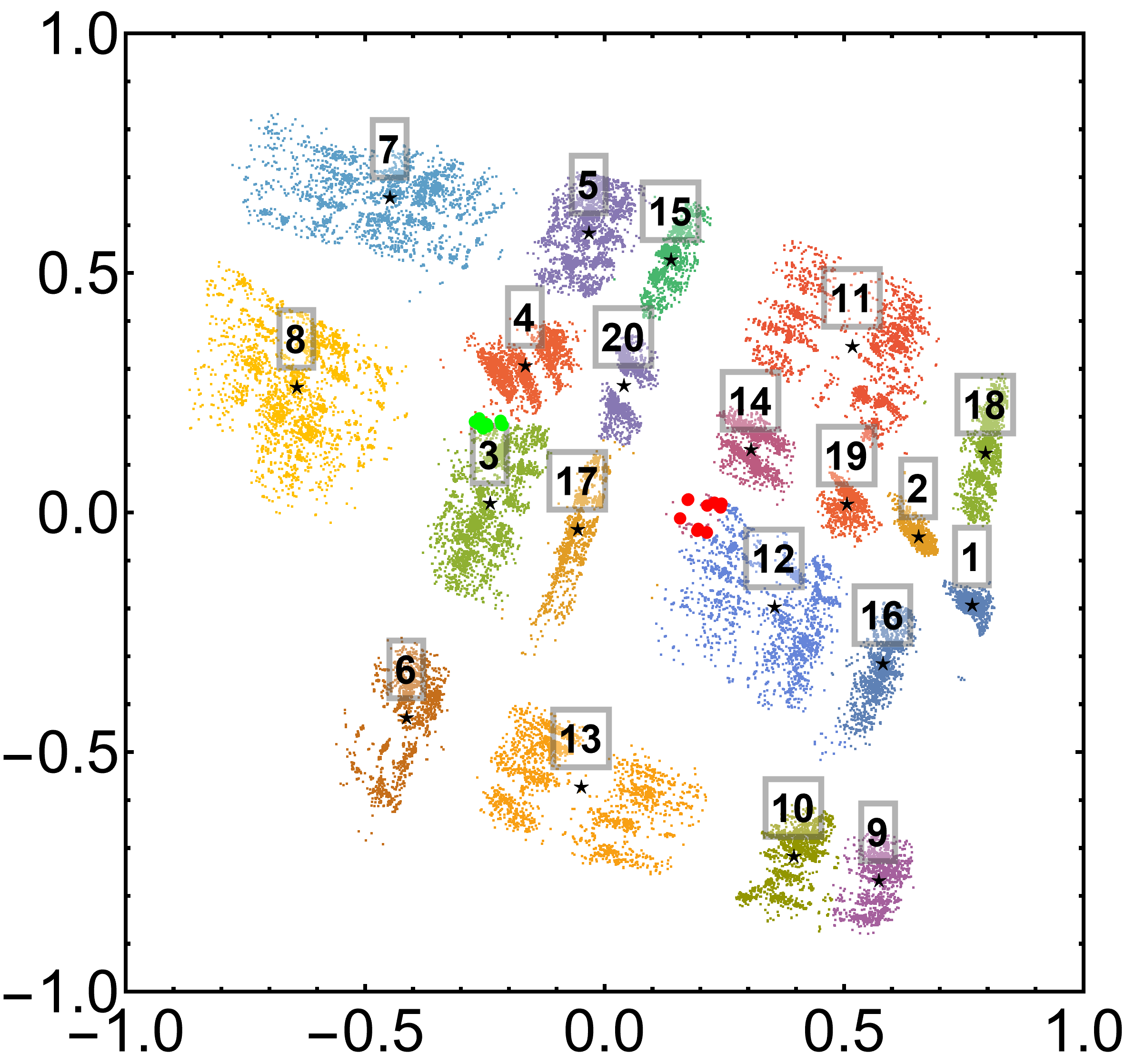}
        \vspace*{-3mm}\caption{The outliers due to the clustering. Initially, the green dots were in Cluster 4, and the red dots were in Cluster 14. We consider them as outliers, and Clusters 4 and 14 are classified into the category not containing the aligned configurations.}
        \label{fig:outliers}
    \end{minipage}
    \end{tabular}
\end{figure}

\begin{figure}[H]
    \centering
    \includegraphics[keepaspectratio, scale=0.35]{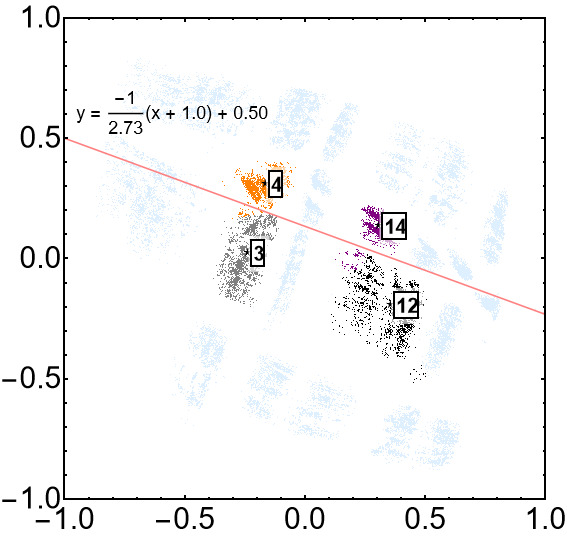}
    \caption{The slight modification by the line shown where $x, y$ are the coordinates for horizontal- and vertical-axes, respectively. For the elements of Clusters 4 (orange) and 14 (purple), we move them to Clusters 3 (gray) and 14 (black) respectively if they are below the line.}
    \label{fig:modification}
\end{figure}

The distributions of the two categories are shown in \Cref{fig:categories}.
Due to the difficulty of the clustering, we observe that 21 aligned configurations are classified into clusters 4 and 14 while different clusters are naively expected.
The situation is illustrated in \Cref{fig:outliers}.
As mentioned, we already tuned parameters and methods to somehow justify the clustering, but it is nearly artificial.
Thus, we would like to modify the clustering slightly by hand
\footnote{We need to reconstruct the clustering since we consider the simple binary classification.
If there are other approaches which can characterize the observed differences between two categories without concrete boundaries of the clusters, we do not need the reconstruction.
It will be better than the modification, but we do not study such approaches here and simply modify the clustering.
We need to show the modification explicitly to ensure transparency.
}.
We separate the clusters by drawing a line as \Cref{fig:modification}.
The numbers of the configurations in modified clusters are shown in \Cref{tab:number of each modified cluster}.
The difference is at most $5.7\%$.

\begin{table}[H]
\centering
\scalebox{0.7}{
    \begin{tabular}{|c||cccccccccc|}\hline
        Cluster No.& 1& 2& 3*& 4*& 5& 6& 7& 8& 9& 10\\ \hline
        Aligned/Total& 0/1426& 137/1449& 147/3140& 0/2462& 297/3211& 297/1649& 0/2593& 0/2745& 0/1318& 0/1529\\ \hline
        Cluster No.& 11& 12*& 13& 14*& 15& 16& 17& 18& 19& 20 \\ \hline
        Aligned/Total& 209/2384& 107/2292& 0/2202& 0/1762& 56/1344& 134/1881& 95/1351& 75/1877& 0/921& 93/2015\\ \hline
    \end{tabular}
    }
    \caption{The numbers of the configurations in each cluster are summarized. Clusters 3, 4, 12, and 14 are modified ones. The notation is the same as that in \Cref{tab:number of each cluster}.}
    \label{tab:number of each modified cluster}
\end{table}

\paragraph{Distributions of the hidden tadpole charges}\mbox{}\\
Having defined the clustering, we then compare the two categories.
To find a feature which the autoencoder weighted, we have to take a brute-force way and examine various quantities, which include the number of hidden gauge groups, their total rank, net chirality \cite{Gmeiner:2005vz}, the generations of quarks or leptons if they are aligned.
Unfortunately, most of the distributions are nearly constant over the clusters and hence they do not characterize differences between the clusters\footnote{It is possible that some combinations of these quantities characterize the dataset. Note that we investigated these quantities over the clusters that contain not only the aligned configurations but the not aligned ones.}.
However, we find that a certain quantity clusters at several islands, and that is the tadpole charge of the hidden sector:
\begin{align}
Q^I_{\rm hid} \coloneqq \sum_{a:\text{hidden}} N_a \hat{X}^{I}_a = 8 - \sum_{a:\text{visible}} N_a \hat{X}^{I}_a, \quad \forall I.    
\end{align}

\begin{figure}[H]
    \centering
    \includegraphics[width=7.5cm]{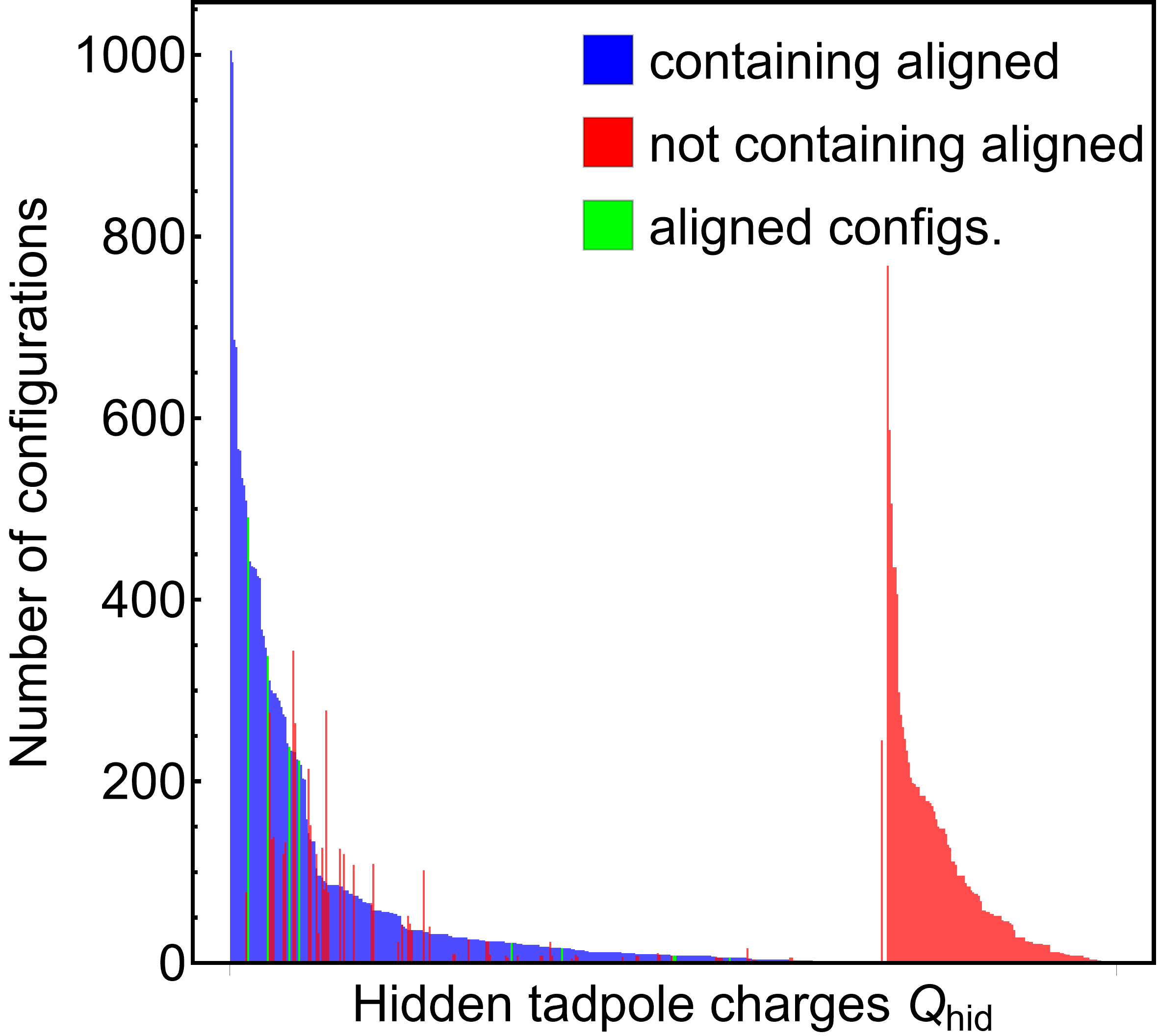}
    \caption{The distribution of the number of configurations for each hidden tadpole charge. The blue bar chart is that of the category containing aligned configurations, while the red bar chart corresponds to the category not containing them. The green one is that of exactly aligned configurations (the green bar chart is a part of the blue one). Note that the hidden tadpole charges of the aligned configurations turn out to be unique for them. The horizontal axis is a set of possible hidden tadpole charges, which is sorted so that a larger number of configurations in the former category comes first.}
    \label{fig:distributionofTC}
\end{figure}

In the dataset, 455 different sets of total tadpole charges of the hidden sector (hidden tadpole charge) are found.
337 of 455 tadpole charges are included in the category containing the aligned configurations, and 206 of 455 are found in the other category.
The number of common tadpole charges is only 88, and we illustrate the observed hidden tadpole charges of each category in \Cref{fig:distributionofTC}.
We can see that the blue bar chart and the red bar chart have their peaks at the different points.
Therefore, we assume that the autoencoder considers the hidden tadpole charge as an important factor.
We obtain the same result with the clustering without the modification.
In this dataset, $69.4\%$ of the configurations in the category containing aligned configurations (blue) have the unique hidden tadpole charges for the category. 
On the other hand, $73.4\%$ of the configurations in the category not containing aligned configurations (red) have the unique hidden tadpole charges for the category.

\begin{figure}[H]
\centering
    \begin{tabular}{cc}
    \begin{minipage}[t]{0.45\hsize}
        \centering
        \includegraphics[keepaspectratio, width=6.5cm]{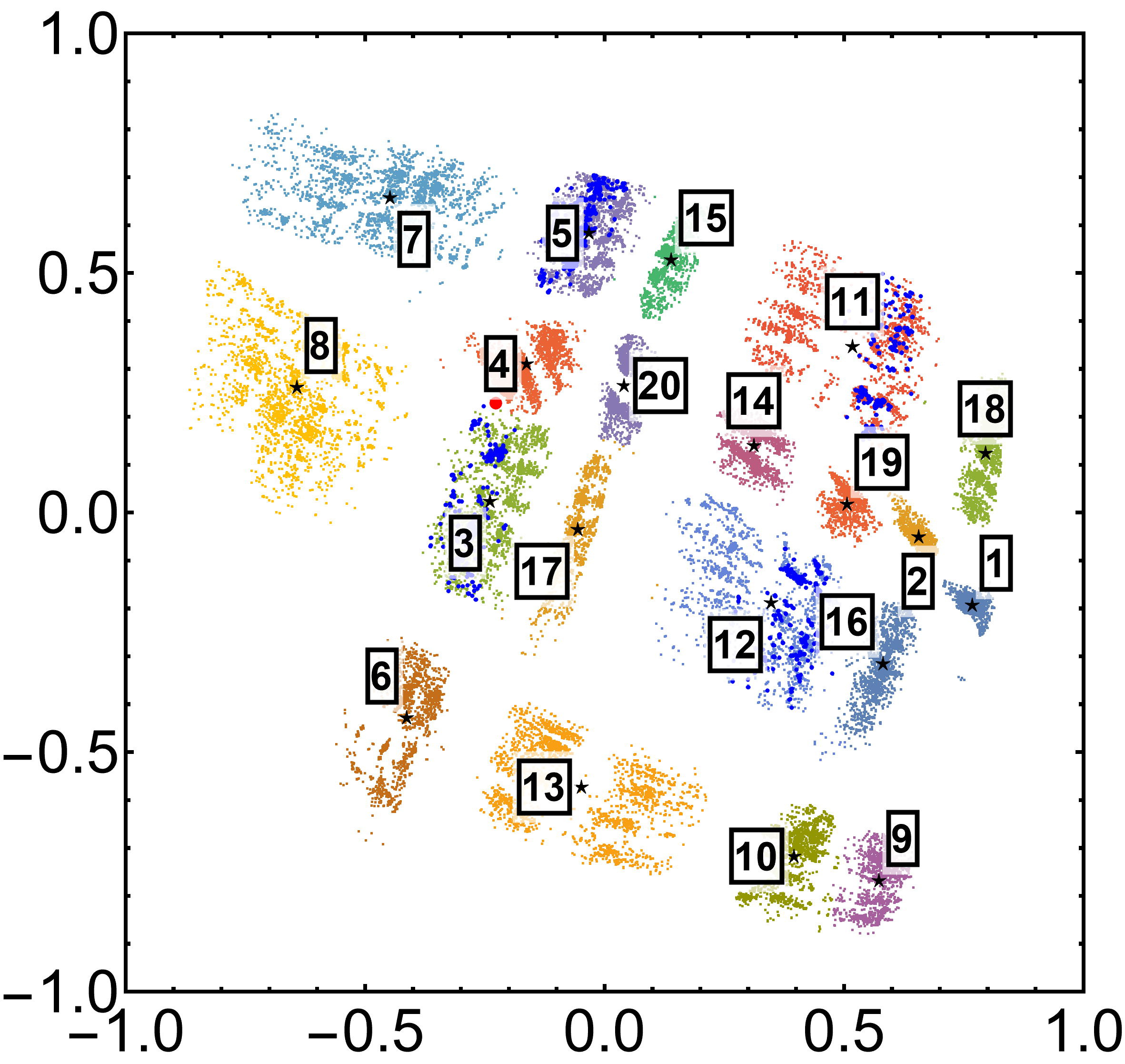}
        \vspace*{-3mm}\caption{Hidden tadpole charge: $Q_{\rm hid}=(1,7,6,5)$. Except for one red point, all of the 1,005 configurations with the charge are in the category containing aligned configurations (blue dots).}
        \label{fig:TC1765}
    \end{minipage} &
    \begin{minipage}[t]{0.45\hsize}
        \centering
        \includegraphics[keepaspectratio, width=6.5cm]{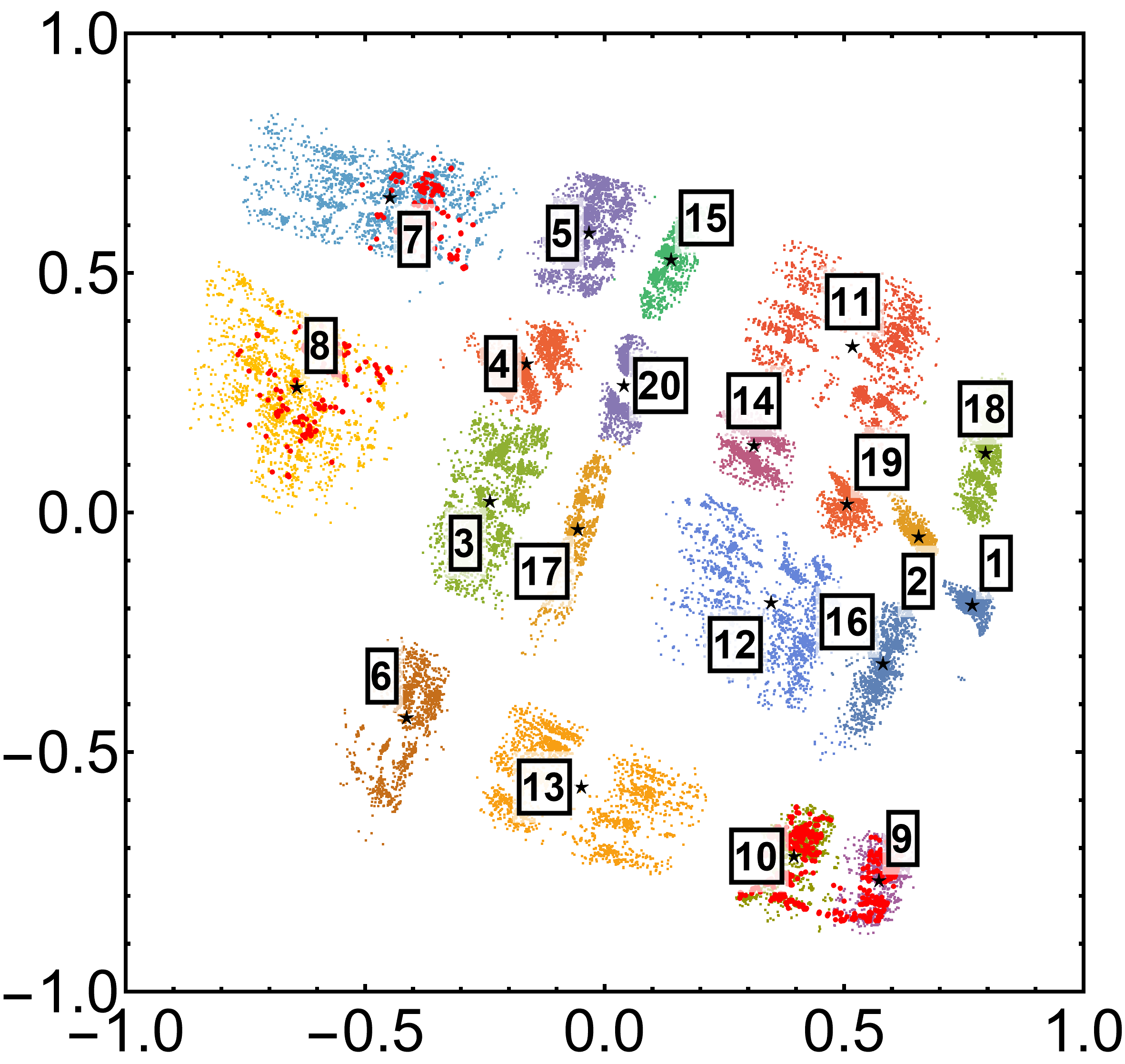}
        \vspace*{-3mm}\caption{Hidden tadpole charge: $Q_{\rm hid}=(2,6,2,7)$. All of the 768 configurations with the charge are in the category not containing aligned configurations (red dots).}
        \label{fig:TC2627}
    \end{minipage}\\
    \begin{minipage}[t]{0.45\hsize}
        \centering
        \includegraphics[keepaspectratio, width=6.5cm]{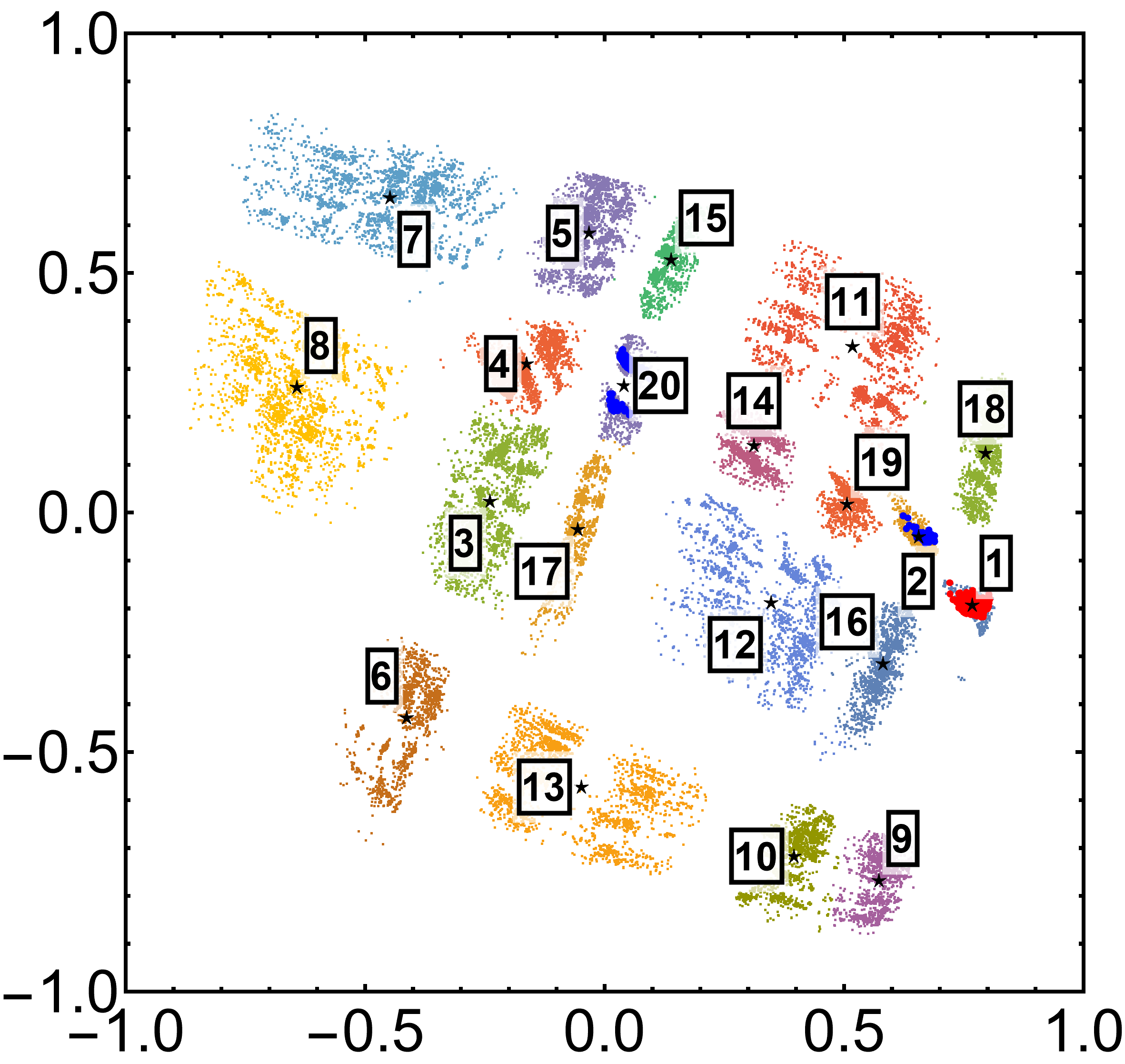}
        \vspace*{-3mm}\caption{Hidden tadpole charge: $Q_{\rm hid}=(6,7,5,1)$. Both two categories have the hidden tadpole charge, but the two distributions colored blue and red are close to each other (Clusters 1 and 2).}
        \label{fig:TC6751}
    \end{minipage}&
    \begin{minipage}[t]{0.45\hsize}
        \centering
        \includegraphics[keepaspectratio, width=6.5cm]{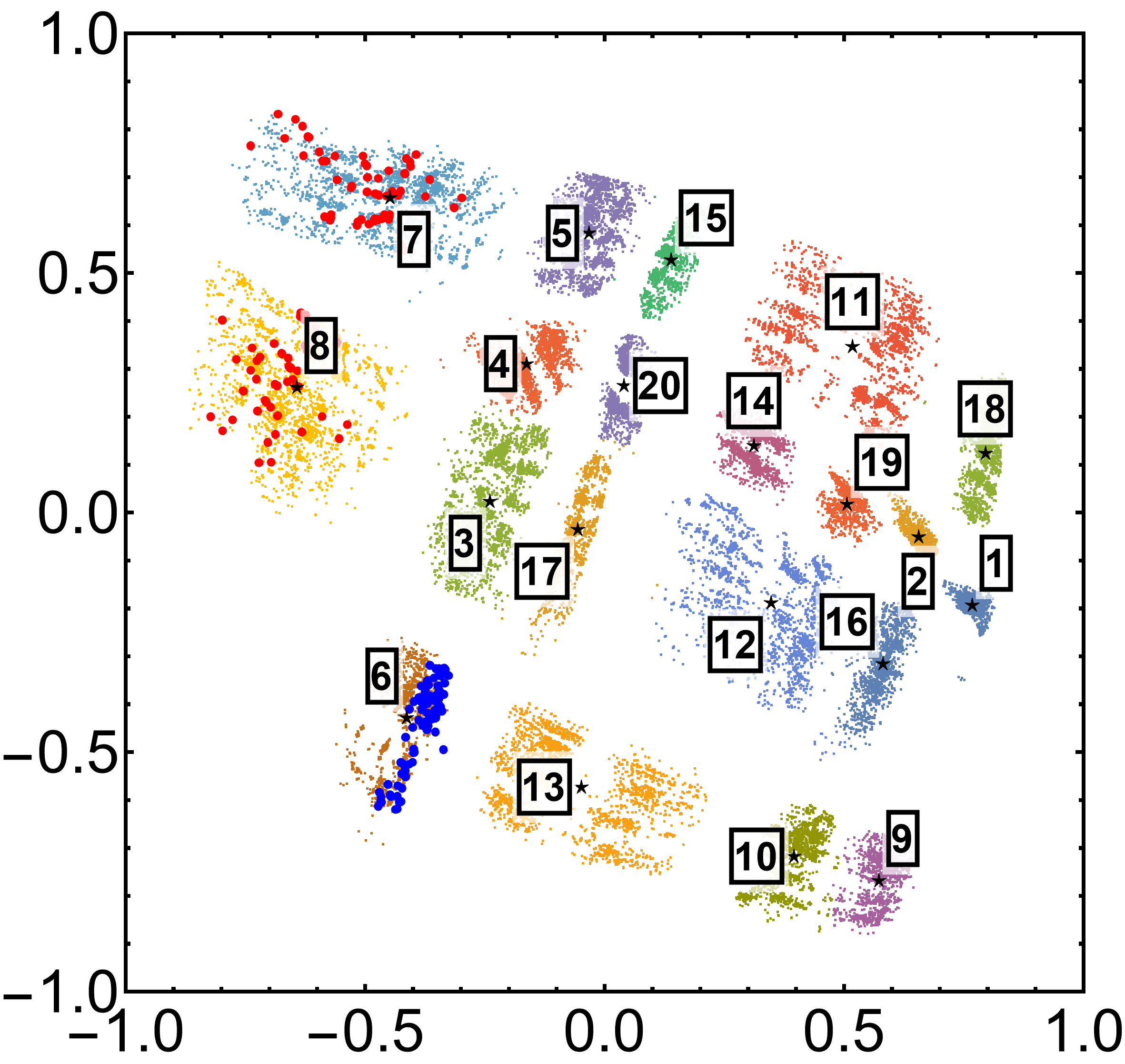}
        \vspace*{-3mm}\caption{Hidden tadpole charge: $Q_{\rm hid}=(3,2,6,6)$. Both two categories have the charge, and the two distributions colored blue and red seem to be independent of each other.}
        \label{fig:TC3266}
    \end{minipage}
    \end{tabular}
\end{figure}

However, as illustrated in \Cref{fig:distributionofTC}, there are 88 overlaps between the two categories.
Taking into account the checkerboard pattern in the latent layer, \Cref{fig:AE-2 PE-ii A0.0}, we expect there is another important factor weighted by the autoencoder.
We could not find the other feature by the brute-force method.
The lack of explanation is one of the crucial difficulties of the usual machine learning approach.
We simply finish the discussion by assuming that there exists another feature as important as the hidden tadpole charge.
Several examples of the clustering of hidden tadpole charges are given in \Cref{fig:TC1765,fig:TC2627,fig:TC6751,fig:TC3266}.
Notably, the configurations with the same hidden tadpole charge are likely to be placed along several straight belts.
This feature will also be observed in other methods and models which we will analyze below.
We expect the appearance of those belts and the checkerboard pattern to be correlated, and it supports the expectation that another important factor exists.
This other feature is expected to explain both the reason why the configurations with the same hidden tadpole charge scatter in several clusters (while they form several lines), and the reason why the configurations with different hidden tadpole charges gather in the same cluster.

Note that we found the aligned configurations have unique hidden tadpole charges compared to other configurations in this dataset, although this phenomenon was not expected.
Indeed, those of aligned configurations are also found in the other configurations in the $U(2)$ cases, as discussed later.

\begin{figure}[H]
\centering
    \begin{tabular}{cc}
    \begin{minipage}[t]{0.45\hsize}
        \centering
        \includegraphics[keepaspectratio, width=6.5cm]{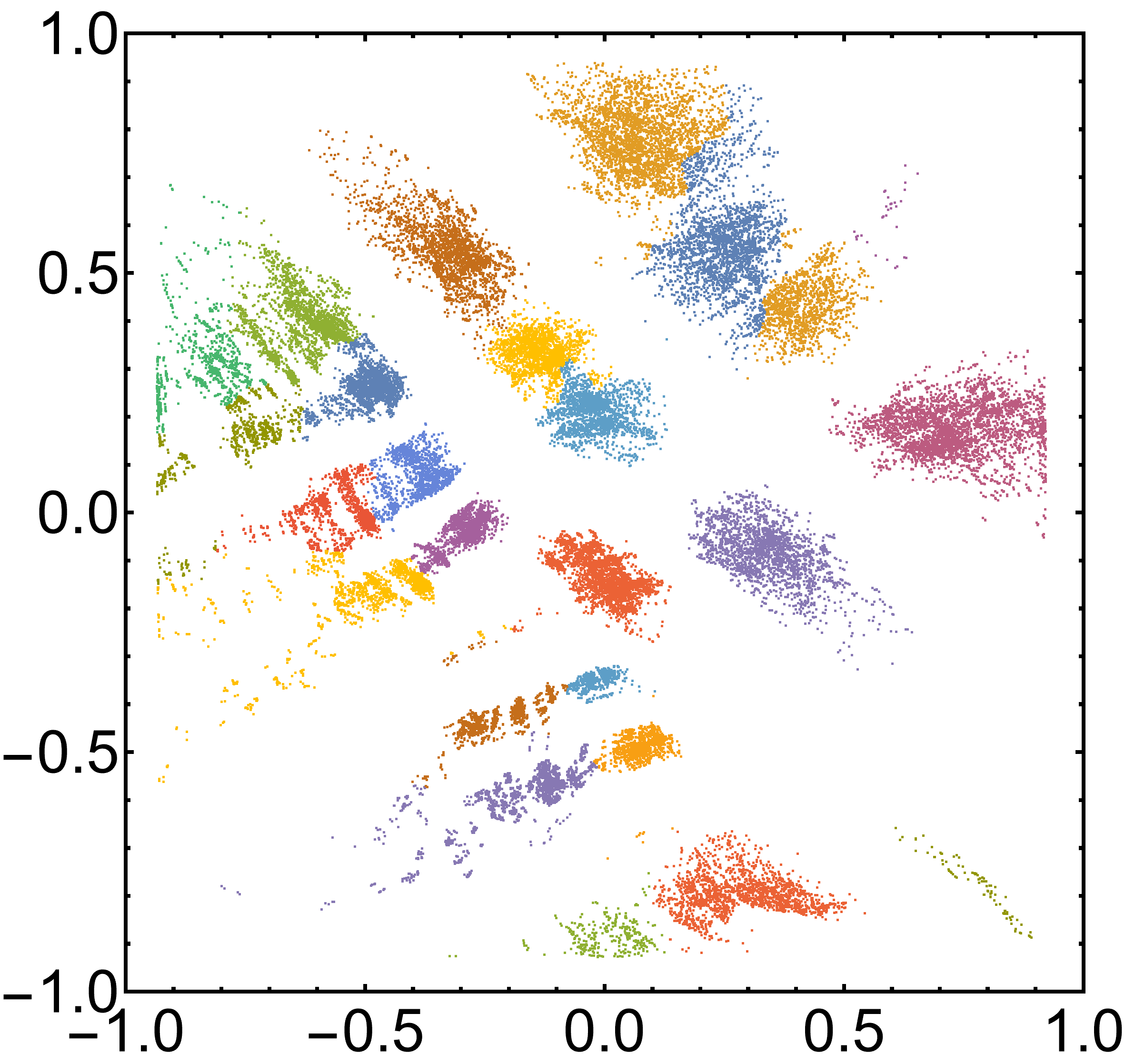}
        \vspace*{-3mm}\caption{\texttt{MeanShift}.}
        \label{fig:MeanShift-A1.0}
    \end{minipage} &
    \begin{minipage}[t]{0.45\hsize}
        \centering
        \includegraphics[keepaspectratio, width=6.5cm]{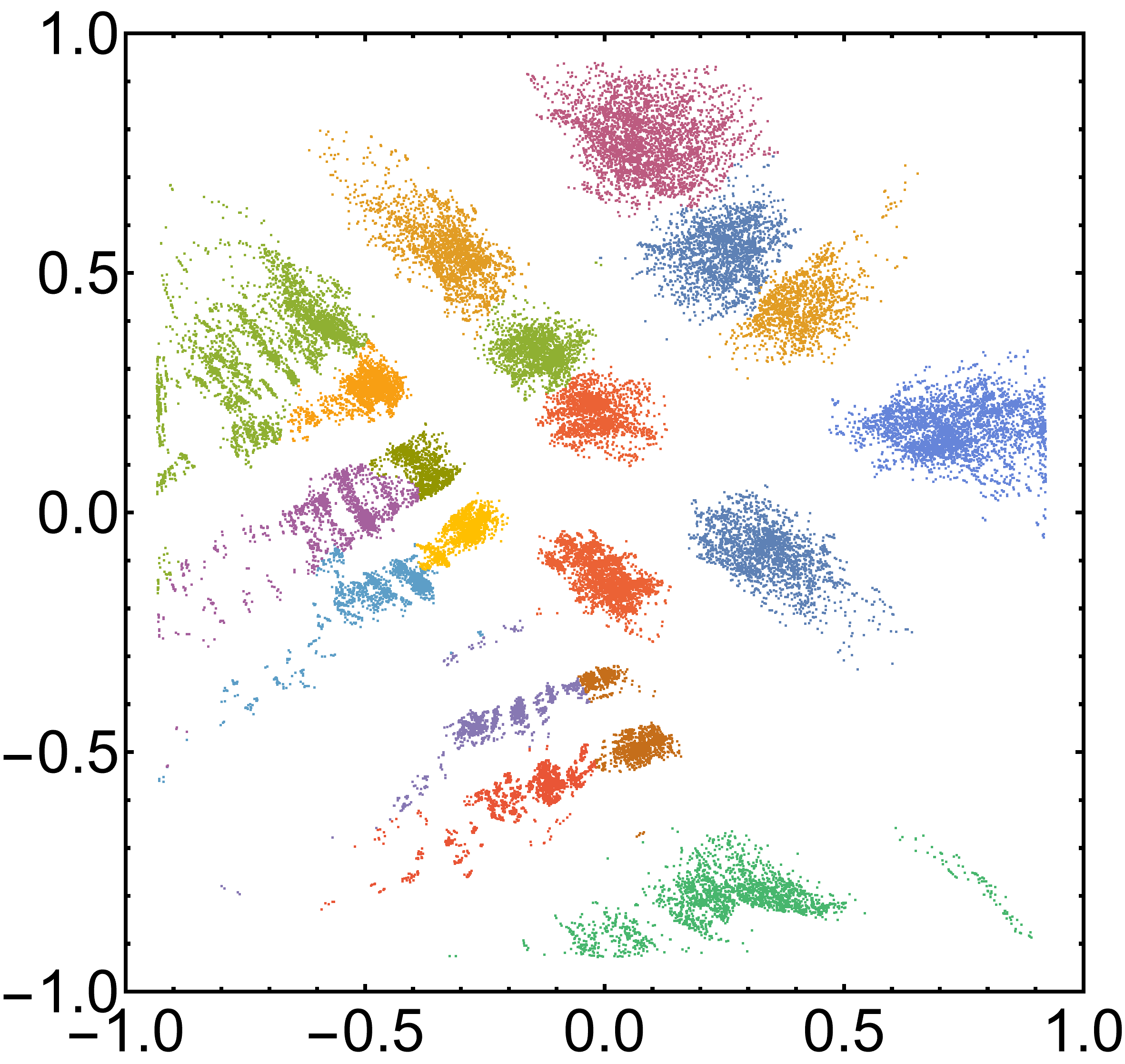}
        \vspace*{-3mm}\caption{\texttt{JarvisPatrick} with radius 0.31.}
        \label{fig:JarvisPatrick-A1.0}
    \end{minipage}
    \end{tabular}
\end{figure}

\paragraph{Robustness: other methods}\mbox{}\\
So far, we have focused on the AE-2 model with tanh-2 configuration and no positional encoding.
Then, it is natural to discuss whether the feature found in the previous case is also possessed by the other models.
For this purpose, let us consider the AE-2 model with the same tanh-2 configuration, but with a large amplitude $A = 1.0$ (\Cref{fig:AE-2 PE-ii A1.0}).
In this case, the aligned configurations seem to cluster again, as a checkerboard pattern is observed.
We adopted two clustering algorithms with fine-tuning and showed the results in \Cref{fig:MeanShift-A1.0,fig:JarvisPatrick-A1.0}.
However, the boundaries between the islands become more vague, and we need to choose clustering algorithms carefully so that they realize a grouping which is the same as expected.
Thus, we should note that the clustering is also artificial in this case.
If we divide the configurations completely by hand, we may draw lines to form a grid to divide the area.
In the following, our discussion is based on the \texttt{JarvisPatrick} algorithm in \Cref{fig:JarvisPatrick-A1.0}.
Then, we have 19 clusters in this case.
Ideally, the number of clusters should be 20, the same as in the previous case.

\begin{figure}[H]
    \centering
    \includegraphics[keepaspectratio, scale=0.35]{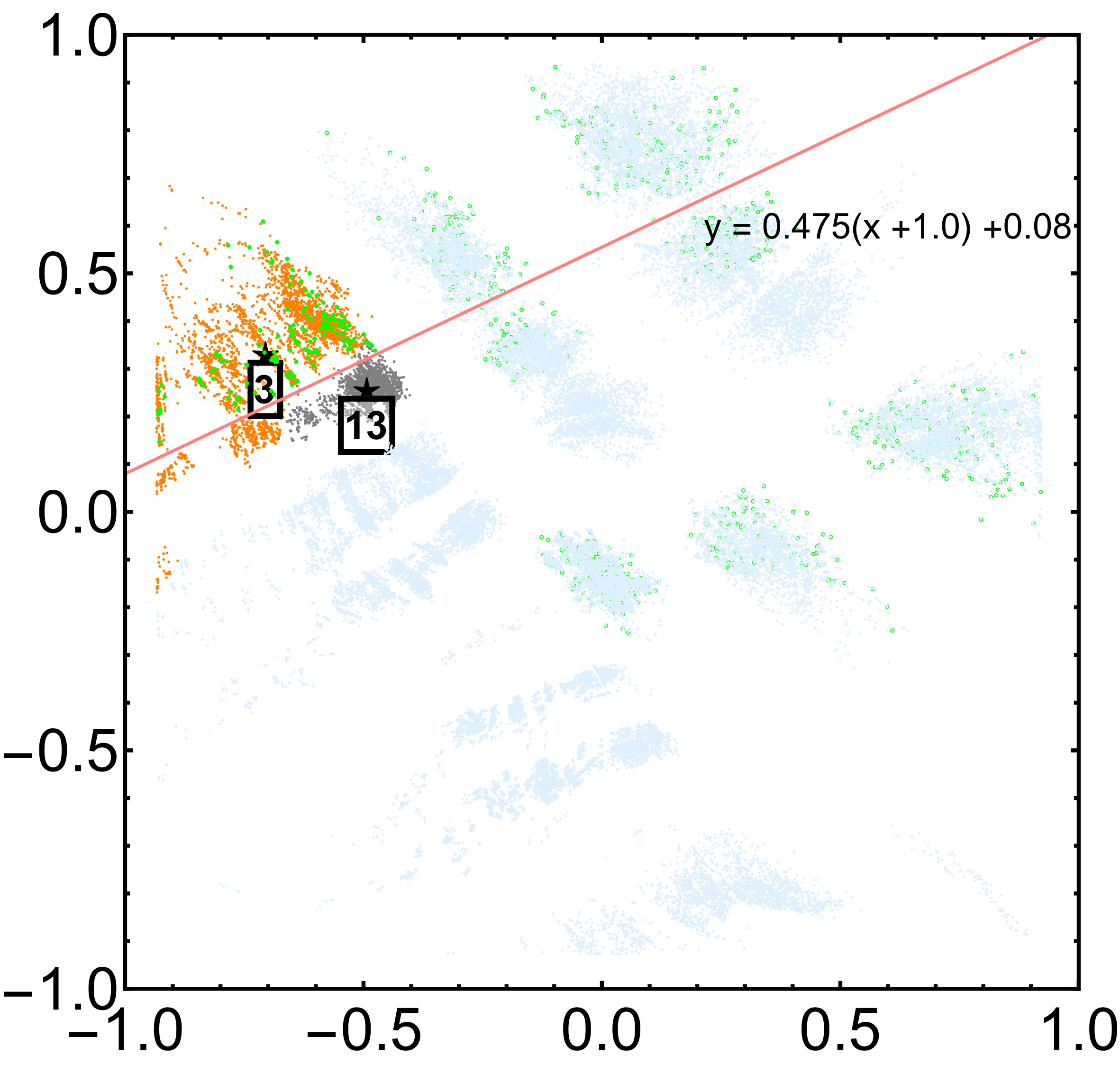}
    \caption{The modification by the line shown where $x, y$ are the coordinates for horizontal- and vertical-axes, respectively. The green dots are the aligned configurations. For the elements of Cluster 3 (orange), we move them to Cluster 3 (gray) if they are below the line. The converse is also done. Note that the line passes through the blank space also between the other clusters.}
    \label{fig:modification-A1}
\end{figure}

As in the previous case, there is a cluster which has only several aligned configurations despite its large number of configurations.
We again reconstruct the relevant two clusters by drawing a line respecting the checkerboard pattern. 
With a modification of the clustering which is illustrated in \Cref{fig:modification-A1}, the numbers of the configurations in each cluster are summarized in \Cref{tab:number of each modified cluster-A1}.

\begin{figure}[H]
    \centering
    \includegraphics[width=7.5cm]{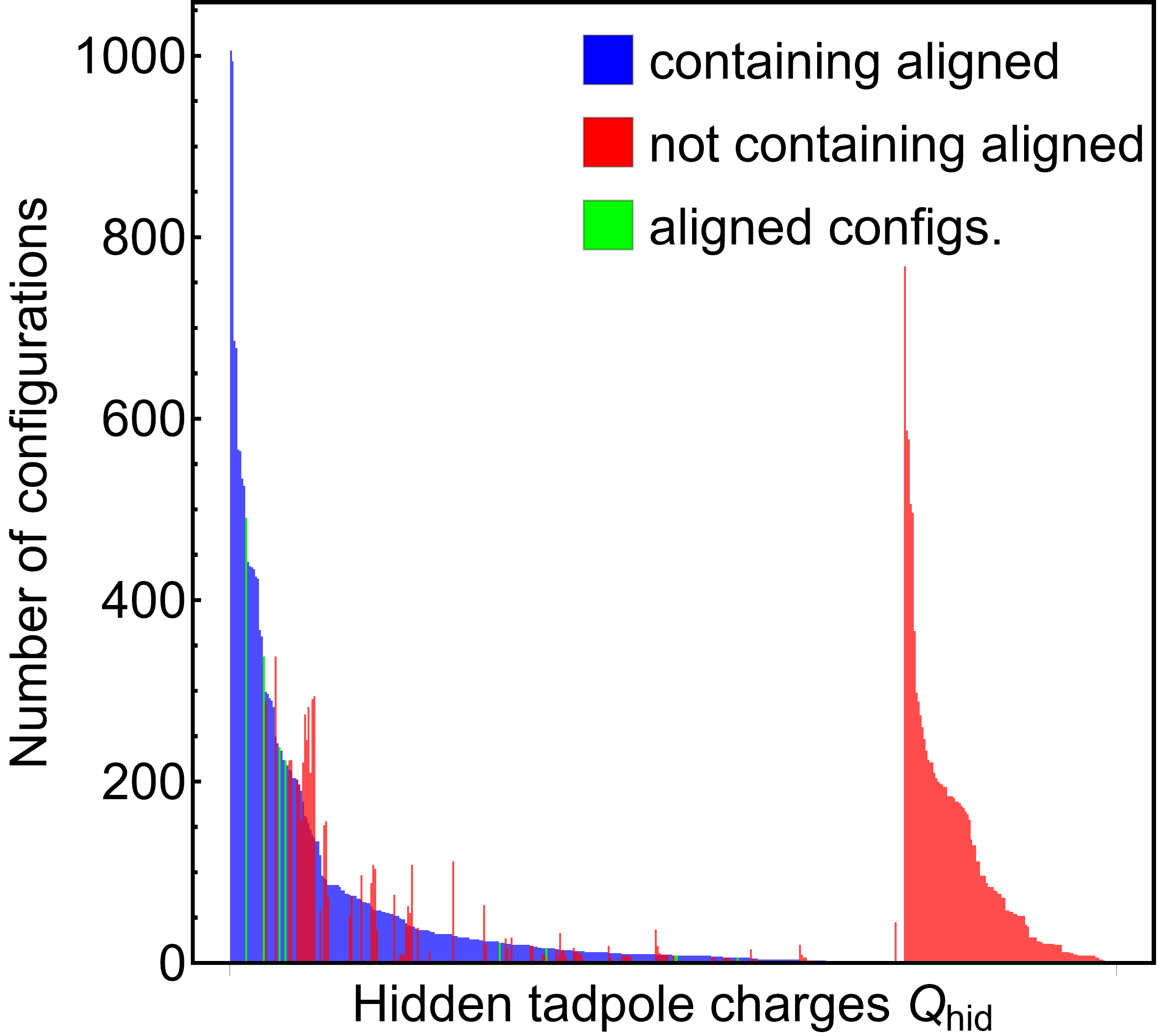}
    \caption{The distributions of the number of configurations for each hidden tadpole charge are shown. Notations are the same as in \Cref{fig:distributionofTC}. We can see the two peaks are located in different places.}
    \label{fig:distributionofTC-A1}
\end{figure}

\begin{figure}[H]
\centering
    \begin{tabular}{cc}
    \begin{minipage}[t]{0.45\hsize}
        \centering
        \includegraphics[keepaspectratio, width=6.3cm]{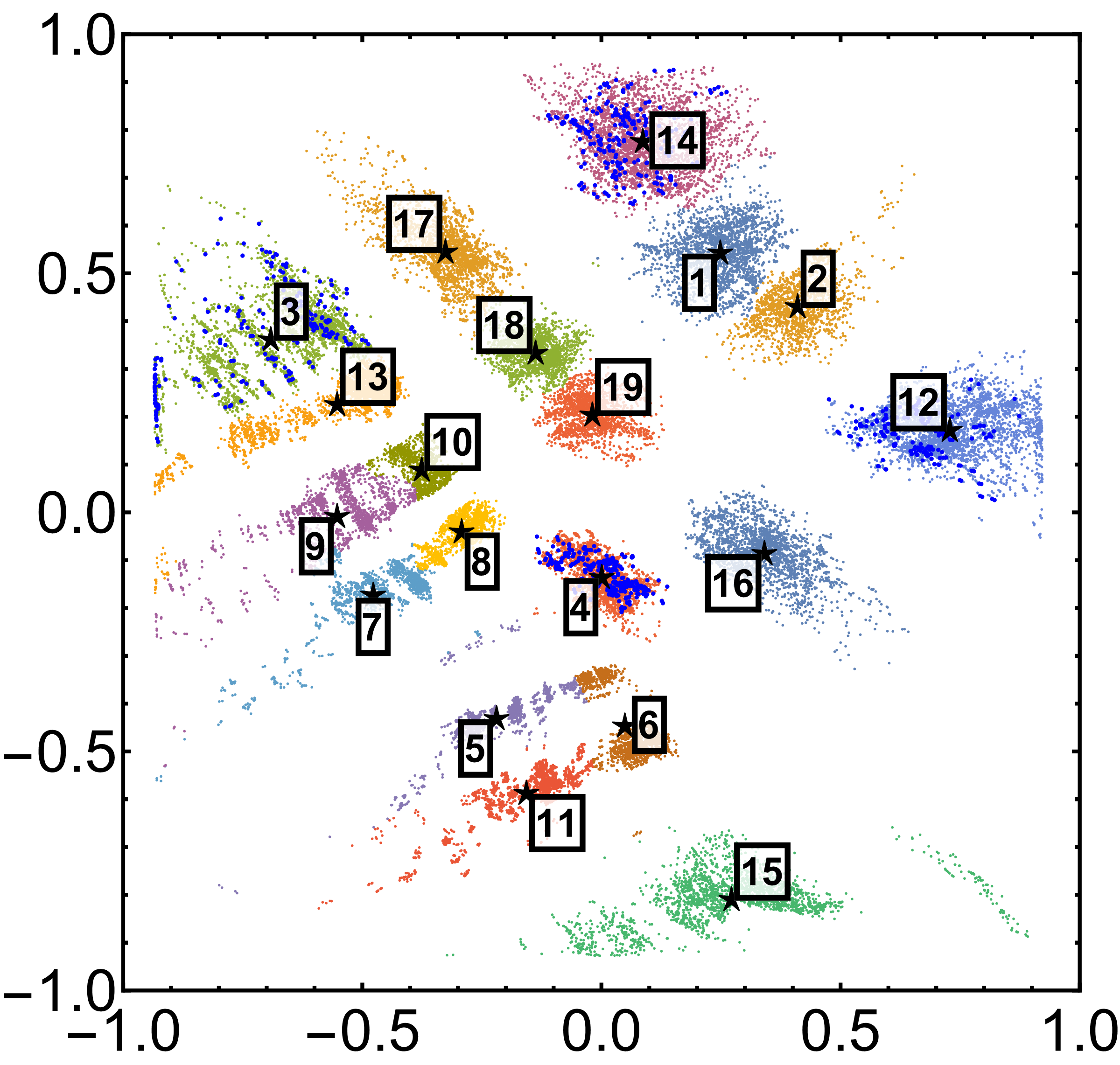}
        \vspace*{-4mm}\caption{Hidden tadpole charge: $Q_{\rm hid} = (1,7,6,5)$. All of the 1,005 configurations with the charge are in the category containing aligned configurations (blue dots).}
        \label{fig:TC1765-A1}
    \end{minipage} &
    \begin{minipage}[t]{0.45\hsize}
        \centering
        \includegraphics[keepaspectratio, width=6.3cm]{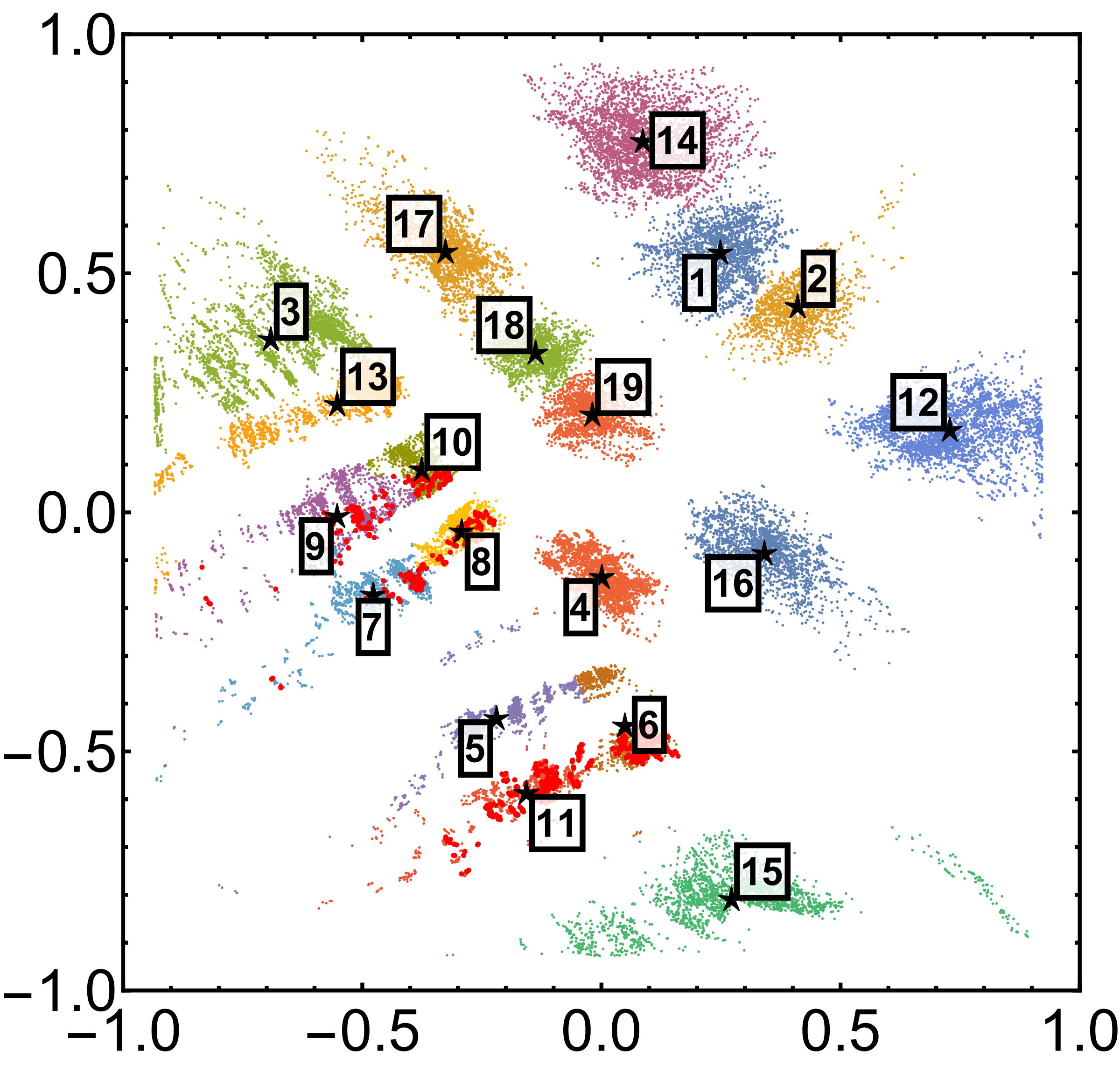}
        \vspace*{-4mm}\caption{Hidden tadpole charge: $Q_{\rm hid} = (2,6,2,7)$. All of the 768 configurations with the charge are in the category not containing aligned configurations (red dots).}
        \label{fig:TC2627-A1}
    \end{minipage}
    \end{tabular}
\end{figure}

Then we classify the data, as in the previous case, into two categories: those that contain the aligned configurations and those that do not.
The corresponding result to \Cref{fig:distributionofTC} is shown in \Cref{fig:distributionofTC-A1}.
As in the previous case, we can see that two distributions of the hidden tadpole charges have their peaks at the different points. 
Thus, we conclude that the hidden tadpole charge is the important factor also in this case with a large positional encoding, and the previous analysis is robust against the encoding. 
We close the analysis of the Model 1 by showing the examples of clustering corresponding to \Cref{fig:TC1765,fig:TC2627} in \Cref{fig:TC1765-A1,fig:TC2627-A1}, respectively.

\begin{table}[H]
\centering
\scalebox{0.7}{
    \begin{tabular}{|c||cccccccccc|}\hline
        Cluster No.& 1& 2& 3*& 4& 5& 6& 7& 8& 9& 10\\ \hline
        Aligned/Total& 136/2329& 0/1494& 147/3173& 297/3171& 0/1132& 0/1681& 0/1326& 0/1332& 0/1606& 0/1048\\ \hline
        Cluster No.& 11& 12& 13*& 14& 15& 16& 17& 18& 19& \\ \hline
        Aligned/Total& 0/1720 & 233/3431& 0/2386& 189/3412& 0/2205& 108/2173& 147/2214& 93/1949& 0/1769 &\\ \hline
    \end{tabular}
    }
    \caption{The numbers of the configurations in each cluster. Clusters 3 and 13 are modified ones.}
    \label{tab:number of each modified cluster-A1}
\end{table}

\subsubsection{Models 2 and 3}
\label{sec:feature extraction of Models 2 and 3}
\begin{figure}[H]
\centering
    \begin{tabular}{cc}
    \begin{minipage}[t]{0.45\hsize}
        \centering
        \includegraphics[keepaspectratio, width=6.3cm]{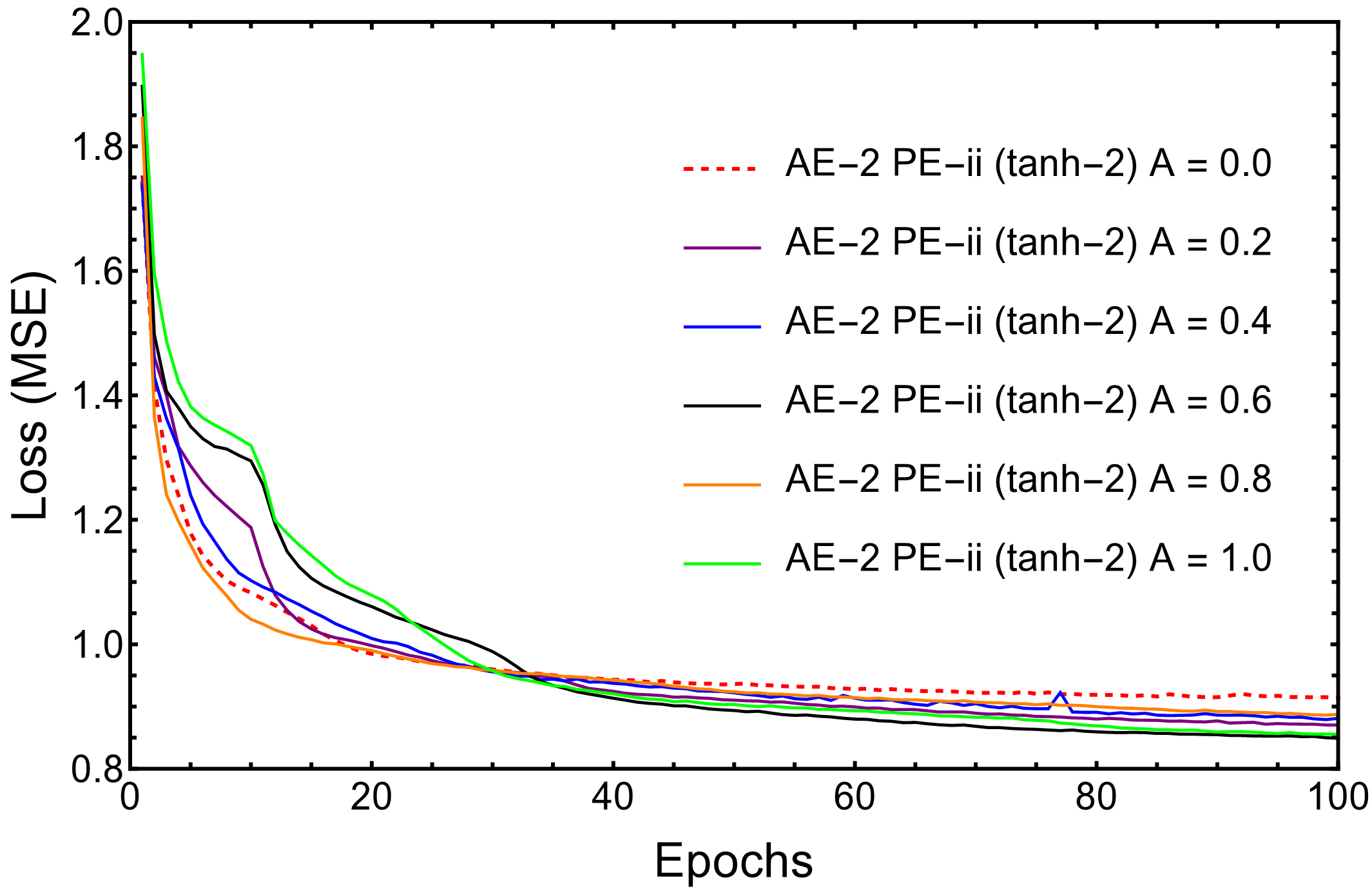}
        \vspace*{-4mm}\caption{Loss in training of AE-2 with the amplitudes for PE-ii. Model 2 is used.}
        \label{fig:losses_AE-2-variousA-u1ys}
    \end{minipage} &
    \begin{minipage}[t]{0.45\hsize}
        \centering
        \includegraphics[keepaspectratio, width=6.3cm]{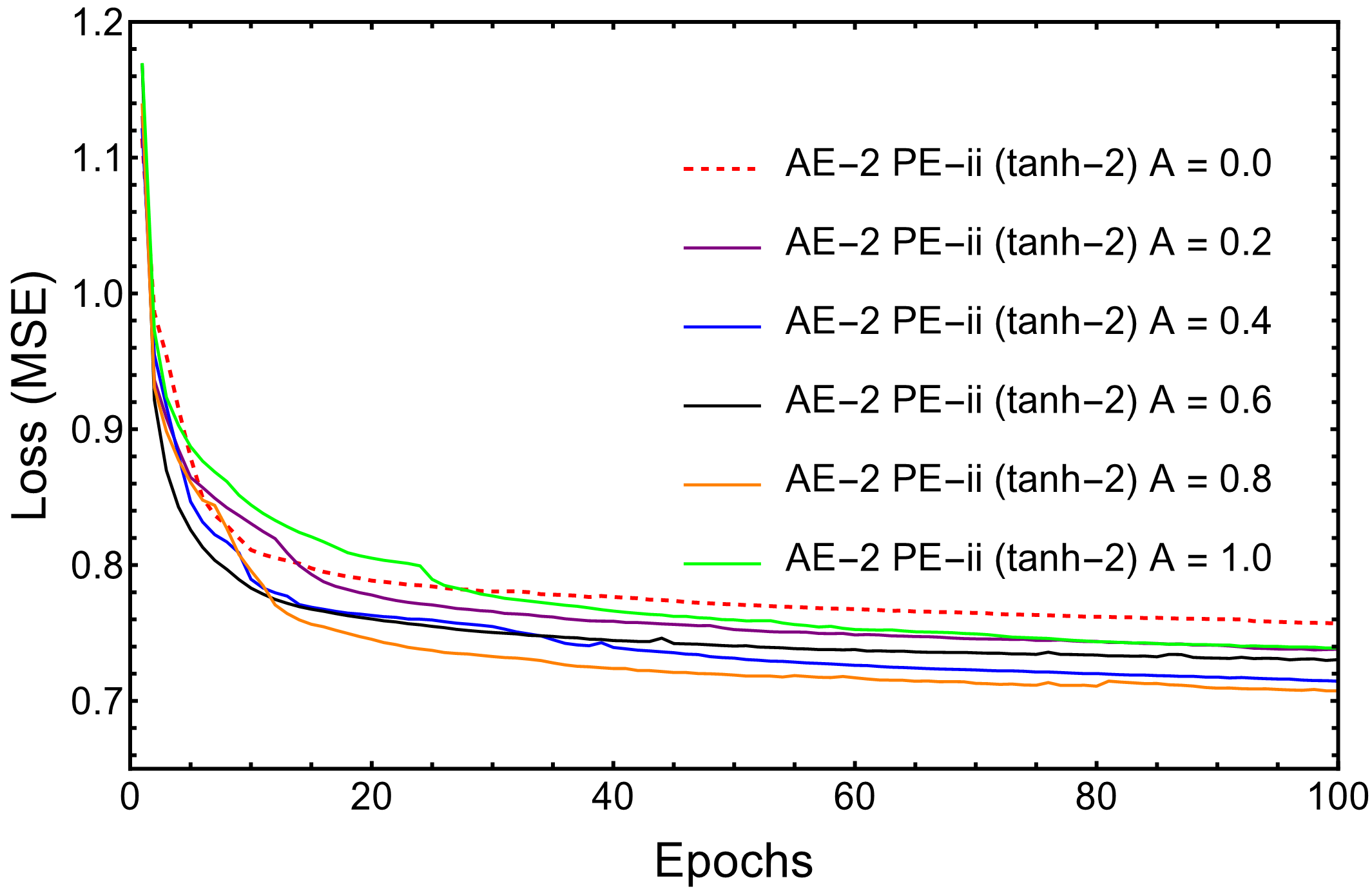}
        \vspace*{-4mm}\caption{Loss in training of AE-2 with the amplitudes for PE-ii. Model 3 is used.}
        \label{fig:losses_AE-2-variousA-u1y1}
    \end{minipage}
    \end{tabular}
\end{figure}

\paragraph{Clustering}\mbox{}\\
In the following, we focus on studying whether the autoencoder models weight the hidden tadpole charges again or not. 
Considering the previous results, we trained only the AE-2 model with $\tanh$-2 and with the amplitudes $A \in \{0.0, 0.2, 0.4, 0.6, 0.8, 1.0\}$. 
Unfortunately, the clustering turns out to be more vague than in the previous case.
The training losses are summarized in \Cref{fig:losses_AE-2-variousA-u1ys,fig:losses_AE-2-variousA-u1y1}.
For Model 2, the losses remain large even after 100 epochs, compared to Model 3.
However, as we studied in the previous section, we would like to extract which features the autoencoder model weight in reproducing the input data.
Since the losses indeed decreased in the training process, we study clustering with these autoencoder models.

\begin{figure}[H]
\hspace*{-10mm}
    \begin{tabular}{ccc}
      \begin{minipage}[t]{0.35\hsize}
        \centering
        \includegraphics[keepaspectratio, scale=0.40]{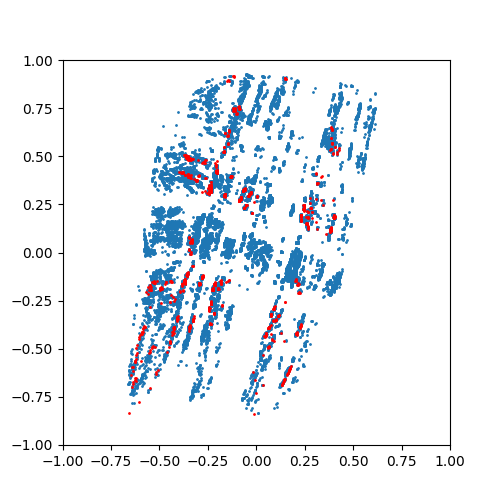}
        \vspace*{-7.0mm}
        \caption{AE-2 PE-ii A=0.0 Model 2.}
        \label{fig:Model2A0}
      \end{minipage} &
      \begin{minipage}[t]{0.35\hsize}
        \centering
        \includegraphics[keepaspectratio, scale=0.40]{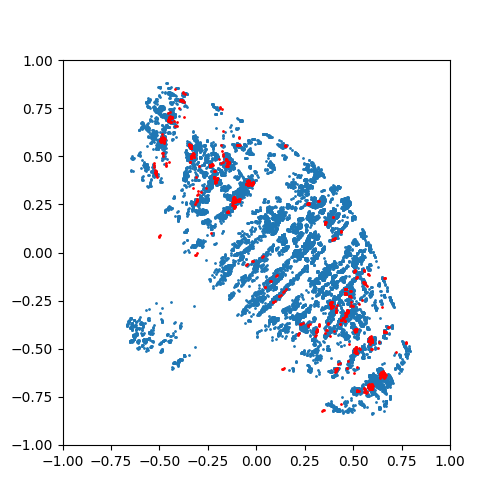}
        \vspace*{-7.0mm}
        \caption{AE-2 PE-ii A=0.2 Model 2.}
        \label{fig:Model2A1}
      \end{minipage} &
      \begin{minipage}[t]{0.35\hsize}
        \centering
        \includegraphics[keepaspectratio, scale=0.40]{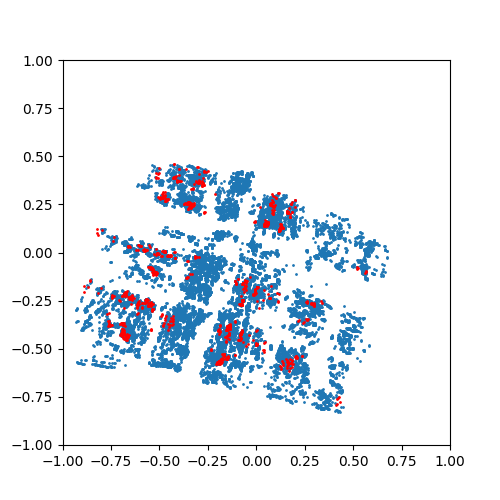}
        \vspace*{-7.0mm}
        \caption{AE-2 PE-ii A=0.4 Model 2.}
        \label{fig:Model2A2}
      \end{minipage}\\
      \begin{minipage}[t]{0.35\hsize}
        \centering
        \includegraphics[keepaspectratio, scale=0.40]{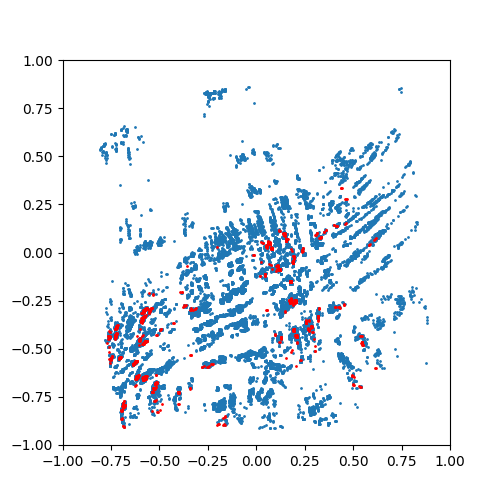}
        \vspace*{-7.0mm}
        \caption{AE-2 PE-ii A=0.6 Model 2.}
        \label{fig:Model2A3}
      \end{minipage} &
      \begin{minipage}[t]{0.35\hsize}
        \centering
        \includegraphics[keepaspectratio, scale=0.40]{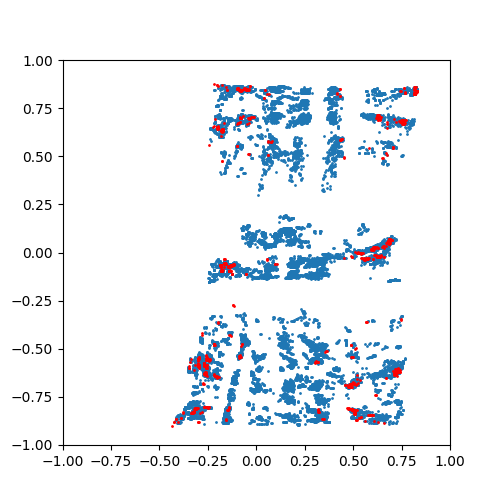}
        \vspace*{-7.0mm}
        \caption{AE-2 PE-ii A=0.8 Model 2.}
        \label{fig:Model2A4}
      \end{minipage} &
      \begin{minipage}[t]{0.35\hsize}
        \centering
        \includegraphics[keepaspectratio, scale=0.40]{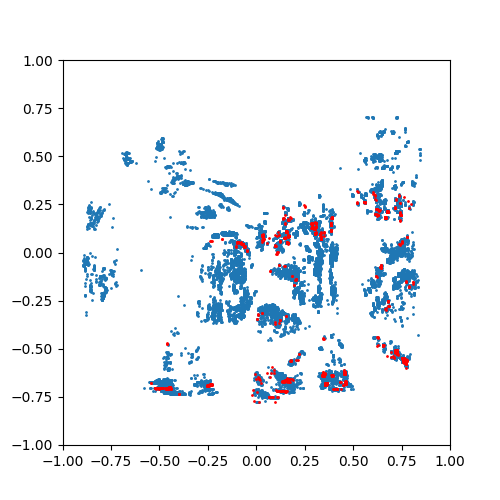}
        \vspace*{-7.0mm}
        \caption{AE-2 PE-ii A=1.0 Model 2.}
        \label{fig:Model2A5}
      \end{minipage} \\ 
      \begin{minipage}[t]{0.35\hsize}
        \centering
        \includegraphics[keepaspectratio, scale=0.40]{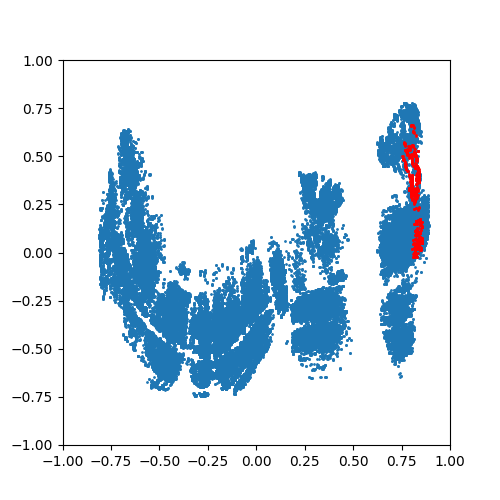}
        \vspace*{-7.0mm}
        \caption{AE-2 PE-ii A=0.0 Model 3.}
        \label{fig:Model3A0}
      \end{minipage} &
      \begin{minipage}[t]{0.35\hsize}
        \centering
        \includegraphics[keepaspectratio, scale=0.40]{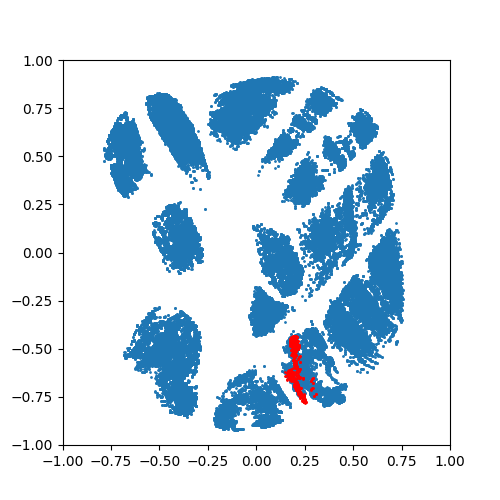}
        \vspace*{-7.0mm}
        \caption{AE-2 PE-ii A=0.2 Model 3.}
        \label{fig:Model3A1}
      \end{minipage} &
      \begin{minipage}[t]{0.35\hsize}
        \centering
        \includegraphics[keepaspectratio, scale=0.40]{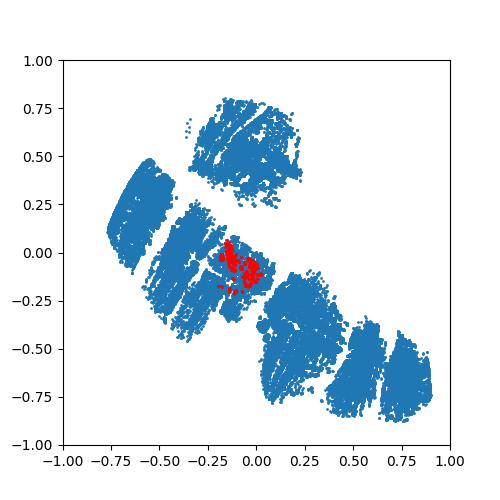}
        \vspace*{-7.0mm}
        \caption{AE-2 PE-ii A=0.4 Model 3.}
        \label{fig:Model3A2}
      \end{minipage}\\
      \begin{minipage}[t]{0.35\hsize}
        \centering
        \includegraphics[keepaspectratio, scale=0.40]{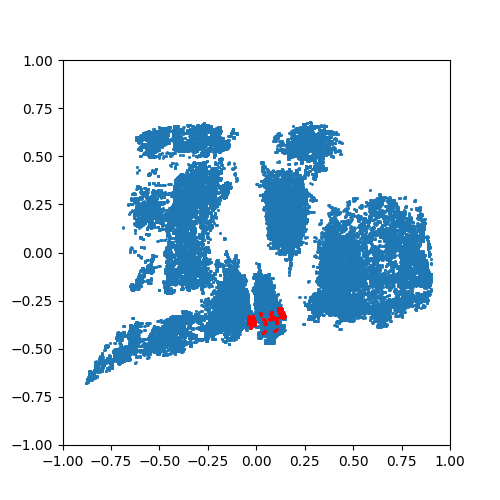}
        \vspace*{-7.0mm}
        \caption{AE-2 PE-ii A=0.6 Model 3.}
        \label{fig:Model3A3}
      \end{minipage} &
      \begin{minipage}[t]{0.35\hsize}
        \centering
        \includegraphics[keepaspectratio, scale=0.40]{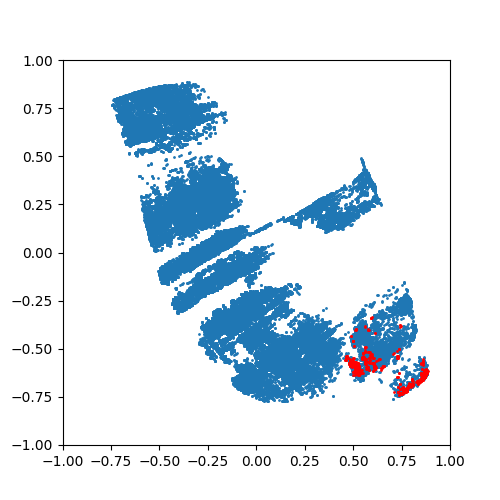}
        \vspace*{-7.0mm}
        \caption{AE-2 PE-ii A=0.8 Model 3.}
        \label{fig:Model3A4}
      \end{minipage} &
      \begin{minipage}[t]{0.35\hsize}
        \centering
        \includegraphics[keepaspectratio, scale=0.40]{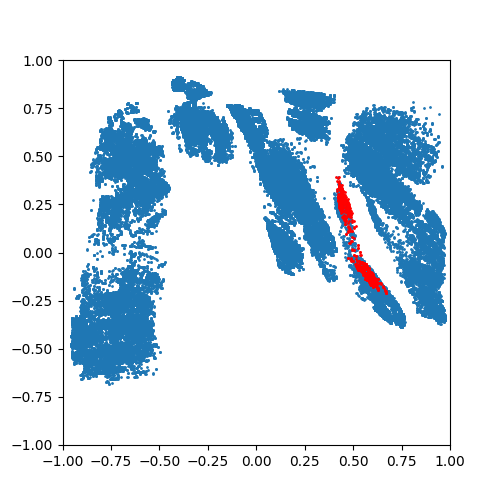}
        \vspace*{-7.0mm}
        \caption{AE-2 PE-ii A=1.0 Model 3.}
        \label{fig:Model3A5}
      \end{minipage}    
    \end{tabular}
    
\end{figure}

The outputs from the latent layers are shown in \Cref{fig:Model2A0,fig:Model2A1,fig:Model2A2,fig:Model2A3,fig:Model2A4,fig:Model2A5} for Model 2 and \Cref{fig:Model3A0,fig:Model3A1,fig:Model3A2,fig:Model3A3,fig:Model3A4,fig:Model3A5} for Model 3.
Although the clustering patterns become vague for these cases, one can see the red points, which represent the aligned configurations, do not scatter universally.
Thus, the autoencoder models distinguish the aligned configurations from the others, at least at the cluster level, as in the previous case.

\newpage

\newpage

\begin{figure}[H]
\centering
    \begin{tabular}{cc}
    \begin{minipage}[t]{0.45\hsize}
        \centering
        \includegraphics[keepaspectratio, width=6.5cm]{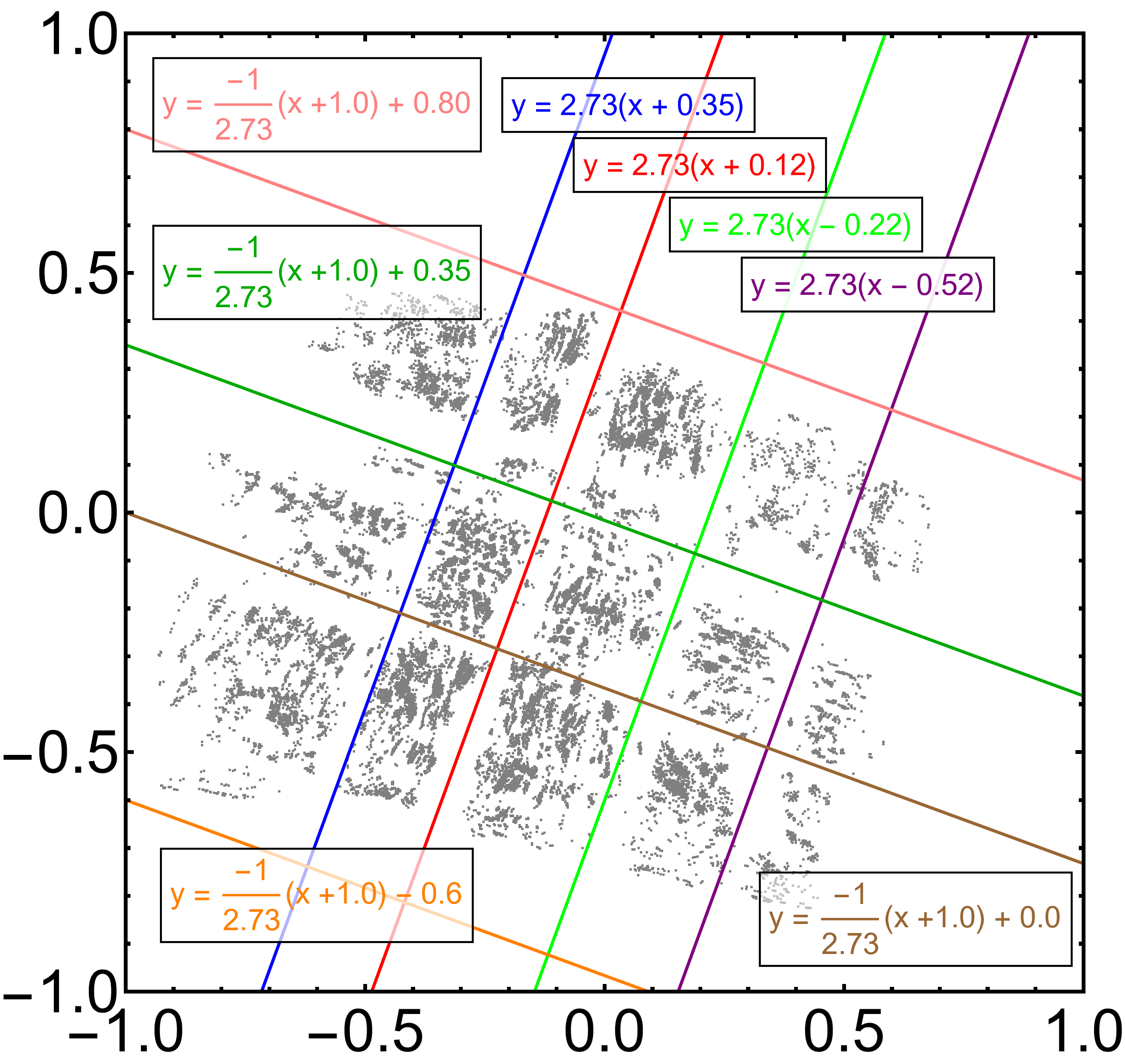}
        \vspace*{-3mm}\caption{Our classification of the data. The output data is separated by the lines shown.}
        \label{fig:clustering-u2-u1ys}
    \end{minipage} &
    \begin{minipage}[t]{0.45\hsize}
        \centering
        \includegraphics[keepaspectratio, width=6.5cm]{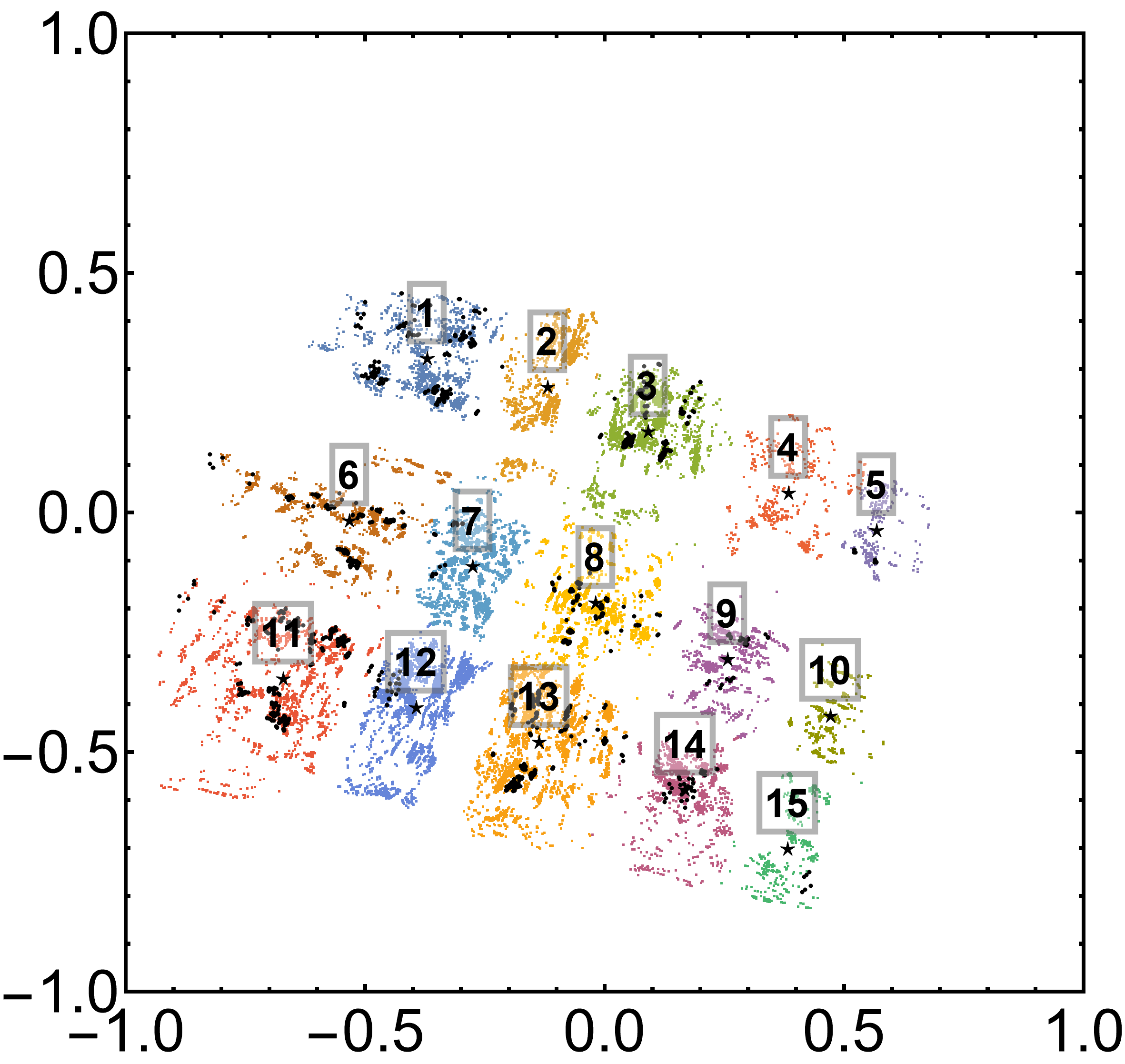}
        \vspace*{-3mm}\caption{The resultant of the classification. Here, the black dots present the aligned configurations.}
        \label{fig:u2-y1s-clusteringoverview}
    \end{minipage}
    \end{tabular}
\end{figure}

\paragraph{Model 2}\mbox{}\\
For Model 2, we choose the $A = 0.4$ case since the clustering pattern is rather clear among the results.
In this case, we observe the checkerboard pattern again, although it is vague.
We assume that several lines drawn by hand classify the clusters appropriately instead of the fine-tuning of clustering algorithms.
In the following, our discussion is based on the clustering illustrated in \Cref{fig:clustering-u2-u1ys,fig:u2-y1s-clusteringoverview}.
We also summarize the number of configurations in each cluster in \Cref{tab:number of each cluster-model 2}.

\begin{table}[H]
\centering
\scalebox{0.7}{
    \begin{tabular}{|c||cccccccccc|}\hline
        Cluster No.& 1& 2& 3& 4& 5& 6& 7& 8& 9& 10\\ \hline
        Aligned/Total& 146/1457& 4/1617& 150/2569& 0/475& 6/316& 122/1512& 14/2436& 132/2667& 24/963& 0/393\\ \hline
        Cluster No.& 11& 12& 13& 14& 15& & & & &  \\ \hline
        Aligned/Total& 179/2072& 18/2883& 190/3347& 39/1270& 6/359& & & & & \\ \hline
    \end{tabular}
    }
    \caption{The numbers of the configurations in each cluster.}
    \label{tab:number of each cluster-model 2}
\end{table}

\begin{figure}[H]
\centering
    \begin{tabular}{cc}
    \begin{minipage}[t]{0.45\hsize}
        \centering
        \includegraphics[keepaspectratio, width=6.5cm]{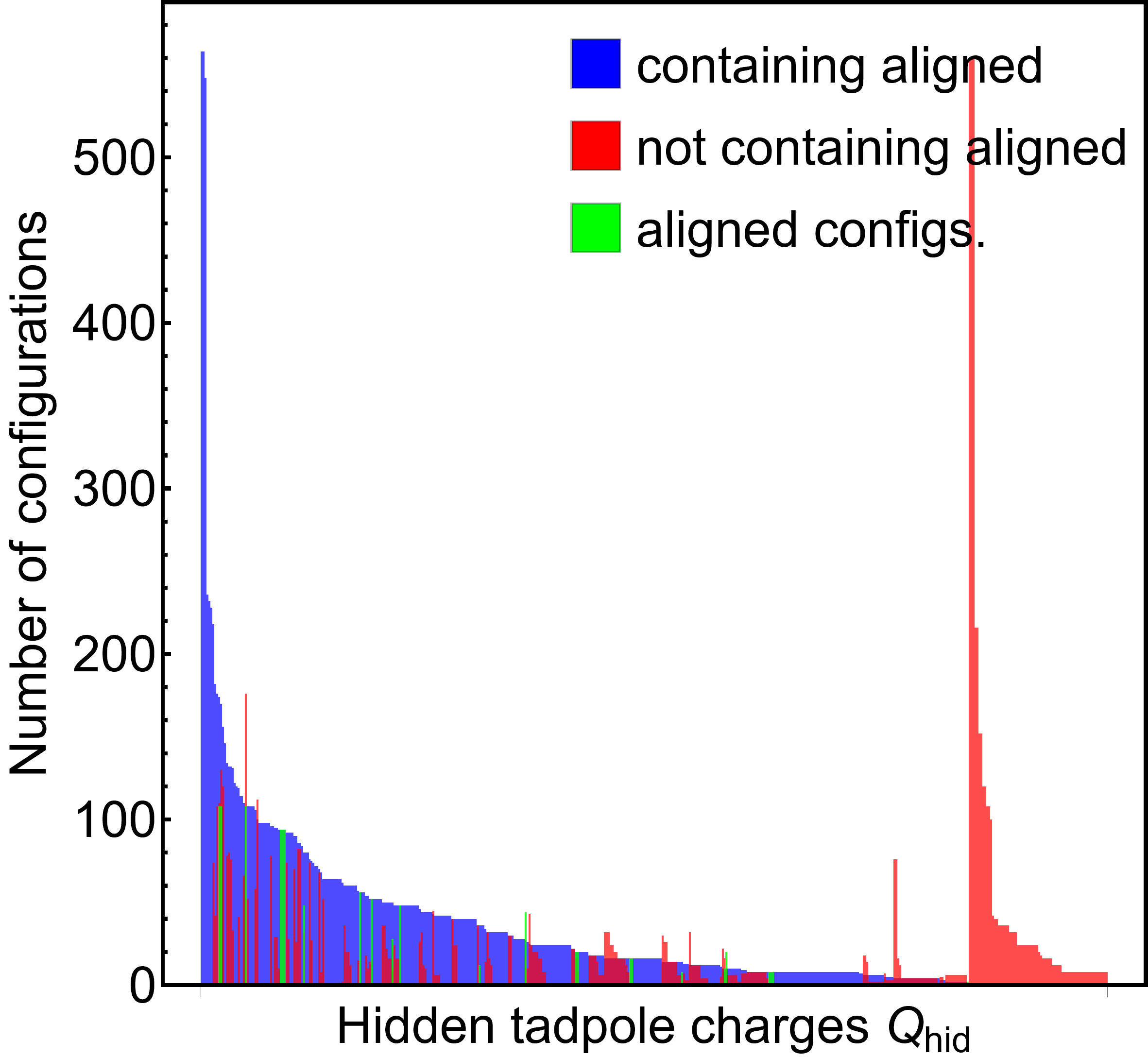}
        \vspace*{-3mm}\caption{The distributions of the number of configurations for each hidden tadpole charge. Here, the ``not containing aligned'' category is defined to include Clusters 2, 4, 5, 7, 10, 12, and 15, where the aligned configurations are less than $2\%$. In this case, the displacement of two distributions can be observed.}
        \label{fig:distributionofTC-model2}
    \end{minipage} &
    \begin{minipage}[t]{0.45\hsize}
        \centering
        \includegraphics[keepaspectratio, width=6.5cm]{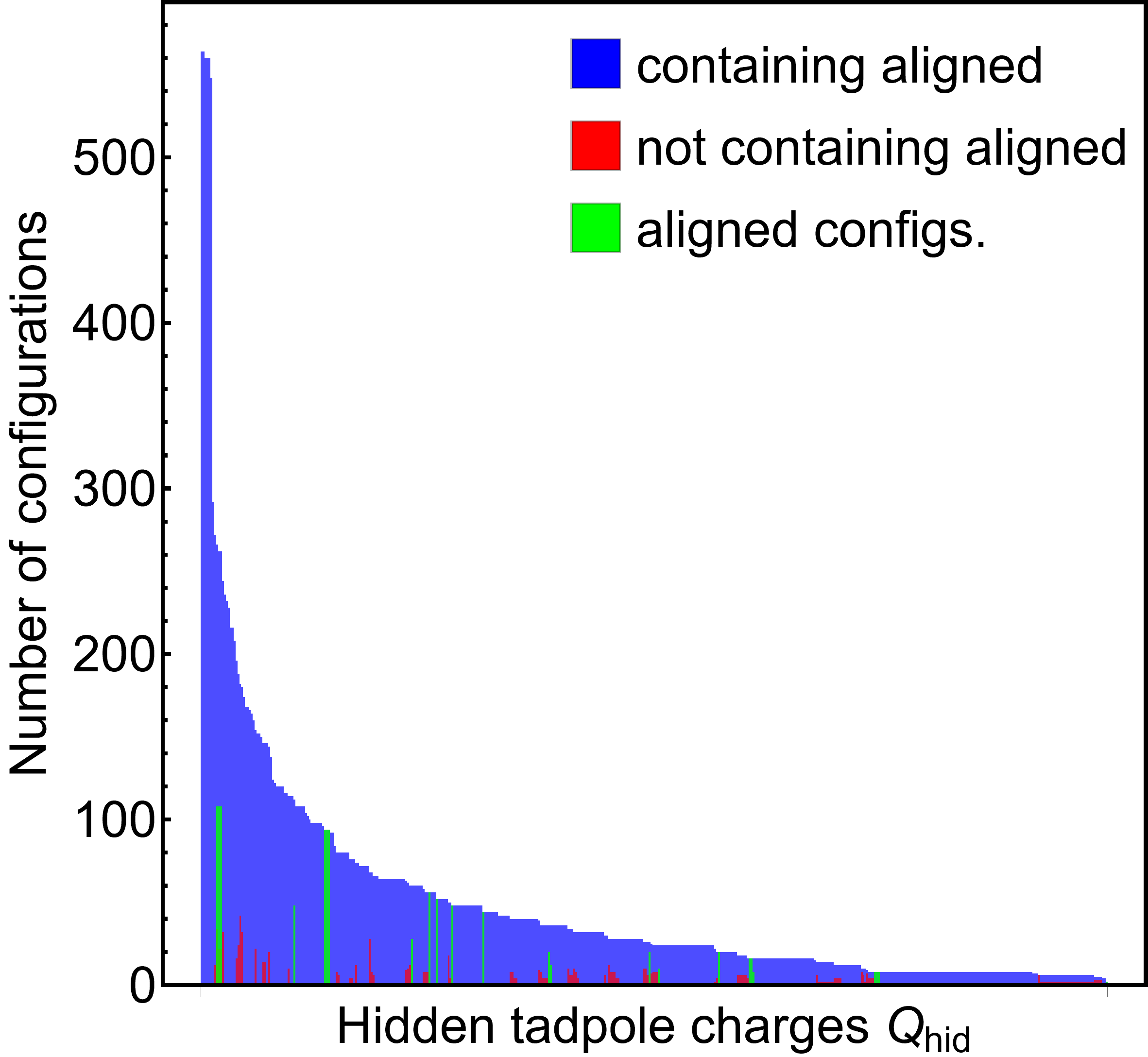}
        \vspace*{-3mm}\caption{As in the left panel, the distributions of the number of configurations for each hidden tadpole charge are shown. However, the ``not containing aligned'' category includes only Clusters 4 and 10, where the numbers of aligned configurations are exactly zero. In this case, we do not observe the displacement.}
        \label{fig:distributionofTC-model2-model2-strong}
    \end{minipage}
    \end{tabular}
\end{figure}

As in the previous case, we would like to distinguish the clusters which contain the aligned configurations from the other clusters.
We see some clusters have few aligned configurations, e.g., they are $\leq 2\%$ in Cluster 5.
Thus it is difficult to perform the binary classification in a natural way.
Indeed, we cannot observe the displacement of two peaks as shown in \Cref{fig:distributionofTC,fig:distributionofTC-A1} in this case.
We summarize this result in \Cref{fig:distributionofTC-model2,fig:distributionofTC-model2-model2-strong}.
As one can see, if we impose a threshold to be $2\%$ in classifying into the two categories, the displacement of two peaks is also seen in this case.
However, if we strictly make the threshold to be zero, the displacement does not appear.
Thus, we can conclude that for Model 2, the previous clustering of the hidden tadpole charges occurs approximately, but the effects of other undiscovered features become stronger than in the previous case.

Indeed, we find the phenomenon of configurations being arranged in lines again in this case.
To quantify this observation, we calculate the angular (cosine) similarity matrix:
\begin{align}
    S_{ij} \coloneqq \arccos{\frac{{\mathbf{n}}_i \cdot {\mathbf{n}}_j}{|{\mathbf{n}_i}| |{\mathbf{n}_j}| }},
\end{align}
where the vector ${\bf n}_i$ is the numbers of different configurations for each hidden tadpole charge; We assign each hidden tadpole charge an index $a$, and the $a$-th element of ${\bf n}_i$ is the number of configurations which have the $a$-th hidden tadpole charge.
The elements of the angular similarity matrix are summarized in \Cref{tab:angular similarity - model 2}.
We colored blue the values which are the smallest to the third in each row to emphasize the near clusters.
Except for (row, column) $=$ (14, 4), one can find that the blued positions are aligned diagonally. For example, 1, 6 and 11 are colored in the first low.
Then, let us remind \Cref{fig:u2-y1s-clusteringoverview}.
These Clusters 1, 6, and 11 are arranged diagonally in the latent layer.
Similarly, one can find from \Cref{tab:angular similarity - model 2} that $(2, 7, 12), (3, 8, 13), (4, 9, 14)$ and $(5, 10, 15)$ are colored blue as triads.
Hence, the angular cosine similarity reflects the phenomenon among configurations, which can also be seen in Model 1.
As mentioned in the previous case, this vertical correlation should be related to the emergence of the checkerboard pattern and implies the existence of at least one other important factor.

The exception at (14, 4) is also noteworthy.
Let us note that the element at (14, 12) has a slightly smaller value compared to the element at (14, 4). 
Similarly, the value at (9, 4) is colored but it is very close to (9, 12). 
This is irregular if clusters on vertical lines are indeed similar to each other, but Cluster 4 has a property that it does not have the aligned configurations.
Thus, we may conclude that the non-existence of the aligned configurations affects the similarity.

\begin{table}[H]
    \centering
    \scalebox{0.6}{
    \begin{tabular}{|c||c|c|c|c|c|c|c|c|c|c|c|c|c|c|c|}
    \hline
    No.& 1 & 2 & 3 & 4 & 5 & 6 & 7 & 8 & 9 & 10 & 11 & 12 & 13 & 14 & 15 \\ \hhline{|=#=|=|=|=|=|=|=|=|=|=|=|=|=|=|=|}
    1& \cellcolor[rgb]{0.9, 0.98, 1.0}\bf{\blue{0}} & 0.476 & 0.361 & 0.418 & 0.466 & \cellcolor[rgb]{0.9, 0.98, 1.0}\bf{\blue{0.340}} & 0.475 & 0.446 & 0.428 & 0.477 & \cellcolor[rgb]{0.9, 0.98, 1.0}\bf{\blue{0.272}} & 0.473 & 0.387 & 0.443 & 0.479 \\ \hline
    2&   0.476 &\cellcolor[rgb]{0.9, 0.98, 1.0}\bf{\blue{0}} & 0.48 & 0.449 & 0.453 & 0.47 & \cellcolor[rgb]{0.9, 0.98, 1.0}\bf{\blue{0.234}} & 0.49 & 0.474 & 0.469 & 0.476 & \cellcolor[rgb]{0.9, 0.98, 1.0}\bf{\blue{0.263}} & 0.482 & 0.478 & 0.469 \\ \hline
    3&   0.361 & 0.48 & \cellcolor[rgb]{0.9, 0.98, 1.0}\bf{\blue{0}} & 0.426 & 0.468 & 0.42 & 0.484 & \cellcolor[rgb]{0.9, 0.98, 1.0}\bf{\blue{0.255}} & 0.457 & 0.477 & 0.399 & 0.477 & \cellcolor[rgb]{0.9, 0.98, 1.0}\bf{\blue{0.194}} & 0.459 & 0.463 \\ \hline
    4& 0.418 & 0.449 & 0.426 & \cellcolor[rgb]{0.9, 0.98, 1.0}\bf{\blue{0}} & 0.49 & 0.44 & 0.466 & 0.444 & \cellcolor[rgb]{0.9, 0.98, 1.0}\bf{\blue{0.399}} & 0.489 & 0.466 & 0.455 & 0.452 & \cellcolor[rgb]{0.9, 0.98, 1.0}\bf{\blue{0.392}} & 0.494 \\ \hline
    5&   0.466 & 0.453 & 0.468 & 0.49 & \cellcolor[rgb]{0.9, 0.98, 1.0}\bf{\blue{0}} & 0.41 & 0.47 & 0.465 & 0.472 & \cellcolor[rgb]{0.9, 0.98, 1.0}\bf{\blue{0.295}} & 0.412 & 0.473 & 0.447 & 0.473 & \cellcolor[rgb]{0.9, 0.98, 1.0}\bf{\blue{0.279}} \\ \hline
    6&   \cellcolor[rgb]{0.9, 0.98, 1.0}\bf{\blue{0.340}} & 0.47 & 0.42 & 0.44 & 0.41 & \cellcolor[rgb]{0.9, 0.98, 1.0}\bf{\blue{0}} & 0.479 & 0.387 & 0.467 & 0.408 & \cellcolor[rgb]{0.9, 0.98, 1.0}\bf{\blue{0.285}} & 0.462 & 0.373 & 0.45 & 0.421 \\ \hline
    7&   0.475 & \cellcolor[rgb]{0.9, 0.98, 1.0}\bf{\blue{0.234}} & 0.484 & 0.466 & 0.47 & 0.479 & \cellcolor[rgb]{0.9, 0.98, 1.0}\bf{\blue{0}} & 0.487 & 0.415 & 0.452 & 0.475 & \cellcolor[rgb]{0.9, 0.98, 1.0}\bf{\blue{0.165}} & 0.481 & 0.422 & 0.463 \\ \hline
    8&   0.446 & 0.49 & \cellcolor[rgb]{0.9, 0.98, 1.0}\bf{\blue{0.255}} & 0.444 & 0.465 & 0.387 & 0.487 & \cellcolor[rgb]{0.9, 0.98, 1.0}\bf{\blue{0}} & 0.464 & 0.468 & 0.427 & 0.478 & \cellcolor[rgb]{0.9, 0.98, 1.0}\bf{\blue{0.213}} & 0.458 & 0.471 \\ \hline
    9& 0.428 & 0.474 & 0.457 & \cellcolor[rgb]{0.9, 0.98, 1.0}\bf{\blue{0.399}} & 0.472 & 0.467 & 0.415 & 0.464 & \cellcolor[rgb]{0.9, 0.98, 1.0}\bf{\blue{0}} & 0.486 & 0.445 & 0.4 & 0.459 & \cellcolor[rgb]{0.9, 0.98, 1.0}\bf{\blue{0.179}} & 0.481 \\ \hline
    10& 0.477 & 0.469 & 0.477 & 0.489 & \cellcolor[rgb]{0.9, 0.98, 1.0}\bf{\blue{0.295}} & 0.408 & 0.452 & 0.468 & 0.486 & \cellcolor[rgb]{0.9, 0.98, 1.0}\bf{\blue{0}} & 0.429 & 0.464 & 0.459 & 0.479 & \cellcolor[rgb]{0.9, 0.98, 1.0}\bf{\blue{0.256}} \\ \hline
    11& \cellcolor[rgb]{0.9, 0.98, 1.0}\bf{\blue{0.272}} & 0.476 & 0.399 & 0.466 & 0.412 & \cellcolor[rgb]{0.9, 0.98, 1.0}\bf{\blue{0.285}} & 0.475 & 0.427 & 0.445 & 0.429 & \cellcolor[rgb]{0.9, 0.98, 1.0}\bf{\blue{0}} & 0.48 & 0.323 & 0.464 & 0.415 \\ \hline
    12& 0.473 & \cellcolor[rgb]{0.9, 0.98, 1.0}\bf{\blue{0.263}} & 0.477 & 0.455 & 0.473 & 0.462 & \cellcolor[rgb]{0.9, 0.98, 1.0}\bf{\blue{0.165}} & 0.478 & 0.4 & 0.464 & 0.48 & \cellcolor[rgb]{0.9, 0.98, 1.0}\bf{\blue{0}} & 0.473 & 0.384 & 0.459 \\ \hline
    13& 0.387 & 0.482 & \cellcolor[rgb]{0.9, 0.98, 1.0}\bf{\blue{0.194}} & 0.452 & 0.447 & 0.373 & 0.481 & \cellcolor[rgb]{0.9, 0.98, 1.0}\bf{\blue{0.213}} & 0.459 & 0.459 & 0.323 & 0.473 & \cellcolor[rgb]{0.9, 0.98, 1.0}\bf{\blue{0}} & 0.46 & 0.449 \\ \hline
    14 &0.443 & 0.478 & 0.459 & 0.392 & 0.473 & 0.45 & 0.422 & 0.458 & \cellcolor[rgb]{0.9, 0.98, 1.0}\bf{\blue{0.179}} & 0.479 & 0.464 & \cellcolor[rgb]{0.9, 0.98, 1.0}\bf{\blue{0.384}} & 0.46 & \cellcolor[rgb]{0.9, 0.98, 1.0}\bf{\blue{0}} & 0.486 \\ \hline
    15 & 0.479 & 0.469 & 0.463 & 0.494 & \cellcolor[rgb]{0.9, 0.98, 1.0}\bf{\blue{0.279}} & 0.421 & 0.463 & 0.471 & 0.481 & \cellcolor[rgb]{0.9, 0.98, 1.0}\bf{\blue{0.256}} & 0.415 & 0.459 & 0.449 & 0.486 & \cellcolor[rgb]{0.9, 0.98, 1.0}\bf{\blue{0}} \\ \hline
    \end{tabular}
    }
    \caption{The angular cosine similarities (rad) between the clusters. Both the rows and the columns represent the clusters, as indicated by the cluster numbers. The blue color indicates the values are the smallest to the third in each row, i.e., the values are colored blue to the third nearest cluster.}
    \label{tab:angular similarity - model 2}
\end{table}

\begin{figure}[H]
\centering
    \begin{tabular}{cc}
    \begin{minipage}[t]{0.45\hsize}
        \centering
        \includegraphics[keepaspectratio, width=6.5cm]{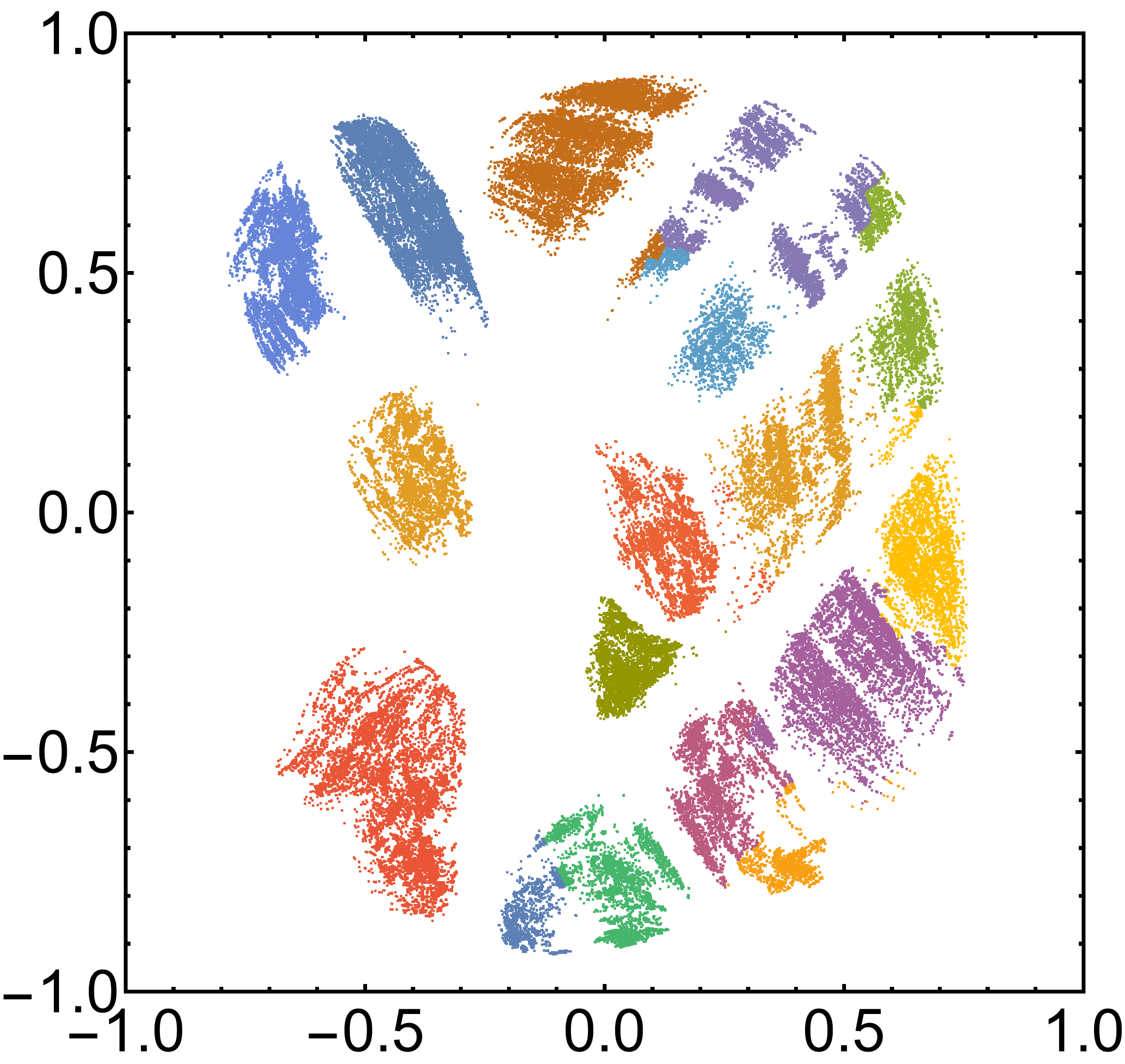}
        \vspace*{-3mm}\caption{\texttt{MeanShift}.}
        \label{fig:MeanShift-Model3}
    \end{minipage} &
    \begin{minipage}[t]{0.45\hsize}
        \centering
        \includegraphics[keepaspectratio, width=6.5cm]{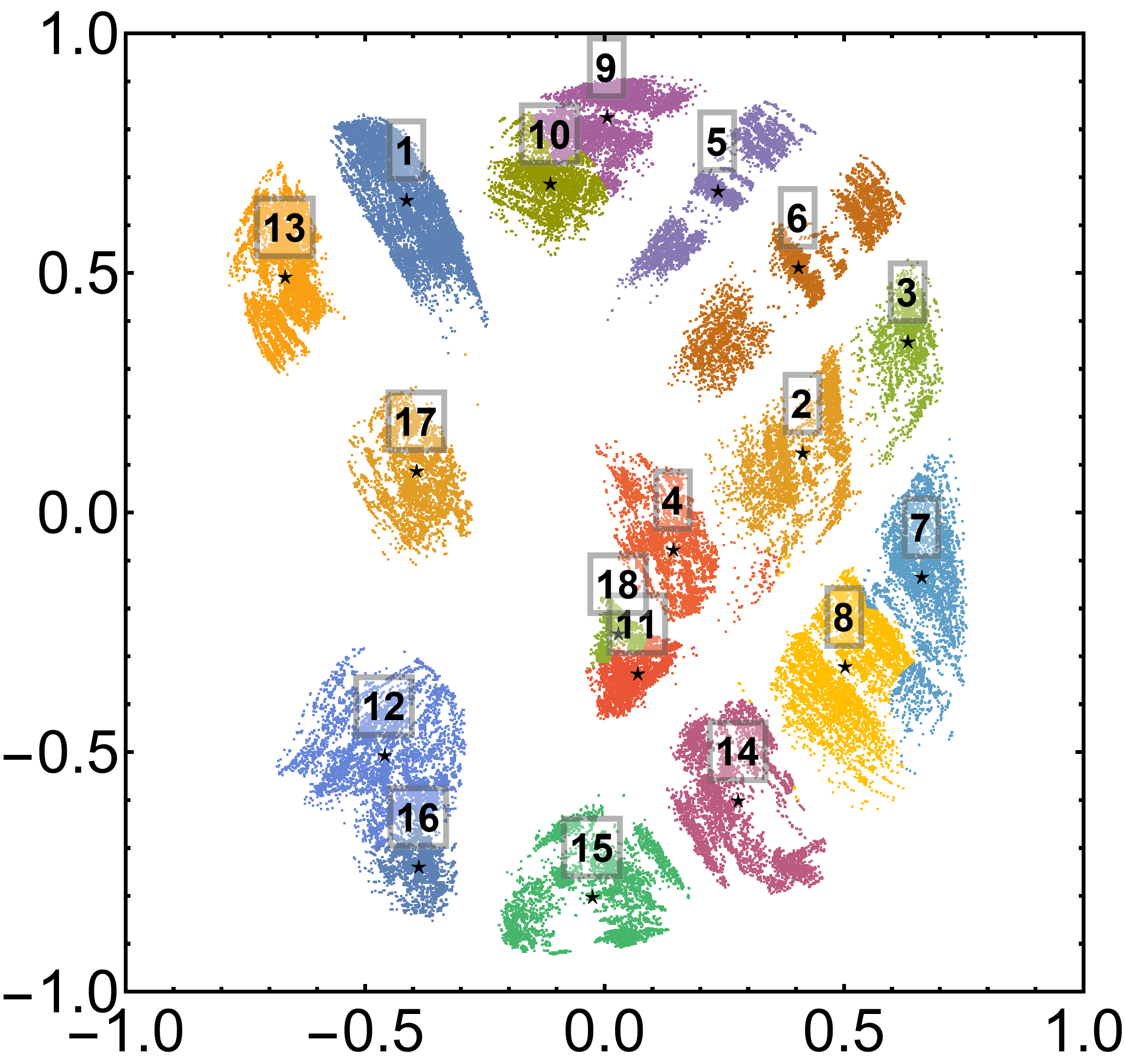}
        \vspace*{-3mm}\caption{\texttt{JarvisPatrick} with radius 0.127, with labels of the clusters.}
        \label{fig:JarvisPatrick-Model3}
    \end{minipage}
    \end{tabular}
\end{figure}

\paragraph{Model 3}\mbox{}\\
For Model 3, we find that the aligned configurations cluster in smaller regions compared with the other models.
On the other hand, there is no example among \Cref{fig:Model3A0,fig:Model3A1,fig:Model3A2,fig:Model3A3,fig:Model3A4,fig:Model3A5} which exhibits a clear checkerboard pattern.
Thus, we need to consider a clustering method which is different from that in the previous case.
We choose the $A = 0.2$ one (\Cref{fig:Model3A1}) since the clustering pattern is rather clear among the results.
We have almost no criteria (except for the apparent boundaries) to make clusters, hence we would like to rely on the clustering methods as we first did in Model 1.
The clustering highly depends on fine-tuning of the algorithm parameters again.
We show two results in \Cref{fig:MeanShift-Model3,fig:JarvisPatrick-Model3} with different algorithms for example.
Both results do not match well with a naive observation, we would like to choose \texttt{JarvisPatrick} method in \Cref{fig:JarvisPatrick-Model3} which is more suitable for our purpose since it distinguishes the cluster containing aligned contributions (which can be seen in \Cref{fig:Model3A1}) from others.

Reflecting the non-appearance of the checkerboard pattern, the angular similarity matrix of the hidden tadpole charge vectors does not exhibit the diagonal pattern which is observed in Model 2.
On the other hand, it is observed that the elements of the angular matrix form clusters for each row.
We summarize the clustering in \Cref{tab:angular similarity - model 3}. 
One can see that each row is classified into two or three clusters, and it implies that similar hidden tadpole charges are assembled in each cluster of clusters.
Although the diagonal pattern becomes vague in this case, we consider this structure to be reminiscent of the pattern.

\begin{table}[H]
    \centering
    \hspace*{-15mm}
    \scalebox{0.6}{
    \begin{tabular}{|c||c|c|c|c|c|c|c|c|c|c|c|c|c|c|c|c|c|c|}
    \hline
        No. & 1 & 2 & 3 & 4 & 5 & 6 & 7 & 8 & 9 & 10 & 11 & 12 & 13 & 14 & 15 & 16 & 17 & 18 \\ \hhline{|=#=|=|=|=|=|=|=|=|=|=|=|=|=|=|=|=|=|=|}
        1 & \cellcolor[rgb]{0.9, 0.98, 1.0}\bf{\blue{0}} & \cellcolor[rgb]{1.0, 1.0, 0.8}\bf{\green{0.264}} & \cellcolor[rgb]{1.0, 1.0, 0.8}\bf{\green{0.282}} & \black{0.482} & \black{0.473} & \black{0.473} & \black{0.468} & \black{0.479} & \black{0.475} & \black{0.475} & \cellcolor[rgb]{1.0, 1.0, 0.8}\bf{\green{0.206}} & \black{0.49} & \black{0.488} & \black{0.499} & \black{0.499} & \black{0.487} & \black{0.485} & \cellcolor[rgb]{1.0, 1.0, 0.8}\bf{\green{0.245}} \\ \hline
        2 & \cellcolor[rgb]{0.9, 0.98, 1.0}\bf{\blue{0.264}} & \cellcolor[rgb]{0.9, 0.98, 1.0}\bf{\blue{0}} & \cellcolor[rgb]{0.9, 0.98, 1.0}\bf{\blue{0.217}} & \black{0.493} & \black{0.464} & \black{0.464} & \black{0.46} & \black{0.478} & \black{0.469} & \black{0.477} & \black{0.459} & \black{0.498} & \black{0.497} & \black{0.499} & \black{0.499} & \black{0.496} & \black{0.497} & \black{0.473} \\ \hline
        3 & \cellcolor[rgb]{0.9, 0.98, 1.0}\bf{\blue{0.282}} & \cellcolor[rgb]{0.9, 0.98, 1.0}\bf{\blue{0.217}} & \cellcolor[rgb]{0.9, 0.98, 1.0}\bf{\blue{0}} & \black{0.489} & \black{0.473} & \black{0.473} & \black{0.467} & \black{0.478} & \black{0.477} & \black{0.468} & \black{0.458} & \black{0.496} & \black{0.495} & \black{0.498} & \black{0.498} & \black{0.494} & \black{0.494} & \black{0.475} \\ \hline
        4 & \black{0.482} & \black{0.493} & \black{0.489} & \cellcolor[rgb]{0.9, 0.98, 1.0}\bf{\blue{0}} & \black{0.499} & \black{0.499} & \black{0.494} & \black{0.496} & \black{0.496} & \black{0.493} & \black{0.482} & \black{0.483} & \black{0.487} & \black{0.500} & \black{0.500} & \black{0.496} & \cellcolor[rgb]{0.9, 0.98, 1.0}\bf{\blue{0.00793}} & \black{0.482} \\ \hline
        5 & \black{0.473} & \black{0.464} & \black{0.473} & \black{0.499} & \cellcolor[rgb]{0.9, 0.98, 1.0}\bf{\blue{0}} & \cellcolor[rgb]{0.9, 0.98, 1.0}\bf{\blue{0.000}} & \black{0.48} & \black{0.467} & \black{0.473} & \black{0.472} & \black{0.492} & \black{0.498} & \black{0.498} & \black{0.491} & \black{0.492} & \black{0.498} & \black{0.499} & \black{0.494} \\ \hline
        6 & \black{0.473} & \black{0.464} & \black{0.473} & \black{0.499} & \cellcolor[rgb]{0.9, 0.98, 1.0}\bf{\blue{0.000}} & \cellcolor[rgb]{0.9, 0.98, 1.0}\bf{\blue{0}} & \black{0.48} & \black{0.467} & \black{0.473} & \black{0.472} & \black{0.492} & \black{0.498} & \black{0.498} & \black{0.491} & \black{0.492} & \black{0.498} & \black{0.499} & \black{0.494} \\ \hline
        7 & \black{0.468} & 0.460 & \black{0.467} & \black{0.494} & 0.480 & \black{0.480} & \cellcolor[rgb]{0.9, 0.98, 1.0}\bf{\blue{0}} & \cellcolor[rgb]{1.0, 1.0, 0.8}\bf{\green{0.166}} & \cellcolor[rgb]{1.0, 1.0, 0.8}\bf{\green{0.131}} & \cellcolor[rgb]{1.0, 1.0, 0.8}\bf{\green{0.117}} & \black{0.488} & \black{0.500} & \black{0.499} & \black{0.491} & \black{0.494} & \black{0.499} & \black{0.495} & \black{0.492} \\ \hline
        8 & \black{0.479} & \black{0.478} & \black{0.478} & \black{0.496} & \black{0.467} & \black{0.467} & \cellcolor[rgb]{1.0, 1.0, 0.8}\bf{\green{0.166}} & \cellcolor[rgb]{0.9, 0.98, 1.0}\bf{\blue{0}} & \cellcolor[rgb]{0.9, 0.98, 1.0}\bf{\blue{0.0639}} & \cellcolor[rgb]{1.0, 1.0, 0.8}\bf{\green{0.165}} & \black{0.49} & \black{0.500} & \black{0.500} & \black{0.484} & \black{0.486} & \black{0.499} & \black{0.496} & \black{0.493} \\ \hline
        9 & \black{0.475} & \black{0.469} & \black{0.477} & \black{0.496} & \black{0.473} & \black{0.473} & \cellcolor[rgb]{0.9, 0.98, 1.0}\bf{\blue{0.131}} & \cellcolor[rgb]{0.9, 0.98, 1.0}\bf{\blue{0.0639}} & \cellcolor[rgb]{0.9, 0.98, 1.0}\bf{\blue{0}} & \cellcolor[rgb]{0.9, 0.98, 1.0}\bf{\blue{0.176}} & \black{0.491} & \black{0.500} & \black{0.500} & \black{0.487} & \black{0.490} & \black{0.499} & \black{0.497} & \black{0.493} \\ \hline
        10 & \black{0.475} & \black{0.477} & \black{0.468} & \black{0.493} & \black{0.472} & \black{0.472} & \cellcolor[rgb]{0.9, 0.98, 1.0}\bf{\blue{0.117}} & \cellcolor[rgb]{0.9, 0.98, 1.0}\bf{\blue{0.165}} & \cellcolor[rgb]{0.9, 0.98, 1.0}\bf{\blue{0.176}} & \cellcolor[rgb]{0.9, 0.98, 1.0}\bf{\blue{0}} & \black{0.487} & \black{0.500} & \black{0.499} & \black{0.490} & \black{0.493} & \black{0.498} & \black{0.493} & \black{0.491} \\ \hline
        11 & \cellcolor[rgb]{0.9, 0.98, 1.0}\bf{\blue{0.206}} & \black{0.459} & \black{0.458} & \black{0.482} & \black{0.492} & \black{0.492} & \black{0.488} & \black{0.490} & \black{0.491} & \black{0.487} & \cellcolor[rgb]{0.9, 0.98, 1.0}\bf{\blue{0}} & \black{0.490} & \black{0.488} & \black{0.500} & \black{0.500} & \black{0.486} & \black{0.483} & \cellcolor[rgb]{0.9, 0.98, 1.0}\bf{\blue{0.174}} \\ \hline
        12 & \black{0.490} & \black{0.498} & \black{0.496} & \black{0.483} & \black{0.498} & \black{0.498} & \black{0.500} & \black{0.500} & \black{0.500} & \black{0.5} & \black{0.490} & \cellcolor[rgb]{0.9, 0.98, 1.0}\bf{\blue{0}} & \cellcolor[rgb]{0.9, 0.98, 1.0}\bf{\blue{0.07}} & \black{0.500} & \black{0.500} & \cellcolor[rgb]{0.9, 0.98, 1.0}\bf{\blue{0.194}} & \black{0.483} & \black{0.488} \\ \hline
        13 & \black{0.488} & \black{0.497} & \black{0.495} & \black{0.487} & \black{0.498} & \black{0.498} & \black{0.499} & \black{0.500} & \black{0.500} & \black{0.499} & \black{0.488} & \cellcolor[rgb]{0.9, 0.98, 1.0}\bf{\blue{0.070}} & \cellcolor[rgb]{0.9, 0.98, 1.0}\bf{\blue{0}} & \black{0.500} & \black{0.500} & \cellcolor[rgb]{0.9, 0.98, 1.0}\bf{\blue{0.124}} & \black{0.487} & \black{0.486} \\ \hline
        14 & \black{0.499} & \black{0.499} & \black{0.498} & \black{0.500} & \black{0.491} & \black{0.491} & \black{0.491} & \black{0.484} & \black{0.487} & \black{0.490} & \black{0.500} & \black{0.500} & \black{0.500} & \cellcolor[rgb]{0.9, 0.98, 1.0}\bf{\blue{0}} & \cellcolor[rgb]{0.9, 0.98, 1.0}\bf{\blue{0.0799}} & \black{0.500} & \black{0.500} & \black{0.500} \\ \hline
        15 & \black{0.499} & \black{0.499} & \black{0.498} & \black{0.500} & \black{0.492} & \black{0.492} & \black{0.494} & \black{0.486} & \black{0.490} & \black{0.493} & \black{0.500} & \black{0.500} & \black{0.500} & \cellcolor[rgb]{0.9, 0.98, 1.0}\bf{\blue{0.0799}} & \cellcolor[rgb]{0.9, 0.98, 1.0}\bf{\blue{0}} & \black{0.500} & \black{0.500} & \black{0.500} \\ \hline
        16 & \black{0.487} & \black{0.496} & \black{0.494} & \black{0.496} & \black{0.498} & \black{0.498} & \black{0.499} & \black{0.499} & \black{0.499} & \black{0.498} & \black{0.486} & \cellcolor[rgb]{0.9, 0.98, 1.0}\bf{\blue{0.194}} & \cellcolor[rgb]{0.9, 0.98, 1.0}\bf{\blue{0.124}} & \black{0.500} & \black{0.500} & \cellcolor[rgb]{0.9, 0.98, 1.0}\bf{\blue{0}} & \black{0.496} & \black{0.485} \\ \hline
        17 & \black{0.485} & \black{0.497} & \black{0.494} & \cellcolor[rgb]{0.9, 0.98, 1.0}\bf{\blue{0.00793}} & \black{0.499} & \black{0.499} & \black{0.495} & \black{0.496} & \black{0.497} & \black{0.493} & \black{0.483} & \black{0.483} & \black{0.487} & \black{0.500} & \black{0.500} & \black{0.496} & \cellcolor[rgb]{0.9, 0.98, 1.0}\bf{\blue{0}} & \black{0.483} \\ \hline
        18 & \cellcolor[rgb]{0.9, 0.98, 1.0}\bf{\blue{0.245}} & \black{0.473} & \black{0.475} & \black{0.482} & \black{0.494} & \black{0.494} & \black{0.492} & \black{0.493} & \black{0.493} & \black{0.491} & \cellcolor[rgb]{0.9, 0.98, 1.0}\bf{\blue{0.174}} & \black{0.488} & \black{0.486} & \black{0.500} & \black{0.500}  & \black{0.485} & \black{0.483} & \cellcolor[rgb]{0.9, 0.98, 1.0}\bf{\blue{0}} \\ \hline
    \end{tabular}
    }
    \caption{The angular cosine similarities (rad) between the clusters. Both the rows and the columns represent the clusters, as indicated by the cluster numbers. In each row, the elements are again classified into new clusters, as their color indicates (the same color in different rows is independent). The clustering method is \texttt{DBSCAN} with radius $= 0.15$.}
    \label{tab:angular similarity - model 3}
\end{table}

\begin{figure}[H]
\centering
    \begin{tabular}{cc}
    \begin{minipage}[t]{0.45\hsize}
        \centering
        \includegraphics[keepaspectratio, width=6.5cm]{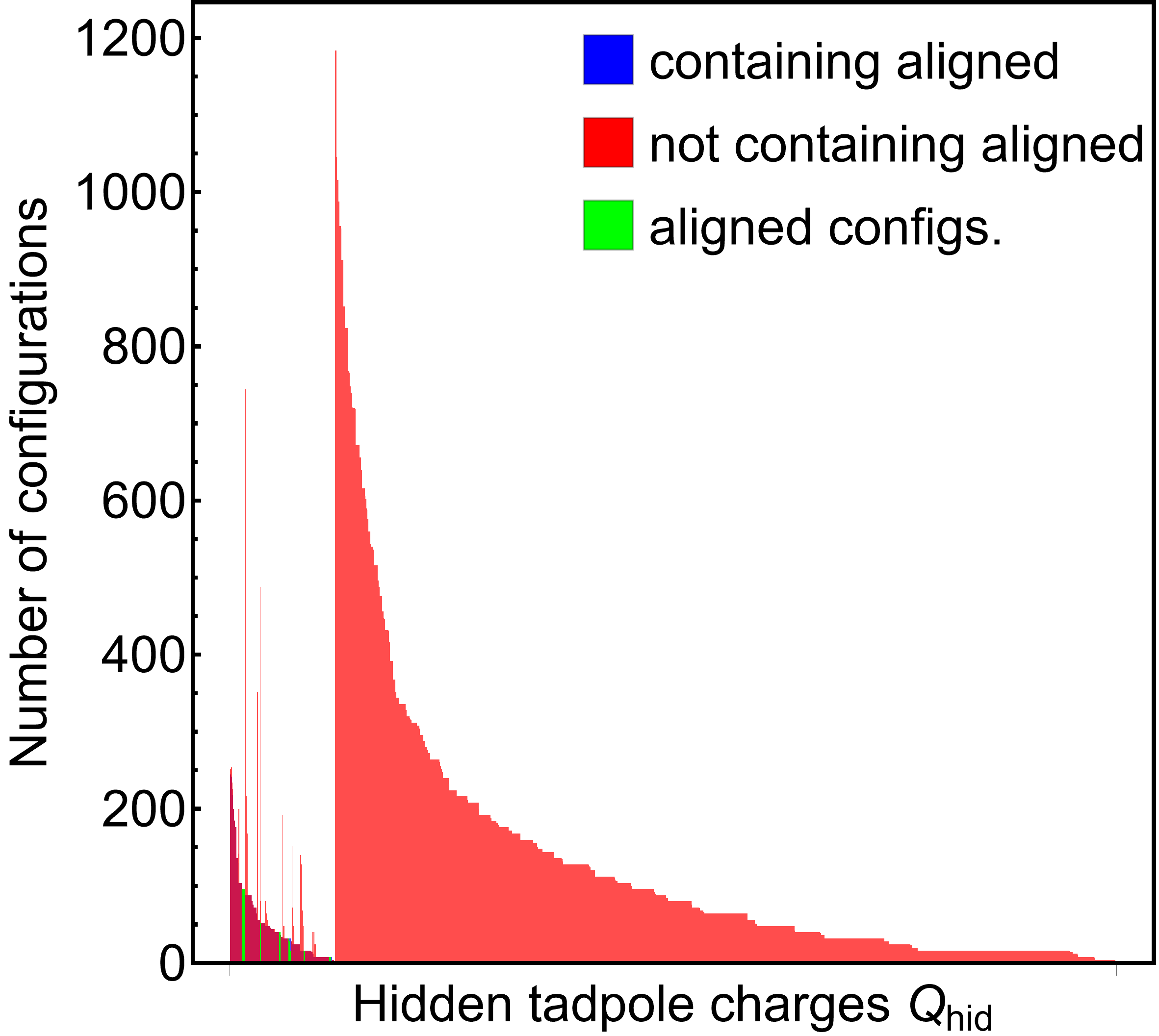}
        \vspace*{-3mm}\caption{The distributions of the number of configurations for each hidden tadpole charge. Here, the ``containing aligned'' category is exactly defined as Cluster 14. In this case, the blue bar chart almost overlaps with the red bar chart, and the displacement of two distributions cannot be observed.}
        \label{fig:distributionofTC-model3-strong}
    \end{minipage} &
    \begin{minipage}[t]{0.45\hsize}
        \centering
        \includegraphics[keepaspectratio, width=6.5cm]{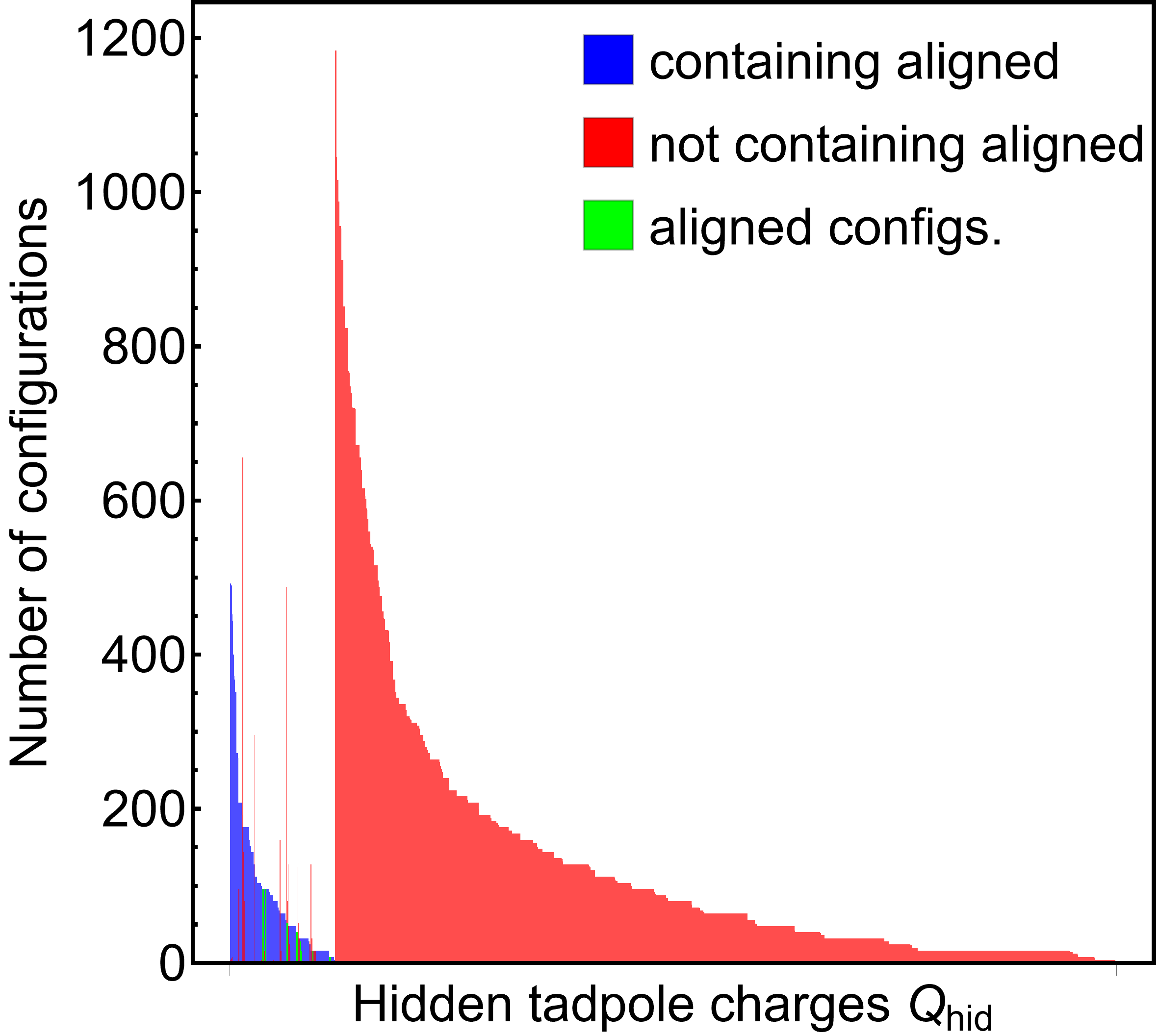}
        \vspace*{-3mm}\caption{As in the left panel, the distributions of the number of configurations for each hidden tadpole charge are shown. However, the ``containing aligned'' category includes Cluster 15 in addition to Cluster 14, which has a very large similarity with Cluster 14. In this case, we observe the displacement again.}
        \label{fig:distributionofTC-model2-model3-weak}
    \end{minipage}
    \end{tabular}
\end{figure}

If we naively draw the distributions of the hidden tadpole charges based on the binary classification corresponding to \Cref{fig:distributionofTC,fig:distributionofTC-A1,fig:distributionofTC-model2,fig:distributionofTC-model2-model2-strong}, we cannot find the displacement of two peaks.
On the other hand, Cluster 14, which is the only cluster containing the aligned configurations, has a very large similarity only with the cluster 15, as one can see in \Cref{tab:angular similarity - model 3}.
If we include Cluster 15, we can see the displacement again.
We summarize this situation in \Cref{fig:distributionofTC-model3-strong,fig:distributionofTC-model2-model3-weak}.
Since the clustering of the clusters with regard to the hidden tadpole charges is observed, and since Cluster 14 forms the cluster with only Cluster 15, we can again conclude that the hidden tadpole charge is a factor weighted by the autoencoder models, but it is expected that at least one more feature is also weighted as important as the hidden tadpole charge, as it is implied by the clustering of the clusters pattern in \Cref{tab:angular similarity - model 3}.
As we can see in \Cref{fig:JarvisPatrick-Model3}, Cluster 14 and Cluster 15 are close to each other.
Hence, the large similarity does not contradict our finding importance of the hidden tadpole charges.
However, the other feature should explain why Cluster 14 and Cluster 15 are finally distinguished from each other.
Furthermore, we find that the aligned configurations of Model 3 themselves have unique hidden tadpole charges as observed in Model 1.
The other feature will explain the reason why the unique tadpole charges and several different hidden tadpole charges of not-aligned configurations are gathered into Cluster 14, but we could not find the feature by the brute-force method.

\section{Conclusions}
\label{sec:con}
In this paper, we have investigated how the neural networks, especially the autoencoder models, learn the intersecting D-brane models.
Focusing on the models in Ref. \cite{Gmeiner:2005vz}, we generated the D-brane configurations on which the generations of quarks and leptons are aligned and not aligned, for each concrete model.
We aimed to train the autoencoder models on the datasets to classify the input data and extract important features hidden among each dataset. 
The important features are ``important'' in the sense that it is necessary to replicate the input data from the latent layer.
We consider that the important features constrain and characterize the configurations satisfying the concrete models severely.
Originally, we expected that the autoencoder models classify each dataset into the aligned configurations and the not-aligned ones.
Indeed, autoencoder models with a small latent layer are expected to extract important features of input data in the latent layer. 
When input data is classified into several different classes, the classification will be reproduced in well-trained autoencoder models.
If the number of features which characterize the differences is sufficiently small, one can observe the classification directly in the latent layer.
However, for the task which we consider in this paper, it would not be the case.
It should be difficult to observe the clean classification directly because the nontrivial and nonlinear relations which we briefly reviewed determine whether the configuration satisfies the algebra and its generation is aligned or not, for each concrete model.
Hence, the autoencoder models with very small latent layers would not classify the dataset in such a way.
Instead, the aligned and the not-aligned configurations will be gathered and classified into several clusters, depending on their macroscopic features.
This is the situation we observed in this paper.
The clustering indeed occurred, and we performed the brute-force analysis to clarify which features were weighted by the autoencoder model. 

As a result, we find that the RR tadpole charge carried by D6-branes in the hidden sector characterizes the clusters.
Indeed, it turned out that the clusters which contain the aligned configurations have characteristic distributions of the hidden tadpole charges compared to other clusters.
Specifically, the aligned configurations have unique hidden tadpole charges for Models 1 and 3.
While those are not unique for the aligned configurations in Model 2, the fact that autoencoder models focus on the hidden tadpole charges seems to be the right direction to characterize the D-brane models.
Moreover, many clusters are indeed observed in the latent layer.
It suggests that the not-aligned configurations are also characterized by the hidden tadpole charges in a nontrivial way.
This classification including the not-aligned configurations is not found by only studying the statistics of the dataset, but the autoencoder models enlighten us on the feature.
Note that we found the same feature with the different autoencoder model in Model 1.
In this way, we concluded that the hidden tadpole charge plays an important role among the datasets with the transparent discussion.

On the other hand, we observed the checkerboard patterns clearly in Models 1 and 2.
Along the vertical lines, the same distribution of hidden tadpole charges is likely to be observed, as suggested by the angular cosine similarity matrix.
While the checkerboard pattern is vague for Model 3, the similarity matrix again shows that the clusters form new clusters of clusters determined by their hidden tadpole charge distributions.
Thus, it is strongly suggested that there is at least one other feature in the dataset.
If the other feature is clarified, we may use the relationship between the hidden tadpole charges and the other feature to find new aligned configurations effectively, instead of solving the nontrivial conditions directly.
While we did not find the other feature, the clusters should exhibit characteristic distributions for the other feature as well as for the hidden tadpole charge. 
The quantity which characterizes the other feature might be constant along the horizontal lines in the checkerboard patterns. 
It is interesting to study the other features and phenomenological applications in future work, but it is better to develop architectures and their interpretability (see for reviews, e.g., Ref \cite{gilpin2018explaining}) of machine learning themselves to attack the problem.
Explainable artificial intelligence (XAI) \cite{gilpin2018explaining} is a candidate for studying this kind of problem, while it is difficult to realize such an expedient model in general.
For example, \textit{Vision Transformer} \cite{Dosovitskiy:2020qjv}, which applies the Transformer to image processing, allows the user to visually check what part of the input data was focused on. 
It would be significant to develop this technology to efficiently find features such as tadpole charges. 
Moreover, it will also be interesting to figure out the role of hidden tadpole charges in the other autoencoder-based studies such as heterotic models \cite{Mutter:2018sra,Otsuka:2020nsk,Deen:2020dlf,Escalante-Notario:2022fik}.

\acknowledgments

This work was supported by JSPS KAKENHI Grant Numbers JP20K14477 (H. O.), JP22J12877 (K. I.), JP23H04512 (H. O), and Kyushu University’s Innovator Fellowship Program (S. N.).

\appendix

\bibliography{ref}{}
\bibliographystyle{JHEP}

\end{document}